\documentclass[aps,pre,twocolumn,amsfonts,amssymb,groupedaddress,showpacs,longbibliography]{revtex4-1}
\usepackage{graphicx}
\usepackage{amsmath}
\usepackage{mathrsfs}
\usepackage{mathbbol}
\usepackage[usenames,dvipsnames]{color}

\begin{document}

\title{One-Loop Fluctuation Entropy of Charge Inversion in DNA}

\author{Matthew D. Sievert} 
   \email[Email: ]{matthew.sievert@rutgers.edu}
   \affiliation{Department of Physics and Astronomy, Rutgers University, Piscataway, NJ 08854, USA}
\author{Marilyn F. Bishop} 
   \email{bishop@vcu.edu} 
   \affiliation{Department of Physics, Virginia Commonwealth University, Richmond, Virginia 23284-2000, USA}
  
\author{Tom McMullen} 
   \email{tmcmulle@vcu.edu}
   \affiliation{Department of Physics, Virginia Commonwealth University, Richmond, Virginia 23284-2000, USA}

\date{\today}

\begin{abstract}

Experiments have revealed correlation-driven behavior of DNA in
charged solutions, including charge inversion and condensation.  This
paper presents calculations of a lattice-gas model of charge inversion for the
adsorption of charged dimers on DNA .   
Each adsorption site is assumed to have either a vacancy or a positively-charged 
dimer attached with the dimer oriented either parallel or perpendicular to the double helix DNA chain.  
The entropy and charge distributions of these three species are calculated including
the lowest order fluctuation corrections to mean-field theory.  We find that the inclusion
of the fluctuation terms has a significant effect on the entropy, primarily in the regime where the dimers are repelled from the DNA molecule and compete with the chemical potential in solution.

\end{abstract}


\maketitle

\section{Introduction}
\label{sec-Introduction}

Polyelectrolytes in solution have been shown to produce a wide array of intriguing phenomena in many different systems.  One example is DNA, which in the presence of a physiological salt solution (a 0.1 molar solution of NaCl) is usually negatively charged, with a double helix DNA molecule strongly repelling another.  However, if DNA is in a dilute salt solution in which a positively charged polyelectrolyte, such as spermine or sperimidine, has been added to the solution, it has been shown to roll up into a tightly-packed torus\cite{vab91, vab97, er99, wmg00}.  In fact, DNA is usually in a very compact state in cells and viruses.   

Another example is F-actin, whose fibrils stack in cross-linked rafts when positive alkaline earth ions are in the solution\cite{gclw03}.  Also in F-actin, counter-ions form one-dimensional charge density waves that have a periodicity equal to twice the actin monomer spacing, coupling to twist distortions of the oppositely-charged actin filaments\cite{tea03}.  These phenomena and others seem to owe their existence to Coulomb interactions between the constituent parts, and the geometry of the underlying structure may also play a vital role.

Determination of the optimal structure of any of these systems requires minimization of the free energy, which involves a competition between the internal energy and entropy.  Often in biological systems the changes in entropy are greater than the changes in internal energy.  For instance, in the hydrophobic effect, an unfolded protein lowers the entropy by ordering the water molecules, and so the protein prefers to be in the folded state\cite{jbu01,hl04}.  In the chloroplast stroma, it has been shown that there is an entropy-driven attraction that determines chloroplast ultrastructure through spontaneous Mg{$^{2+}$}-induced stacking of membranes\cite{ts10}.  In sickle hemoglobin, the aggregation of monomers into polymers is also entropy driven, with the internal energy and entropy in a delicate balance\cite{cf97}.  

Historically, the most common description of charged solutions has been
Poisson-Boltzmann theory, but continued research has indicated that ions in
solution can have far more complex and counterintuitive effects than simple
charge screening \cite{wmg00}.  In 1969 Manning proposed \cite{gsm69a,gsm69b}
that a portion of the ambient ions condense onto (i.e., attach to) the surface of
a charged macro-ion, partially neutralizing the bare charge.  This occurs up to
the point at which the energy required to condense another ion equals the
available thermal energy $k_B T$ at temperature $T$, where $k_B$ is the
Boltzmann constant.  The proposed effect, later termed ``Manning condensation,"
marked a significant departure from linearized Poisson-Boltzmann theory that
predicted only exponential screening.  In Manning's treatment of
polyelectrolytes, the long chains of charged subunits, the ion condensation
was addressed \cite{wmg00} separately from the ions that remain in solution, which were
treated using linearized Poisson-Boltzmann theory \cite{gsm69a,gsm69b}. 

Further refinements in the treatment of these ambient-ion solutions have been
motivated in part by surprising effects observed in DNA that cannot be accounted
for using Poisson-Boltzmann theory.  By increasing the charge on the salt-ions in
the DNA solution from $+e$ (monovalent) to $+2e$ (divalent) and higher, it
is possible to cause the DNA to undergo a radical structural transition into a variety of new geometries, including rod-like bundles and toroids
\cite{vab96}.  The toroidal structures, in particular, have received considerable
attention in the biological community because such toroidal packing motifs are
employed by spermidine and other molecules to contain their own DNA
in small volumes \cite{hv05}.  A variety of techniques, including
cryo-transmission electron microscopy \cite{hv05} and UV spectroscopy
\cite{rpfl06}, have enabled direct observation of the formation of these toroidal
condensates, and similar studies on other biological polyelectrolytes like
F-actin \cite{gclw03} indicate that condensation is a general phenomenon.  If
the electrical interactions between the polyelectrolyte, such as DNA, and the
ambient-ion solution are purely those predicted by the Poisson-Boltzmann
description, then two like-charged polyelectrolytes always repel one another,
and such condensates will not form.  For these condensates to be
stable, the net electrical force between like-charged polyelectrolytes must be
{$\it attractive$}, which is incompatible with the Poisson-Boltzmann theory of
ions in solution.

Furthermore, under the same conditions of divalent solution-ions, a radical change in the electrical properties of a
single DNA molecule is observed.  In
gel electrophoresis experiments \cite{rpfl06} under these conditions, DNA is
observed to drift toward the {$\it negative$} electrode under the influence of
an electric field, indicating that the net charge is {$\it positive$}.  This
difference is quantified as a change in sign of the electrophoretic mobility,
which occurs for some minimum concentration of solution-ions with valence $\geq
2$.  This change in sign cannot be explained by the Poisson-Boltzmann description
of charge screening, and, although Manning condensation can reduce the effective
charge, it cannot invert it \cite{gns02}.  As with polyelectrolyte condensation,
charge inversion requires a fuller explanation of the ambient-ion solution than
Poisson-Boltzmann theory can provide.  Understanding the nature of these
phenomena is important not only for the insight into the natural role of DNA in a
cell, but also for use in medical techniques, such as gene therapy \cite{ls00}.

\begin{figure}
\includegraphics[width=1.5in]{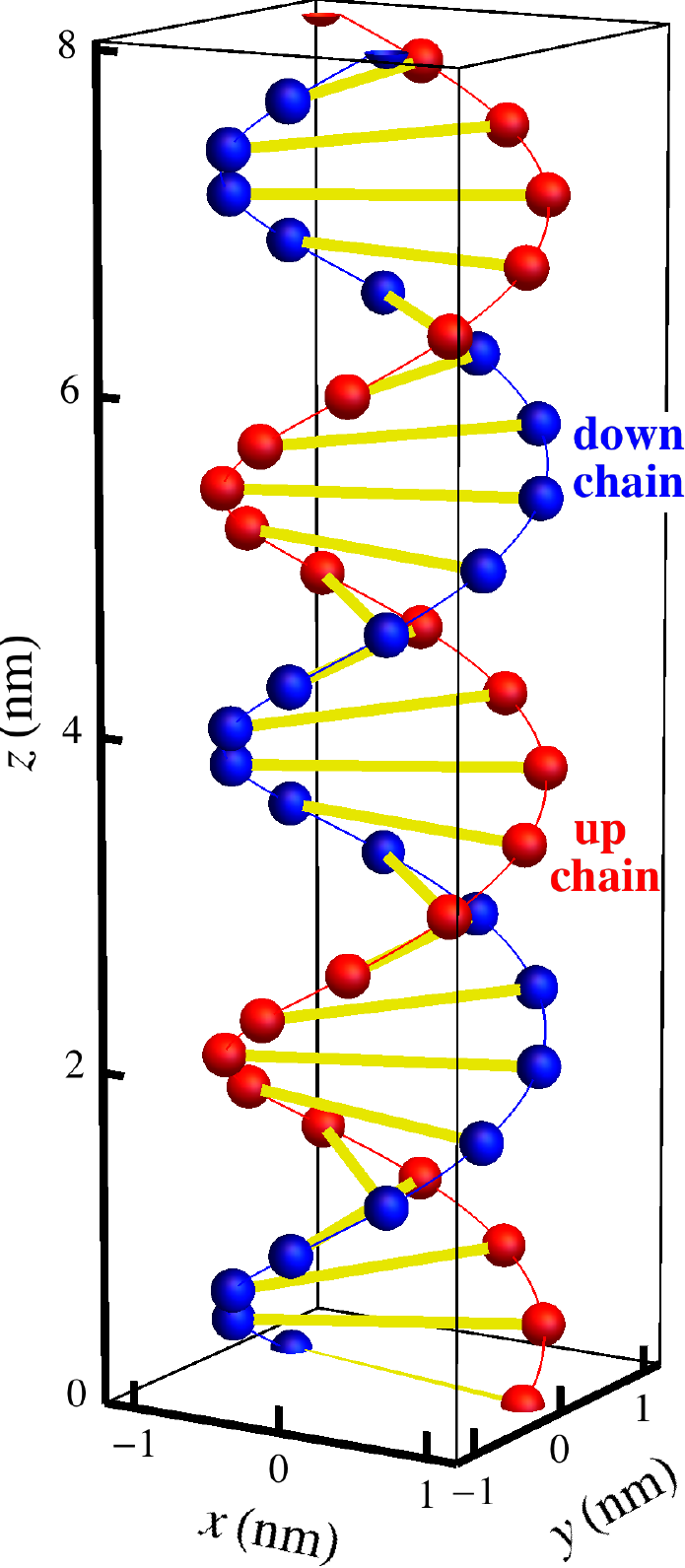}
\caption{Model of DNA as two helical chains of phosphates, shown as red and blue balls, with the yellow lines representing the base pairs.  These have been labeled ``up" and ``down" chains, corresponding to the direction of the carbon atoms in the sugar backbone.}
\label{fig-DNAPhosphateChains}
\end{figure}

\begin{figure}
\includegraphics[width=3in]{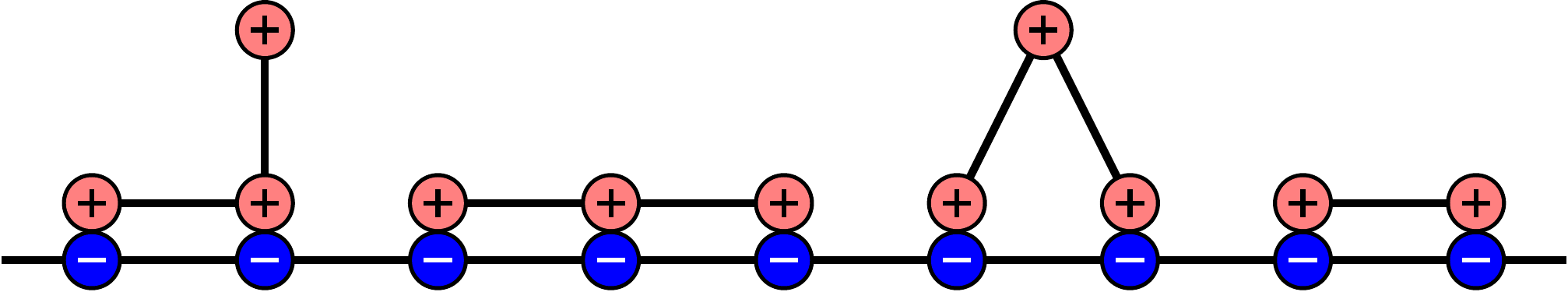}
\caption{Charge fractionalization occurs when only a fraction of the positive charges of the polyelectrolyte attach to the negatively-charged surface.  The charges that are not attached then cause the surface to become positively charged.}
\label{fig-fractionalization}
\end{figure}

Many of the approaches to understanding the role of polyelectrolytes in biological systems have adopted models of continuum electrostatics, and some of these have focussed on solving the Poisson-Boltzmann equation for a cylindrical model of DNA in the presence of multivalent ions\cite{wmg00}.  However, it has been recognized that including charge correlations is essential to understanding these systems\cite{bis99b,gns02}.  Nguyen and Shklovskii \cite{ns02a,ns02d}  proposed a simple theory of charge inversion that considers the geometrical structure of the polyelectrolyte together with its electrostatic interaction with the substrate.  The basic theory could be applied to arbitrary-length polyelectrolytes adsorbed on various surfaces from linear biomolecules to biomembranes.   The idea was that there is fractionalization of charge in which a polyelectrolyte can either neutralize the charge or reverse it depending on how it attaches.  In that model, a single
double-helix strand of DNA was represented by a rigid cylinder with two one-dimensional lattices of negative charges $-e$ in helices around the surface to represent DNA's double helix of negatively-charged phosphates, as shown in Fig.~\ref{fig-DNAPhosphateChains}.  Since the DNA is modeled to be in a polyelectrolyte solution, each strand is surrounded by positively-charged species that are long and possibly flexible.  These adsorbing species have multiple charges that may partially attach to the surface, with excess charge protruding into the solution, as shown in Fig.~\ref{fig-fractionalization}.  By this means, since there is excess charge dangling into the solution, the charge on the surface can be not only neutralized but reversed. 

The results of these generalized theories indicate that the correlation effects
in the solution are also strongly-dependent on the system geometry. 
Thermodynamically, the distribution of charge in the solution is
governed by a competition between energy and entropy.  The bound state in which
the ions are condensed on the surface of the macroion/polyelectrolyte is
energetically-favored, and the continuum state, in which the ions are free to
drift in any direction, is entropically-favored.  The balance between these two
factors in minimizing the free energy has been shown to vary significantly based
on the geometry of the macroion \cite{wmg00}.

To understand this, consider the electric potential outside spherical,
cylindrical, and planar charged surfaces in vacuum.  For spheres, the potential
decays as the inverse of the distance; for the cylinder, the potential varies as
the logarithm of the distance; and for the plane, the potential increases linearly
with the distance.  Gelbart et al.\ \cite{wmg00} assert that, in an ambient-ion
solution, the entropy of the point ions varies logarithmically with their
concentration, and therefore with the distance from the cylinder. Thus they
conclude that, in a crude comparison, for spherical geometry, the $r^{-1}$
potential is dominated by the logarithmic entropy; for planar geometry, the
entropy is dominated by the linear potential; and for cylindrical geometry, both
the energy and the entropy vary logarithmically.\cite{kas95a}  In this paper, we will incorporate the impact of the ions in solution on the free energy through the chemical potential of these ions, as detailed in Appendix~\ref{sec-ChemicalPotentialDimers}.

The helical geometry of DNA, then, sits precisely balanced on the fulcrum
between energy- and entropy-driven processes under physiological conditions. 
The geometry of the biomolecule plays an even more crucial role when the
highly-charged ions in solution are not point charges but have geometries of
their own.  Such is the case for DNA immersed in a solution of
polyelectrolytes.  These polyelectrolytes can be proteins or other fragments
of biomolecules which are routinely found in the nucleus \cite{al82}, so
that, again, the central biological processes occur in precisely the
most difficult regimes to model.

\begin{figure}
\includegraphics[width=2in]{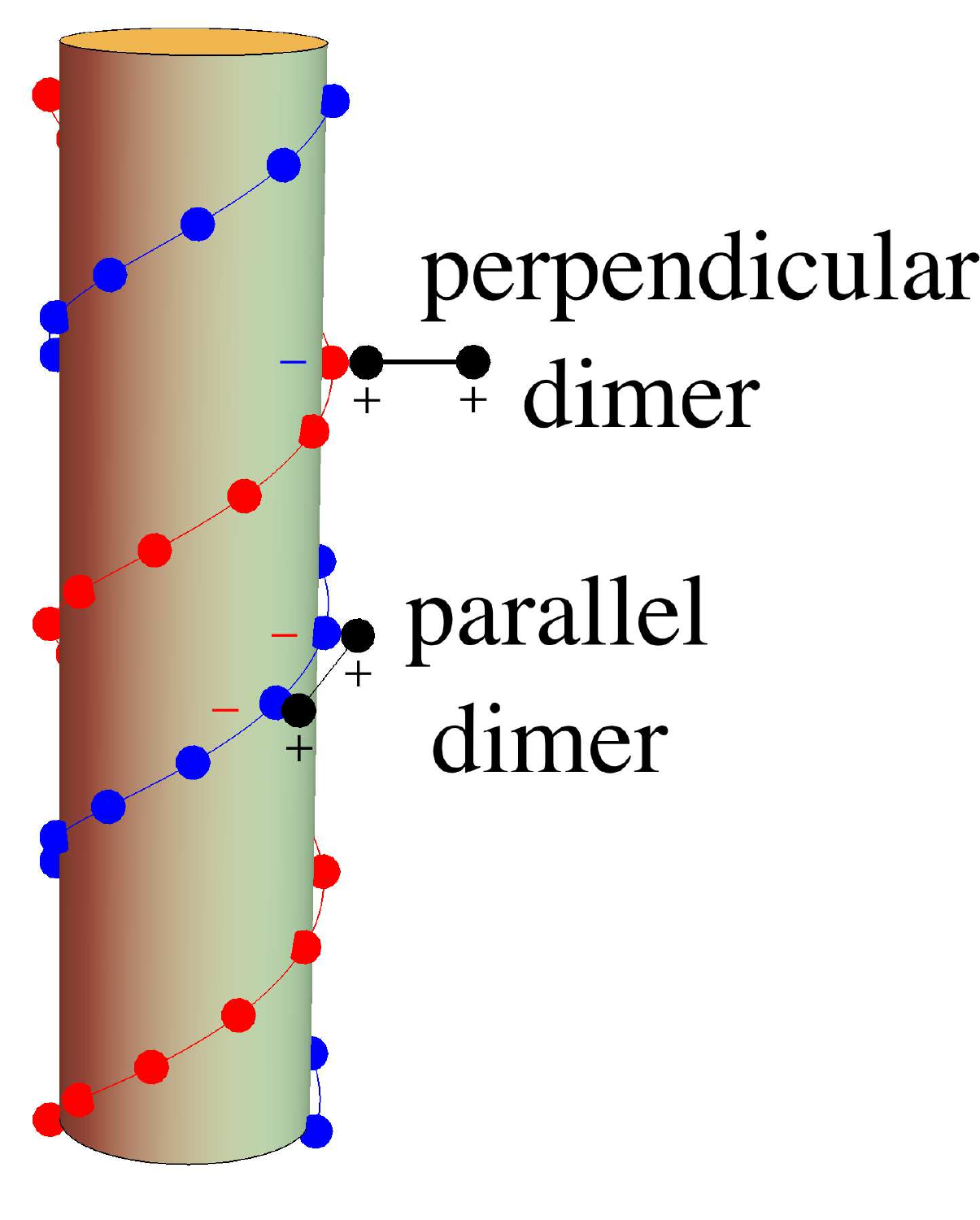}
\caption{Double helix of negatively-charged phosphate chains of red and blue balls are wrapped around a cylinder.  A parallel dimer attaching to one of the chains would neutralize two sites, while a perpendicular dimer would have one excess charge dangling into the solution, causing that site to have a net positive charge.}
\label{fig-DimerAdsorption}
\end{figure}

A special case of the the work of Nguyen and Shklovskii\cite{ns02a} studied by
Bishop and McMullen\cite{bm06} considers the case of dimers (polyelectrolytes that are
two units in length) on DNA.  The choice to model dimers is motivated by the
wealth of studies (refs. \cite{mef66} and \cite{ac03}, among others) in the
literature about dimer models in other branches of physics, as well as for
geometrical simplicity.  Dimers, the shortest polyelectrolytes, have only
two possible orientations when adsorbed on DNA, assuming that the length of the dimer is
comparable to the helical spacing between charged sites on DNA, as shown in Fig.~\ref{fig-DimerAdsorption}.  A dimer lying on
the cylindrical surface of the molecule must lie parallel to the helical
strands, neutralizing the charge on two adjacent sites.  Otherwise, the dimer
must adsorb perpendicular to the cylindrical surface, extending one end radially
out from the central axis of the cylinder and inverting the charge on a single
site.  Charge inversion by dimers, then, is quite similar to charge inversion
due to two species of point ions: one monovalent species representing
parallel-adsorbed dimers and one divalent species representing
perpendicular-adsorbed dimers.  In this way, Bishop and McMullen\cite{bm06}
modeled the adsorption of dimers in a lattice-gas model as a two-component
solution of point ions, allowing the possibility of vacancies, and used field-theoretic methods to describe the
thermodynamics of the system.  They carried their calculations to a
mean-field level of approximation of the inverted charge on DNA, but did not
calculate the entropy.  Their work confirmed the possibility of charge inversion within this model. 
 
For charge inversion on DNA with polyelectrolytes, ``physiological conditions"
require incorporating the combined effects of charge correlation, thermodynamic
fluctuations, crowding, and geometrical considerations all at once.  As we have
discussed in this introduction, there has been considerable work in addressing all of
these issues.  Some approaches treat only the geometry, with no interactions\cite{mwb06}.  
Others include both geometry and interactions, but use a
continuum model that neglects discreteness effects\cite{kn02}.  The lattice
gas dimer model is unique in its simultaneous consideration of all these
effects, and the thermodynamic and geometrical idealizations it does make can be systematically improved.  

While the formalism of Bishop and McMullen\cite{bm06} included both a mean field theory and corrections due to fluctuations or charge correlations, the computed results were obtained only at the mean field level. Those computations gave the charge per site on the DNA helix as a function of the chemical potential, or equivalently the concentration of the polyelectrolyte in solution.  While the Nguyen-Shklovskii\cite{ns02a} calculation assumed complete filling of the lattice, the Bishop-McMullen approach allowed for vacancies, represented as negatively-charged sites, in addition to the neutral or positively-charged sites arising from dimer adsorption.  However, the lattice-gas model, which assumes all sites are equivalent, does not take into account that the parallel dimer occupies two sites.  Instead, the occupation of two sites is mimicked by making the binding energy of the parallel dimer twice as large as that of the perpendicular dimer, and we will use the same approach here.   In this paper, this simple model is extended to calculate the fluctuation corrections and the entropy of the system in this simple model.  The purpose is to determine the importance of the fluctuation terms for inverting the charge and to see whether these terms have a significant impact on the entropy of the system.  A preliminary version of this work was in Sievert's master's thesis\cite{mds07}.  This paper does not attempt to include all the possible effects of coions, counterions, and the nonzero hard-core radius of the background ions of the solution, as was done in the work by Solis and Olvera de la Cruz \cite{so01a,fjs02}.  We also do not attempt to study complicated geometric effects of multivalent
polyelectrolyte macroions, such as the ``ion-bridging" model of Olvera de la Cruz \cite{mo95}, which is used as an explanation of the data of Raspaud et al. \cite{er98} on spermine-induced
aggregation of DNA, nucleosome, and chromatin.  However, this approach could certainly be extended to include all of those possibilities, and in this model, some of these effects can considered to be contained in the binding energies of the polyelectrolytes to the surface and the screened Coulomb interaction between charged species.

In Sec.~\ref{sec-NILatticeGas} of this paper, we outline the model of the charged binding sites on DNA and present the computation of the entropy due only
to the hard-core repulsion that prevents multiple binding on the same site.  In Sec.~\ref{sec-DNAgeometry} we explain the geometry of the double helix and show how it can be represented as a one-dimensional lattice and in Sec.~\ref{sec-DNApotential} derive the form of the potential and determine the orthogonal transformation that diagonalizes it.
In Sec.~\ref{sec-PartitionFunctionInteracting}, we use functional integral techniques to derive the partition function, which uses a Gaussian integral identity to perform the sum over configurations exactly, at the expense of an integral over a new auxiliary field.  In Sec~\ref{sec-expansion}, we show how the grand canonical potential can be computed order by order, in which the first two terms are the mean field and the one-loop correction to the mean field.  In Sec.~\ref{sec-saddlepoint}, we examine the saddle point equation that defines the mean field, and this is used to calculate the entropy, the number densities of all species, and the charge density.  In Sec.~\ref{sec-HartreePropagator}, we find the explicit form for the inverse-Hartree-field-fluctuation propagator.   In Sec.~\ref{sec-Fluctuations}, we include the one-loop order terms in the calculations to reveal the effects of fluctuations.  Finally, in Sec.~\ref{sec-discussion}, we compare the results of the various approximations.  Details concerning the Gaussian integral identity and the chemical potential of dimers in solution are given in Appendices \ref{sec-GaussianIntegralIdentity} and \ref{sec-ChemicalPotentialDimers}.


\section{The Noninteracting DNA Lattice Gas}
\label{sec-NILatticeGas}

In this section, we will derive the entropy and the average occupations for the three species of vacancies, parallel dimers, and perpendicular dimers for the lattice-gas model of DNA in the absence of electrostatic interactions between the species on different sites of the lattice.  Although this could be done using combinatorics, we will use an approach analogous to what be used later when interactions are included, and this will make the procedure for the interacting case clearer.
In the lattice-gas model that we consider here, one feature of the adsorption of species must necessarily be neglected.  When a dimer is adsorbed parallel to a phosphate chain it occupies two adjacent sites, but the lattice-gas model treats all sites independently and cannot take into account the blocking of an adjacent site by a parallel dimer.  This would be an even more complicated issue for longer polymers, and is generally difficult to incorporate.  As
a first step toward building a comprehensive model of such systems, here we neglect the adjacent blocking effect for parallel adsorption.  Nevertheless, we retain the different binding energies 
$\varepsilon^{(\parallel)}$ and $\varepsilon^{(\perp)}$, since the two-site occupation implies $\varepsilon^{(\parallel)} \approx 2\varepsilon^{(\perp)}$. 
 This allows us to describe adsorption of either species independently for each site.  Any configuration of the DNA-dimer
complex in this model can then be described by identifying the type of dimer (parallel, perpendicular, or none) adsorbed on each site on the DNA molecule.  Such a model resembles the lattice gas model of condensed matter physics \cite{pb89}, which treats the ways of distributing particles of different types onto a regular lattice of
sites.  If the particles on different lattice sites do not interact with one another, then the structure of the lattice does not matter.  

It is convenient to consider a vacancy as a third species $\gamma$ of particle, because then we can impose the constraint that each site is singly occupied, either by a parallel dimer $(\gamma = \: \parallel)$, a perpendicular dimer ($\gamma = \: \perp$), or a vacancy ($\gamma = v$).   For each our three species of ``particles" that can reside on our lattice sites, we define a relative charge $q_\gamma$ in units of the magnitude $e$ of the charge of an electron.  These relative charges are then $q_v=-1$ for vacancies, $q_\parallel=0$ for parallel dimers, and $q_\perp=+1$ for perpendicular dimers.  We will assume that the binding energy $\varepsilon_\gamma$ of species $\gamma$ to the lattice depends only on the type of species and not the location.  We will be specifying each lattice point by its location $\ell$ along helix $b$, with the pair $(\ell,b)$ specifying that lattice position.  Then the quantity $n_{\ell,b}^\gamma$ will be the number of particles of species $\gamma$ on the lattice site at $\ell$ on chain $b$.  For each chain, $\ell$ extends from $-{\mathcal N}$ to ${\mathcal N}$, such that the total number of sites is ${\mathcal N}_{\text{sites}} =2(2{\mathcal N}+1)$, and we employ periodic boundary conditions.   

We begin by considering the Hamiltonian ${\mathcal{H}}_{\text{NI}}$ for this noninteracting system (that is, with no interactions between different sites aside from the hard-core repulsion that blocks double-occupancy), which is
\begin{equation}
{\mathcal{H}}_{\text{NI}}  = \sum_{\ell,b} \sum _{\gamma }\varepsilon _\gamma  n_{\ell,b}^{(\gamma )}. 
\label{eq-HNI}
\end{equation}
Because we have a system that exchanges particles with its surroundings, specifically the ions in the solution surrounding the DNA, which attach to the surface, we work in the grand canonical ensemble.  If the system contains  $N_\gamma$ particles of type $\gamma$ in equilibrium with its surroundings, with an average internal energy $E$, the grand canonical potential $\Omega_G$, which is a function of the temperature $T$, volume $V$, and chemical potentials $\mu_\gamma$ for each species $\gamma$, is written as\cite{pb89, fr65},
\begin{equation}
\Omega_G(T, V, \mu) = E-TS-\sum_\gamma \mu_\gamma N_\gamma \;  ,
\label{eq-GrandCanonicalPotential}
\end{equation}
where the number of particles of type $\gamma$ is given by
\begin{equation}
     N_\gamma = \sum_{\ell,b} n_{\ell,b}^{(\gamma)} \; .
\label{eq-NiDef}
\end{equation}
Then the entropy can be written in terms this potential as
\begin{equation}
     S = -\left( \frac{\partial \Omega_G}{\partial T} \right)_{V, \{\mu_\gamma\}}
     = k_B \beta^2 \left( \frac{\partial \Omega_G}{\partial \beta} \right)_{V, \{\mu_\gamma\}} \;   ,
\label{eq-EntropyDefinition}
\end{equation}
where $\beta=\frac{1}{k_B T}$ and $k_B$ is Boltzmann's constant.   In this partial derivative, the volume and all the chemical potentials $\mu_\gamma$ of all species $\gamma$ are held constant.  It is convenient to define the grand canonical partition function $Z_G$ in terms of the grand canonical potential $\Omega_G$ as
\begin{equation}
     Z_G =\sum_{\text{configurations}}
     e^{-\beta (E-\sum_\gamma \mu_\gamma N_\gamma)}  
     = e^{-\beta \Omega_G}   \;  ,
\label{eq-ZG1}
\end{equation}
where the sum over configurations includes all the accessible states of the system.  The grand canonical potential can alternatively be written in terms of the partition function as
\begin{equation}
     \Omega_G =-\frac{\ln Z_G}{\beta}  \;  ,
\label{eq-OmegafromZ}
\end{equation}
and this allows us to write an expression for the entropy in terms of $Z_G$ as:
\begin{equation}
   S =  k_B \ln Z_G - \frac{k_B \beta}{Z_G} \left( \frac{\partial Z_G}{\partial \beta} \right)_{V, \{\mu_\gamma\}} \;  .
\label{eq-EntropyfromZG}
\end{equation}

For the noninteracting lattice, the average internal energy is represented here by the Hamiltonian \eqref{eq-HNI}, $E={\mathcal{H}}_{NI}$, and the noninteracting grand canonical partition function $Z_{\text{NI}} $ becomes
\begin{equation}
Z_{\text{NI}}  = \sum_{\text{configurations}} 
e^{-\beta \sum_{\ell,b} \sum _{\gamma } 
\left(\varepsilon_\gamma - \mu_\gamma \right) n_{\ell,b}^{(\gamma)} } 
\;  ,
\label{eq-ZNI1}
\end{equation}
where we have suppressed the $G$ subscript for simplicity.

Since a vacant site does not really correspond to a particle, we recognize that energy and chemical potential of the vacancy must be related such that $\varepsilon^{(v)}-\mu_v =0$.
In addition, because the dimers adsorbed on the surface and those in solution are in equilibrium, 
$\mu _{\parallel }=\mu _{\perp}=\mu _{\text{dimer}}$ is the chemical potential of dimers in solution at the appropriate concentration.
We thus need an estimate for $\mu_{\text{dimer}}$.  
Approximating the dimer as a uniformly-charged cylinder, this value is shown in Appendix \ref{sec-ChemicalPotentialDimers} to be $\beta \mu_{\text{dimer}} \simeq 0.79$ at the physiological temperature of $T \simeq 310$~K. 

The average occupancy $\langle n^{(\gamma)} \rangle_{\text{NI}}$ for species $\gamma$ per site is found by taking the derivative of this partition function with respect to $\mu_\gamma,$
\begin{equation}
\langle n^{(\gamma)} \rangle_{\text{NI}}  
=\frac{1}{{\mathcal{N}}_{\text{sites}} \beta Z_{\text{NI}} }
\frac{\partial Z_{\text{NI}} }{\partial \mu_\gamma}
   \;  .
\label{eq-ngammaderivZ}
\end{equation}
This can be verified by taking the derivative of Eq.~(\ref{eq-ZNI1}).
\begin{eqnarray}
\langle n^{(\gamma)} \rangle_{\text{NI}}  
&=& 
\frac{1}{{\mathcal{N}}_{\text{sites}}
Z_{\text{NI}}} 
\sum_{\text{configurations}} \sum_{\ell,b}  \times  
   \nonumber \\    && \times   n_{\ell,b}^{(\gamma)} 
e^{-\beta \sum_{\ell,b} \sum _{\gamma } \left(\varepsilon_\gamma -
\mu_\gamma \right) n_{\ell,b}^{(\gamma)} } 
   \; ,
\label{eq-ngammaderivZNI1}
\end{eqnarray}
which is by definition the average number per site of species $\gamma$.

Continuing with the evaluation of the partition function, we note that the exponential of a sum can be written as the product of exponentials, enabling us to rewrite the partition function \eqref{eq-ZNI1} as
\begin{equation}
Z_{\text{NI}}  =
 \sum_{\text{configurations}} 
\prod_{\ell,b} \prod_{\gamma } 
\left[e^{-\beta  \left(\varepsilon_\gamma -
\mu_\gamma \right)}\right]^{n_{\ell,b}^{(\gamma)}} 
\;  .
\label{eq-ZNI1a}
\end{equation}
In order to simplify and to appreciate the physical meaning of this expression, it is useful to define the relative activity \cite{wjm72} of species $\gamma$ in the
noninteracting lattice-gas model, given by 
\begin{equation}
a_{\text{NI}}^{(\gamma)}
=e^{-\beta
\left(\varepsilon _\gamma
-\mu _{\gamma }\right)}    \;  .
\label{eq-aNI}
\end{equation}
Note that this is independent of the lattice site $\ell$ or chain $b$.    
The partition function can then be written in the simple form
\begin{equation}
Z_{\text{NI}} =\sum_{\text{configurations}}  
\prod _{\gamma}\prod_{\ell,b}
\left[a_{\text{NI}}^{(\gamma)}\right]^{n_{\ell,b}^{(\gamma)}}    \;  .
\label{eq-ZNIProductForm}
\end{equation}
Because we have assumed that there can be only one species per site, parallel dimer, perpendicular dimer, or vacancy, the sum over configurations can now be done over each site separately, where there are three possible configurations
\begin{equation}
\{ n_{\ell,b}^{(\perp)},n_{\ell,b}^{(\parallel)},n_{\ell,b}^{(v)} \} = \{1,0,0\} \; ,  \;  \{0,1,0\} \; , \; 
{\mbox{or}}  \;  \{0,0,1\}   \;  .
\label{eq-gammaConfigs}
\end{equation}
This is the same as saying that $n_{\ell,b}^{(\gamma)}=1$ for one and only one of $\gamma=\perp,\parallel$, or $v$ and is zero otherwise.  This means that
\begin{equation}
\sum_{\substack{{\text{single-site}} \\ {\text{configurations}}  }}
\prod_{\gamma=\perp, \parallel, v}
\left[a_{\text{NI}}^{(\gamma)}\right]^{n_{\ell,b}^{(\gamma)}} 
= \sum_\gamma a_{\text{NI}}^{(\gamma)} \; ,
\label{eq-SingleSiteNIConfigs}
\end{equation}
and so the grand partition function is
\begin{equation}
Z_{\text{NI}} =\prod_{\ell,b} \left( \sum_\gamma 
a_{\text{NI}}^{(\gamma)} \right) \; .
\label{eq-ZNIProd}
\end{equation}
Since every term in the product is the same, the expression in parentheses is simply raised to the power ${\mathcal{N}}_{\text{sites}}$, giving
\begin{equation}
Z_{\text{NI}} = \left(\sum_\gamma 
a_{\text{NI}}^{(\gamma)} \right)
^{{\mathcal{N}}_{\text{sites}}} \; .
\label{eq-ZNIPower}
\end{equation}

\begin{figure}
\includegraphics[width=3in]{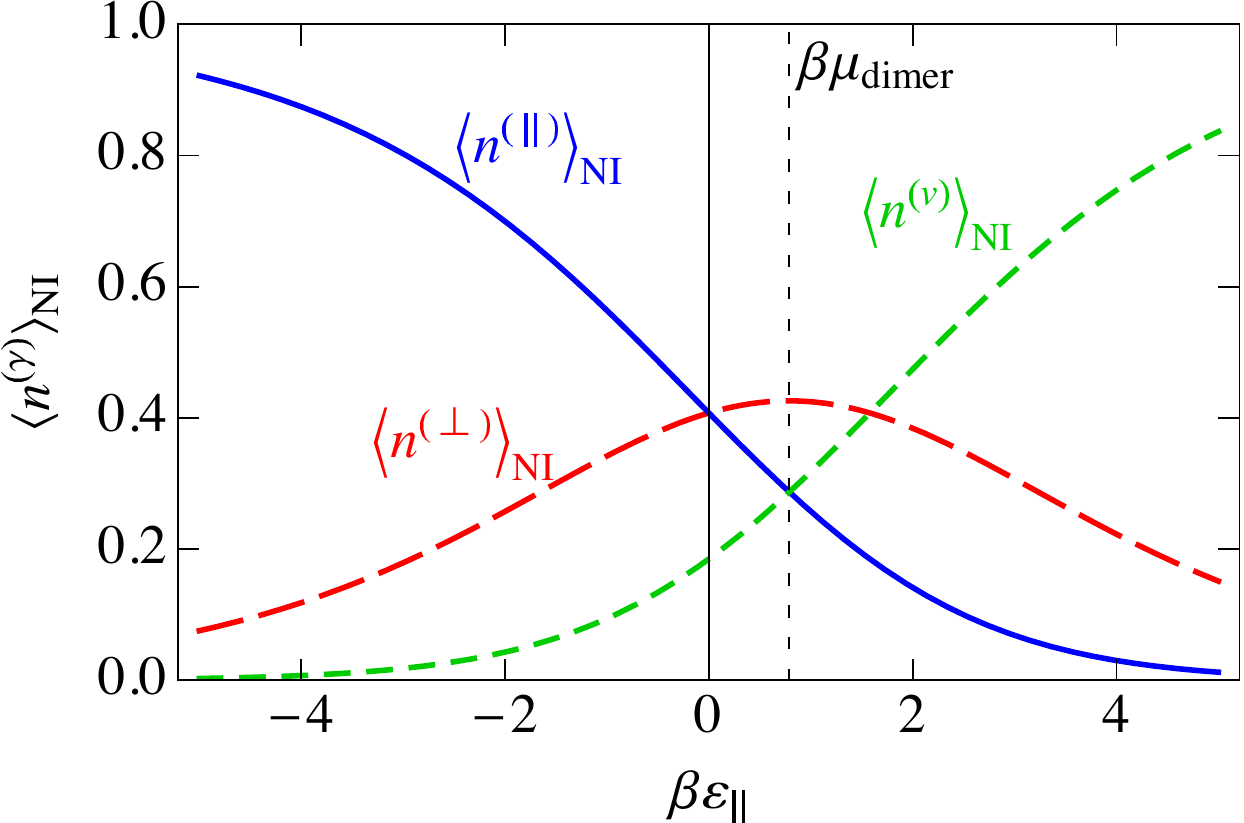}
\caption{Plots of the mean occupation numbers $\langle n^{(\gamma)}\rangle_{NI}$
in the noninteracting lattice gas model.  Curves are shown for $\gamma = \parallel$ (blue,solid), $\perp$
(red,long-dashed), and $v$ (green,short-dashed).  The parameters used in the calculation are
$\beta \mu_{\text{dimer}}=0.79$ and $\varepsilon_{\parallel}=2\varepsilon_{\perp}$.
For negative  $\varepsilon_{\parallel}-\mu_{\text{dimer}}$, dimers are attracted to the lattice, while for positive values of this quantity, they are repelled.}
\label{fig-NIOccupation}
\end{figure}

At this point, we can easily see that we can obtain the average number per site using our derivative form from Eq.~(\ref{eq-ngammaderivZ}) as
\begin{equation}
\langle n^{(\gamma)} \rangle_{\text{NI}}  
=\frac{1}{{\mathcal{N}}_{\text{sites}} \beta Z_{\text{NI}} }
\frac{\partial}{\partial \mu_\gamma}
 \left(\sum_{\gamma'} a_{\text{NI}}^{(\gamma')} \right)
^{{\mathcal{N}}_{\text{sites}}} 
   \;  .
\label{eq-ngamNIav1}
\end{equation}
Taking the derivative and substituting the expression for $Z_{\text{NI}}$ in the denominator, we have
\begin{eqnarray}
\langle n^{(\gamma)} \rangle_{\text{NI}}  
&=&\frac{{\mathcal N}_{\text{sites}} }{{\mathcal{N}}_{\text{sites}} \beta \left(\sum_{\gamma'' }a_{\text{NI}}^{(\gamma'')} \right)
^{{\mathcal{N}}_{\text{sites}}}  }  \times  \nonumber \\  
&& \times
 \left(\sum_{\gamma'} a_{\text{NI}}^{(\gamma')} \right)
^{{\mathcal{N}}_{\text{sites}}-1} 
\frac{\partial}{\partial \mu_\gamma}
a_{\text{NI}}^{(\gamma)}
   \; ,
\label{eq-ngamNIav2}
\end{eqnarray}
where the derivative of the activity is
\begin{equation}
\frac{\partial}{\partial \mu_\gamma}  a_{\text{NI}}^{(\gamma)}  = 
\frac{\partial}{\partial \mu_\gamma}  
e^{-\beta \left(\varepsilon _\gamma -\mu _{\gamma }\right)}   
= \beta  a_{\text{NI}}^{(\gamma)}
 \; .
\label{eq-aderivmu}
\end{equation}
Using this in the expression for the mean site occupancy in Eq.~(\ref{eq-ngamNIav2}) and cancelling 
${\mathcal{N}}_{\text{sites}} - 1$ factors of the sum over 
$\gamma$, the mean occupancy for species $\gamma$ is given by
\begin{equation}
\langle n^{(\gamma)} \rangle_{\text{NI}}  
=\frac{a_{\text{NI}}^{(\gamma)}}
{  \sum_{\gamma'}a_{\text{NI}}^{(\gamma')}  }
   \;  .
\label{eq-ngamNIfinal}
\end{equation}
In Fig.~\ref{fig-NIOccupation}, we show the mean occupancies for the three species
versus $\varepsilon_{\parallel}$, with $\varepsilon_{\parallel}=2\varepsilon_{\perp}$. Negative $\varepsilon_{\parallel}$ corresponds to an attraction of dimers to the lattice, and we see that parallel adsorption of dimers dominates because of the stronger binding, followed by perpendicular adsorption, with vacancies becoming nonexistent.  
Positive $\varepsilon_{\parallel}$ corresponds to a repulsion of dimers from the lattice, so that at large $\varepsilon_{\parallel}$, the ordering is reversed, for the same reasons.  At small positive $\varepsilon_{\parallel}$, the dimers are repelled from the lattice, but this competes with the chemical potential, which tries to put dimers back onto the lattice.  In this low-coverage regime, perpendicular adsorption dominates, while vacancies dominate for large $\beta \varepsilon_{\parallel}$ because the dimers would prefer to stay in solution.  For $\varepsilon_\parallel, \varepsilon_\perp \rightarrow 0$, the parallel and perpendicular mean occupancies become the same.
Similarly, where $\varepsilon_{\parallel}-\mu_{\text{dimer}}=0$, 
$\langle n^{(\parallel)} \rangle_{\text{NI}}= \langle n^{(v)} \rangle_{\text{NI}}$
because, as mentioned earlier, $\varepsilon_v-\mu_{\text{dimer}}$ is always zero.
Also, $\langle n^{(\perp)} \rangle_{\text{NI}}= \langle n^{(v)} \rangle_{\text{NI}}$ where 
$\varepsilon_{\parallel}=2\mu_{\text{dimer}}$ because that is where 
$\varepsilon_{\perp}=\mu_{\text{dimer}}$.
It is at this point where $\langle n^{(\parallel)} \rangle_{\text{NI}}$ becomes a maximum.

\begin{figure}
\includegraphics[width=3in]{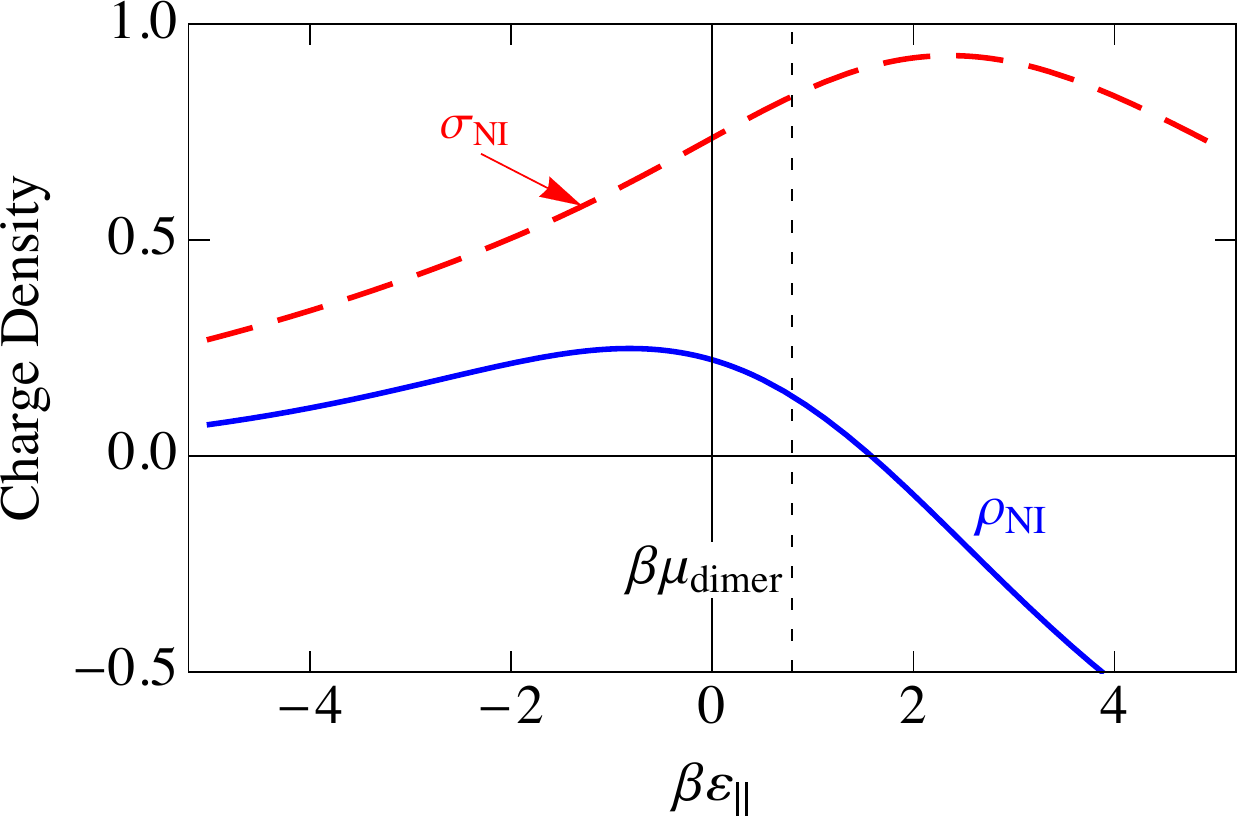}
\caption{Plots of the charge densities for the noninteracting lattice gas model.  The charge density $\rho_{NI}$ is shown as the solid blue curve, and the standard deviation of the charge density, $\sigma_c$, is shown as the dashed red curve.  The parameters used in the calculation are$\beta \mu_{\text{dimer}}=0.79$ and $\varepsilon_{\parallel}=2\varepsilon_{\perp}$.  As in Fig.~\ref{fig-NIOccupation}, dimers are attracted to the lattice when $\varepsilon_{\parallel}-\mu_{\text{dimer}}<0$, and dimers are repelled from the lattice when $\varepsilon_{\parallel}-\mu_{\text{dimer}}>0$.}
\label{fig-NIchargeDensities}
\end{figure}

The average charge per site can now be determined by multiplying $q_\gamma$, the charge of species $\gamma$, by $n_{\ell,b}^{(\gamma)}$ the occupation of species $\gamma$ on site $\ell$ of chain $b$,
\begin{equation}
\rho_{\ell b}=\sum_\gamma q_\gamma n_{\ell b}^{(\gamma)}
   \;  .
\label{eq-rhoNIsitedef}
\end{equation}
Since we assume that every site has either a parallel dimer, a perpendicular dimer, or a vacancy with single occupancy, then for a given site and species, $n_{\ell b}^{(\gamma)}$ is either 0 or 1.  The average charge density is then the sum the averages of the individual terms,
\begin{equation}
\rho_{\text{NI}} \equiv  \langle \rho_{\ell b} \rangle_\text{NI} = \sum_\gamma q_\gamma \langle n^{(\gamma)} \rangle_{\text{NI}}
   \;  .
\label{eq-rhoNIdef}
\end{equation}
This charge density is plotted as the solid blue curve in Fig.~\ref{fig-NIchargeDensities}.  Because a parallel dimer has no charge, $q_{\parallel}=0$, a perpendicular dimer has a positive unit charge, $q_{\perp}=+1$, and a vacancy has a negative unit charge, $q_v = -1$, the total mean field charge per site $\rho_c$ in 
Fig.~\ref{fig-NIchargeDensities} is the difference between the curves $\langle n^{(\perp)}\rangle_\text{NI}$ and 
$\langle n^v\rangle_\text{NI}$ in Fig.~\ref{fig-NIOccupation} and goes to zero where those two curves cross.  

A measure of the magnitude of the charge fluctuations is given by the average of the square of the charge on a site minus the square of its average, which is the charge variance $\sigma_{NI}^2$, given by
\begin{equation}
\sigma_{\text{NI}}^2=\langle\rho^2 \rangle_{\text{NI}} - \rho_{\text{NI}}^2    \; ,
\label{eq-sigmaNIdef}
\end{equation}
where
\begin{equation}
\langle\rho^2\rangle_{\text{NI}} =\langle \rho_{\ell b}^2 \rangle_\text{NI}    
   \;  ,
\label{eq-rhosqNIdef}
\end{equation}
and
\begin{equation}
\rho_{\ell b}^2 = \left[\sum_\gamma q_\gamma n_{\ell b}^{(\gamma)} \right]^2  
=  \sum_{\gamma,\gamma'} q_\gamma q_{\gamma'} n_{\ell b}^{(\gamma)} n_{\ell b}^{(\gamma')}
   \;  .
\label{eq-rhosqNIdef1}
\end{equation}
The product $n_{\ell b}^{(\gamma)} n_{\ell b}^{(\gamma')}$ describes a double occupancy of site $(\ell,b)$ by species $\gamma$ and $\gamma'$.  Since we have required single occupancy, then we must have $\gamma=\gamma'$.   Also, since $n_{\ell b}^{(\gamma)}$ can only take the values $0$ or $1$, 
\begin{equation}
[n_{\ell b}^{(\gamma)}]^2=n_{\ell b}^{(\gamma)}  \;.
\label{eq-nsqeqn}
\end{equation} 
Then the charge density per site reduces to
\begin{equation}
\rho_{\ell b,\text{NI}}^2 =\sum_\gamma q_\gamma^2 n_{\ell b,\text{NI}}^{(\gamma)}
   \;  .
\label{eq-rhosqsiteANI}
\end{equation}
Taking the thermal average of both sides, we have
\begin{equation}
\langle\rho^2\rangle_{\text{NI}} \equiv \langle \rho_{\ell b}^2 \rangle_{\text{NI}} =
\sum_\gamma q_\gamma^2 \langle n^{(\gamma)}\rangle_{\text{NI}}
   \;  .
\label{eq-rhosqsiteNI}
\end{equation}
In Fig.~\ref{fig-NIchargeDensities}, we have plotted the standard deviation in the charge density, $\sigma_{\text{NI}}$, which is the square root of the charge variance, for the noninteracting model. 
As we saw in Fig.~\ref{fig-NIOccupation}, there are three distinct regions, 
$\varepsilon_{\parallel} \ll 0$, $\varepsilon_{\parallel} \gg 0$ and $\varepsilon_{\parallel}$ near $\mu_{\text{dimer}}$.  These three regions are also reflected in Fig.~\ref{fig-NIchargeDensities}.  At large negative $\varepsilon_{\parallel}$, parallel adsorption dominates, which leads to $\rho \approx 0$.  At large positive $\varepsilon_{\parallel}$, vacancies dominate, leading to $\rho < 0$, and for $\varepsilon_{\parallel}$ near $\mu_{\text{dimer}}$, there is a small window where perpendicular adsorption of dimers dominates, leading to positive values of $\rho$.  Correspondingly, the fluctuations, represented by the standard deviation $\sigma_{NI}$, become small when 
$|\varepsilon_{\parallel}|$ becomes large, and the fluctuations are largest at small 
$|\varepsilon_{\parallel}|$, when there are comparable numbers of all species.  The phenomenon of charge inversion is demonstrated in Fig.~\ref{fig-NIchargeDensities} because the average charge is positive, indicating that sufficiently many dimers adsorb in a
perpendicular configuration to invert the charge on the molecule from negative to
positive.  The magnitude of charge inversion increases in the weak-binding limit
$\varepsilon_\perp,\varepsilon_\parallel \rightarrow 0$.

The noninteracting entropy $S_{\text{NI}} $ can be obtained from the partition function using Eq.~(\ref{eq-EntropyfromZG}) as
\begin{eqnarray}
S_{\text{NI}}   &=&   k_B  \ln { \left[  \left(\sum_\gamma 
a_{\text{NI}}^{(\gamma)}\right)
^{{\mathcal{N}}_{\text{sites}}} \right] }
     \nonumber \\  
&&  - \frac{k_B \beta}
{ \left(  \sum_{\gamma'} a_{\text{NI}}^{(\gamma')} \right) 
^{{\mathcal{N}}_{\text{sites}} }}
\frac{\partial}{\partial \beta}\left(  \sum_{\gamma} a_{\text{NI}}^{(\gamma)} \right) 
^{{\mathcal{N}}_{\text{sites}} }
 \label{eq-EntropyfromZNI}
\end{eqnarray}
Because the derivative of the activity with respect to $\beta$, can be written
in terms of its logarithm as
\begin{equation}
\beta \frac{\partial}{\partial \beta}  a_{\text{NI}}^{(\gamma)}  = 
a_{\text{NI}}^{(\gamma)}
\ln{a_{\text{NI}}^{(\gamma)} }\; ,
\label{eq-aderivbeta2}
\end{equation}
the entropy can be written in the simple form
\begin{equation}
\frac{S_{\text{NI}}}{k_B {\mathcal{N}}_{\text{sites}}}   
= 
- \sum_\gamma
\frac{a_{\text{NI}}^{(\gamma)}}
{  \sum_{\gamma''} a_{\text{NI}}^{(\gamma'')}}
\ln{ \left( \frac{a_{\text{NI}}^{(\gamma)}}
{  \sum_{\gamma'} a_{\text{NI}}^{(\gamma')}}
\right)}
 \label{eq-EntropyfromZNI4}
\end{equation}
Substituting the mean occupancy for the noninteracting model from Eq.~(\ref{eq-ngamNIfinal}) allows us to write the dimensionless entropy per site for the noninteracting model in a simplified form as
\begin{equation}
\frac{S_{\text{NI}}}{k_B {\mathcal{N}}_{\text{sites}}}   
= 
-\sum_\gamma  \langle n^{(\gamma)} \rangle_{\text{NI}}  
\ln{ \langle n^{(\gamma)} \rangle_{\text{NI}} } 
   \;  ,
\label{eq-SNIasnlogn}
\end{equation}
which agrees with the standard result for the entropy of mixing of an
ideal solution with species $\gamma =(\parallel ,\perp ,v)$
\cite{wjm72}.   

\begin{figure}
\includegraphics[width=3in]{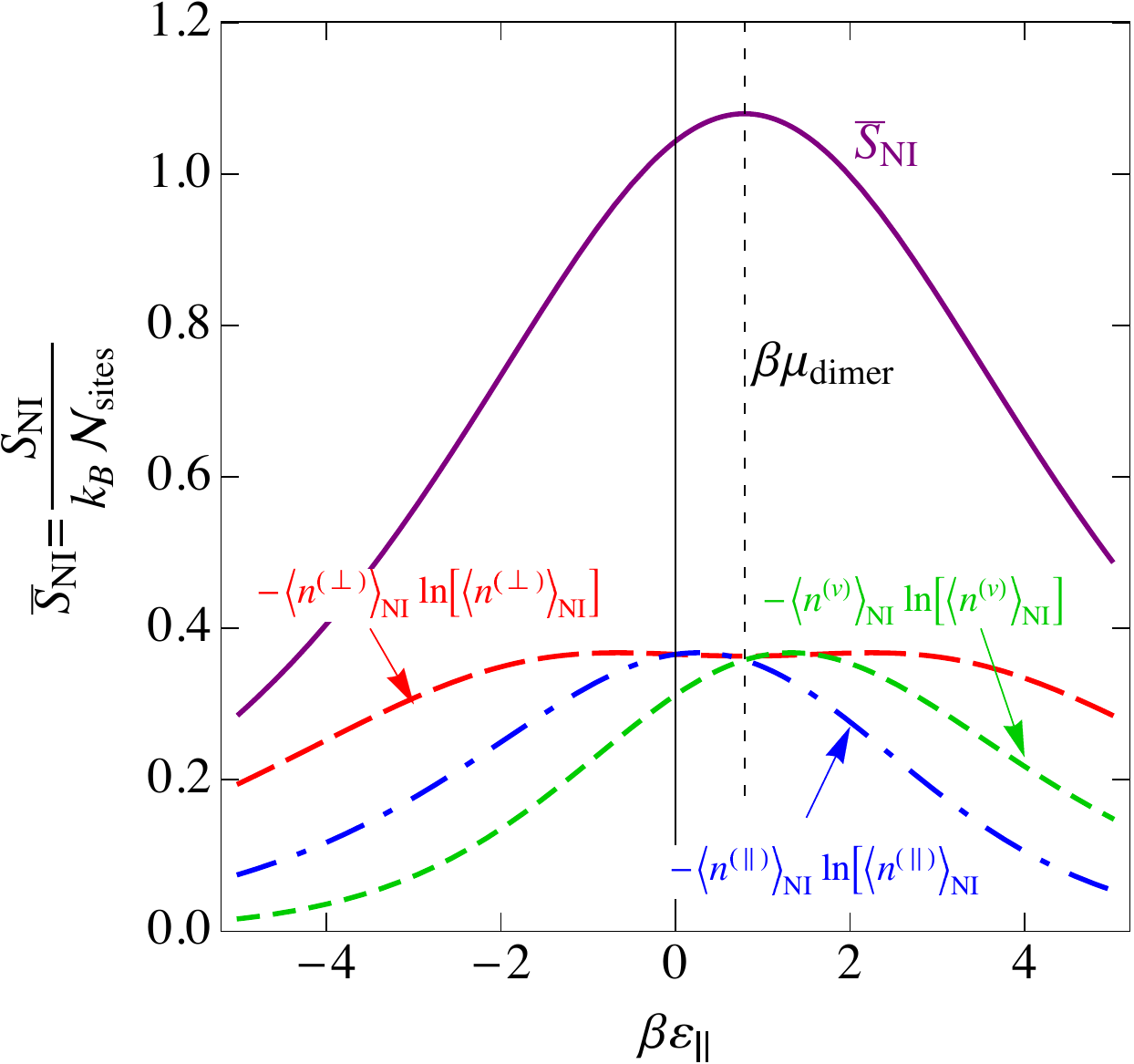}
\caption{The entropy $S_{NI}$ per site in the
noninteracting lattice gas model.  The parameters used in the calculation are
$\beta \mu_{\text{dimer}}=0.79$ and
$\varepsilon_{\parallel}=2\varepsilon_{\perp}$.
The dashed curves are the individual contributions in Eq.~(\ref{eq-SNIasnlogn}).}
\label{fig-NIEntropy}
\end{figure}

In Fig.~\ref{fig-NIEntropy}, we show the entropy
$S_{NI}$ of the noninteracting model as a function of the binding energy 
$\beta \varepsilon_\parallel$, assuming $\varepsilon_\parallel=2\varepsilon_\perp$.  We also show the individual contributions of Eq.~(\ref{eq-SNIasnlogn}) to the entropy.   The entropy is a maximum when disorder is greatest, and this occurs when the numbers of each of the species are as close to equal as possible, which occurs
at $\varepsilon_\parallel= \mu_{\text{dimer}}$, the maximum of $\langle n^{(\perp)}\rangle $ in Fig.~\ref{fig-NIOccupation}.

\section{Geometry of the Charged Double Helix of DNA}
\label{sec-DNAgeometry}

While stored in the nucleus of a cell, DNA is wrapped compactly both
around histone protein complexes and around itself, but on
sufficently small scales ($\approx$ 150 base pairs \cite{jm07}, or 15
turns \cite{al82}, or $50$~nm \cite{wmg00}), DNA's dominant
geometrical structure is the familiar double-helix structure shown in
Fig.~\ref{fig-DNAPhosphateChains}.

\begin{figure}
\includegraphics[width=3in]{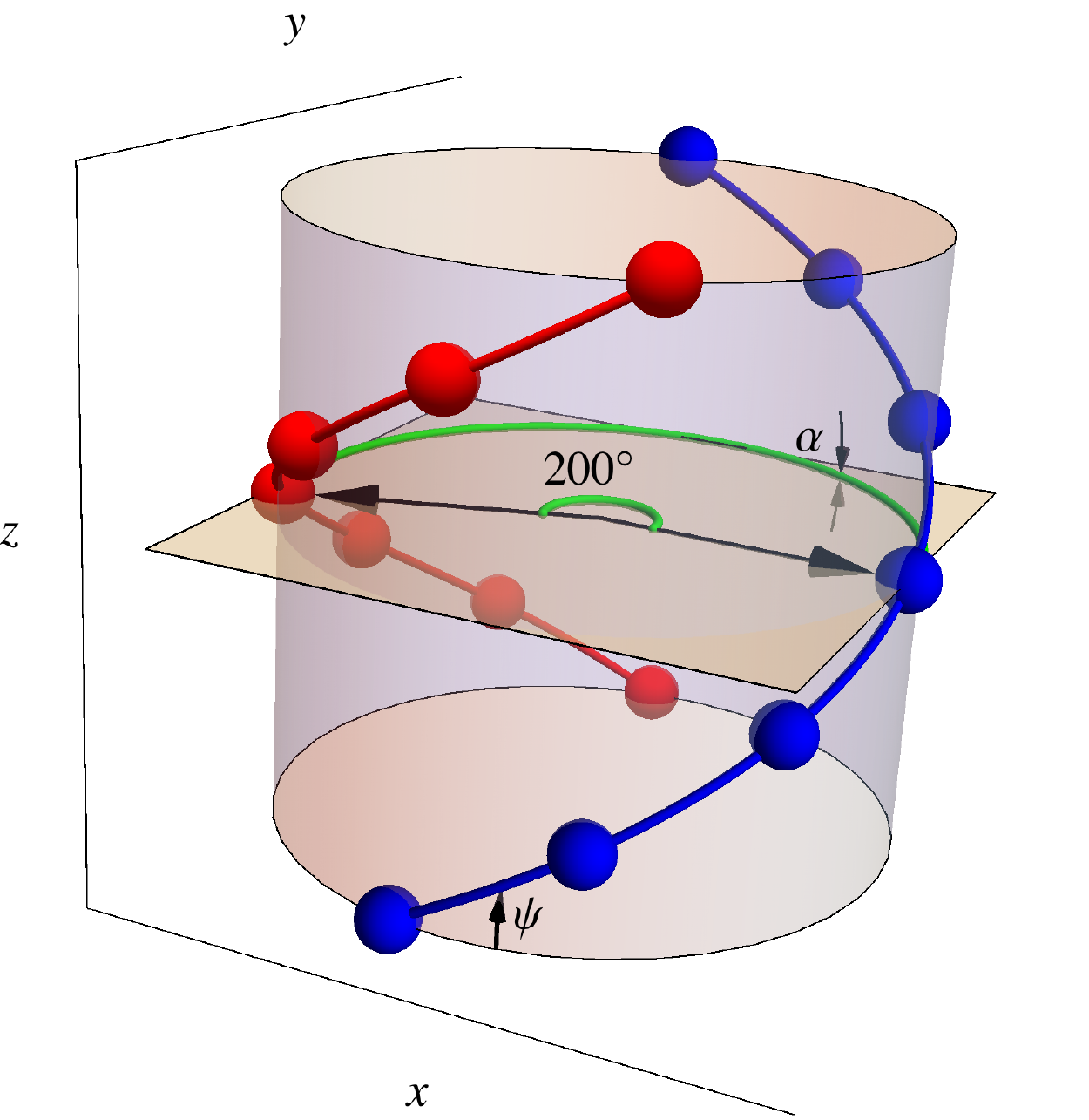}
\caption{A segment of a single DNA strand.  The helix winds along the cylindrical
surface with a pitch angle $\psi$, and a helix of pitch angle $\alpha$ connects
a site with its neighbor on the other chain.  The helix has been stretched in the $z$ direction so
that the angle $\alpha$, which is only $\approx 0.4^{\circ}$, can be clearly labeled on the diagram.}
\label{fig-Helices}
\end{figure}

Each strand of the DNA takes the shape of a helix with a
characteristic radius $R_{\text{DNA}} = 0.946$~nm and 
pitch angle $\psi =
29.6^{\circ}$, as shown in Fig.~\ref{fig-Helices}.  Here the pitch angle $\psi$ denotes the angle with
respect to the $xy$-plane that gives the appropriate altitude per
unit circumferential winding; in cylindrical coordinates, 
$\tan \psi = \Delta z / R_{\text{DNA}}\Delta \phi$, where $ \Delta z$ is the distance along the $z$ axis.  Each strand of DNA has a
``direction", identified by a particular carbon on the backbone
structure, corresponding to the chirality of the helix.  In the DNA
double-helix, the two strands are antiparallel and therefore have
opposite chiralities.  As a consequence, the azimuthal angle 
between the two helices is always a constant, 
$\Delta \phi=160^{\circ}$.  Because this phase shift is not exactly $180^{\circ}$, the
chains have unequal separation in the clockwise and counterclockwise azimuthal
directions.  The larger gap is referred to as the {\it major groove}, and the
smaller as the {\it minor groove}.

When the hydrogen atoms dissociate under physiological conditions,
the resulting negative charges ($-e$, where $e$ is the magnitude of the
charge of the electron), can be regarded as located at the mean position of the oxygen atoms
on the backbone, as shown in Fig.~\ref{fig-DNAPhosphateChains}.  This figure was
constructed using geometrical data taken from Bishop and McMullen
\cite{bm06}, which is based on experimental x-ray diffraction data
\cite{ah69,ah72} as input to the SYBYL molecular modeling program\cite{syb06}.  The oxygen atoms occur at
regular intervals along each strand, separated by a helix segment of arc length
$a^{(1)}=0.684$~nm.  It is these sites located at regular intervals along the
helix to which dimers will adsorb.  These negatively-charged sites do not occur
at the same altitudes on both strands, however.  Rather, there is a vertical
separation $\Delta z  = 0.023$~nm between corresponding sites on the
two strands.  With the relative phase of the strands and the vertical
separation between sites on those strands taken together,
corresponding sites on the two strands may be viewed as connected by
a helical segment of arc length $a^{(2)} =3.34$~nm at a pitch angle of
$\alpha = 0.394^{\circ}$.  This geometry is shown in Fig.~\ref{fig-Helices}.

When positively-charged dimers approach the DNA molecule, they will
be attracted to the negative charges at the sites on the
double-helix.  We consider dimers with a length
comparable to the spacing $a^{(1)}$ between sites on a strand
and having positive charges $+e$ at either end.  
The dimers can then adsorb onto the surface in two possible
orientations, either parallel to the strand or perpendicular to the helix
axis. (See Fig.~\ref{fig-DimerAdsorption}.)  If the dimer adsorbs parallel
to the strand, the positive charges from the dimer lie directly over
the negative charges on the strand, neutralizing the charge on two
adjacent sites.  If the dimer adsorbs perpendicular to the surface of the
bounding cylinder, one end of the dimer sits atop the site, while the other
extends radially outward.  This perpendicular adsorption effectively
inverts the charge on the site from $-e$ to $+e$.  In order to use a lattice-gas model, 
this geometric constraint is loosened by having the parallel dimer block only a single site. 
 This deficiency can be somewhat compensated by making binding energy of the parallel dimer twice as large, 
 $\varepsilon_\parallel \approx 2 \varepsilon_\perp$.

Note that because the length of the dimer (equal to the same-chain site spacing
$a^{(1)} =0.684$~nm) is much smaller than both the cross-chain site spacing
$a^{(2)} =3.34$~nm and the vertical separation $\Delta z  =3.4$~nm between turns of the helix, other orientations of the dimer are not possible. 

The problem thus described is a complex one, but the similarities with
the lattice gas models of condensed matter physics provide guidelines
for how to proceed.  These prescriptions, however, are aimed at the
treatment of a periodic crystalline lattice, and, although the DNA
sites exhibit helical symmetry, they do not constitute a
periodic lattice in the strict sense of the term.  However, an
appropriate choice of coordinates can take advantage of the helical
symmetry, so that, in these new coordinates, the positions of the sites will
fall on a regular, one-dimensional lattice.

\begin{figure}
\includegraphics[width=2in]{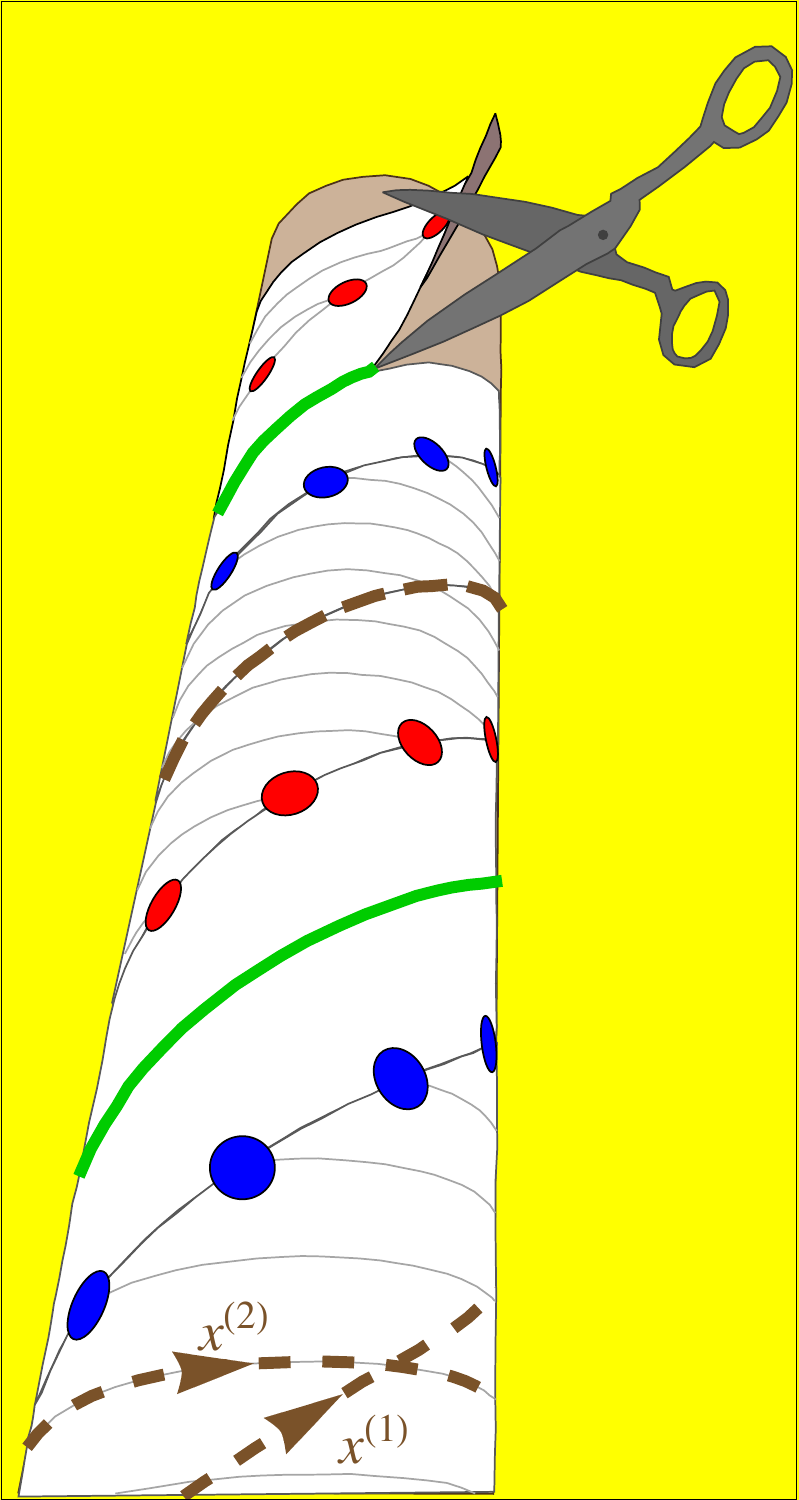}
\caption{DNA double helix wrapped around a cylindrical tube.  If the scissors cut along the solid green line, which is the center of the minor groove, then the paper can be put on a flat surface, with the double helix forming a one-dimensional chain.  The coordinate $x^{(1)}$ lies along the dashed brown line in the major groove, midway between the up and down chains, and $x^{(2)}$ lies along a path on the surface of the cylinder that connects the two phosphates on the two sub-chains that are connected by base pairs.}
\label{fig-DNAWrappedOnTube}
\end{figure}

\begin{figure}
\includegraphics[width=2.5in]{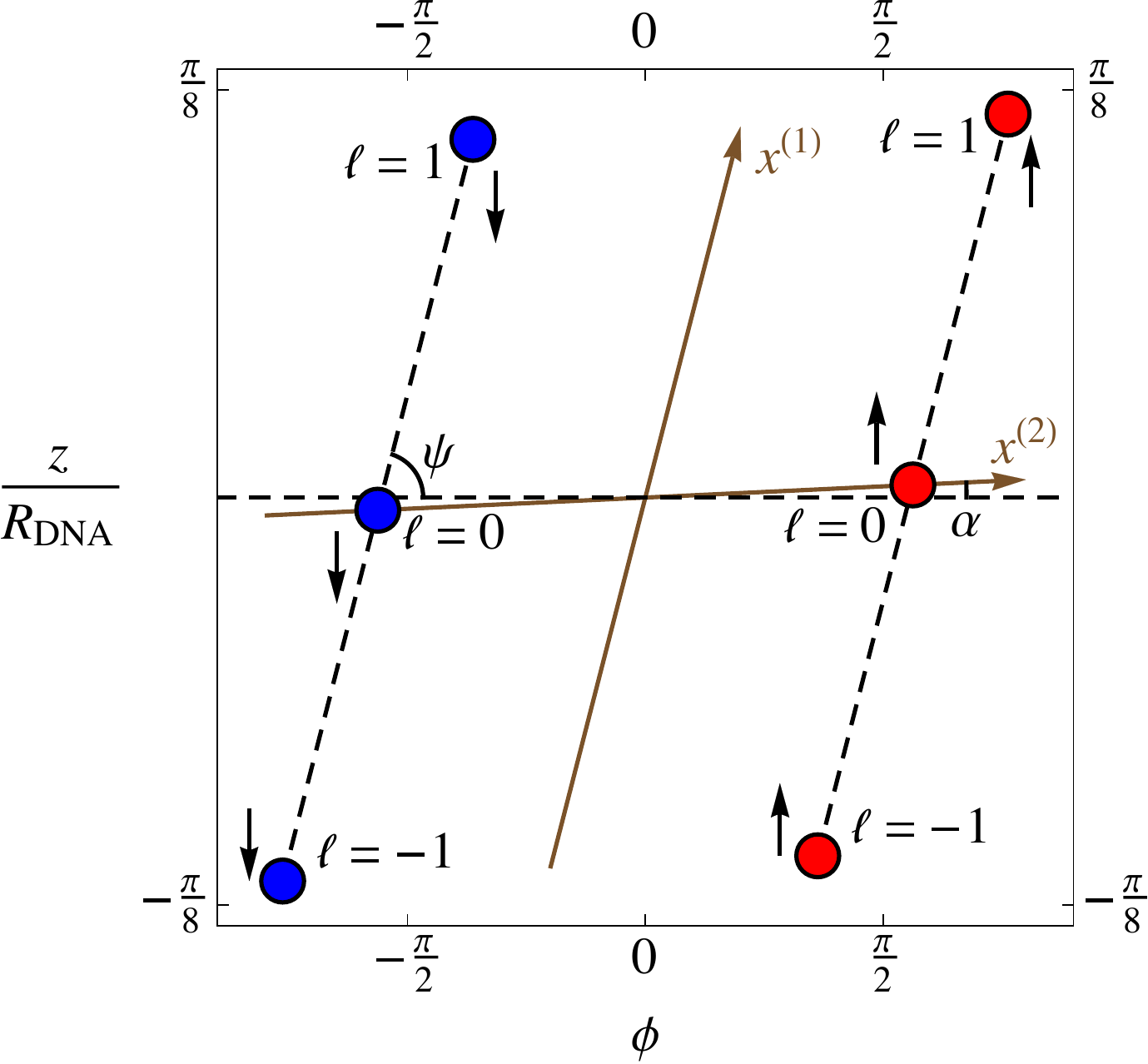}
\caption{Diagram of the new coordinates $x^{(1)}$ and $x^{(2)}$ in terms of the
cylindrical coordinates $(\phi,z)$.}
\label{fig-Coordinates}
\end{figure}

We will define these coordinates on a cylinder of radius $R_{DNA}$, as shown in Figs.~\ref{fig-Helices} and \ref{fig-DNAWrappedOnTube}.  The first
coordinate $x^{(1)}$ traces out a path with pitch angle $\psi$ along a single
helical strand, and the other coordinate
$x^{(2)}$ traces out a path with pitch angle $\alpha$ that connects
corresponding sites on the two strands.  Geometrically, a cylinder can be regarded as flat in the sense that it has no curvature.  In Fig.~\ref{fig-DNAWrappedOnTube}, we show the way that the cylinder can be cut with scissors and unwrapped so that this lattice can be mapped on a flat surface as shown in Fig.~\ref{fig-Coordinates}.  If we define the origin of coordinates $x^{(1)}$ and $x^{(2)}$ to be halfway along the helical path between the partners on the two chains, 
as shown in Figs.~\ref{fig-DNAWrappedOnTube} and \ref{fig-Coordinates}, then the positions of the sites on both strands form a one-dimensional lattice in the coordinates
$(x^{(1)},x^{(2)})$.

These coordinates can be written simply in terms of
cylindrical coordinates $(\phi,z)$ in matrix form as
\begin{equation}
\left[\begin{array}{c}
R_{\text{DNA}} \phi  \\
z
\end{array}\right] =
\left[\begin{array}{cc}
\cos\psi & \cos\alpha \\
\sin\psi & \sin\alpha
\end{array}\right]
\left[\begin{array}{c}
x^{(1)}  \\
x^{(2)}
\end{array}\right].
\end{equation}
Inverting this gives definitions of the two coordinates as
\begin{equation}
x^{(1)} = \left(\frac{\sin \alpha}{\sin(\alpha - \psi)}\right)
  R_{\text{DNA}}\phi- \left(\frac{\cos \alpha}{\sin(\alpha - \psi)}\right)z
\label{eq-coords1}
\end{equation}
and
\begin{equation}
x^{(2)} = - \left( \frac{\sin \psi}{\sin(\alpha - \psi)} \right)
 R_{\text{DNA}}\phi + \left(\frac{\cos \psi}{\sin(\alpha - \psi)}\right)z.
\label{eq-coords2}
\end{equation}
With these definitions, the difference in coordinates between
adjacent sites on the same strand is $\Delta x^{(1)} = a^{(1)}$, and the
difference in coordinates between corresponding sites on the two strands
is $\Delta x^{(2)} = a^{(2)}$.  That is, $a^{(1)}$ is the distance along the helical path of a single chain from one phosphate ion to the next, and $a^{(2)}$ is the distance along a helical path from a phosphate ion on one chain to its partner phosphate ion on the other chain.  

\begin{figure}
\includegraphics[width=2in]{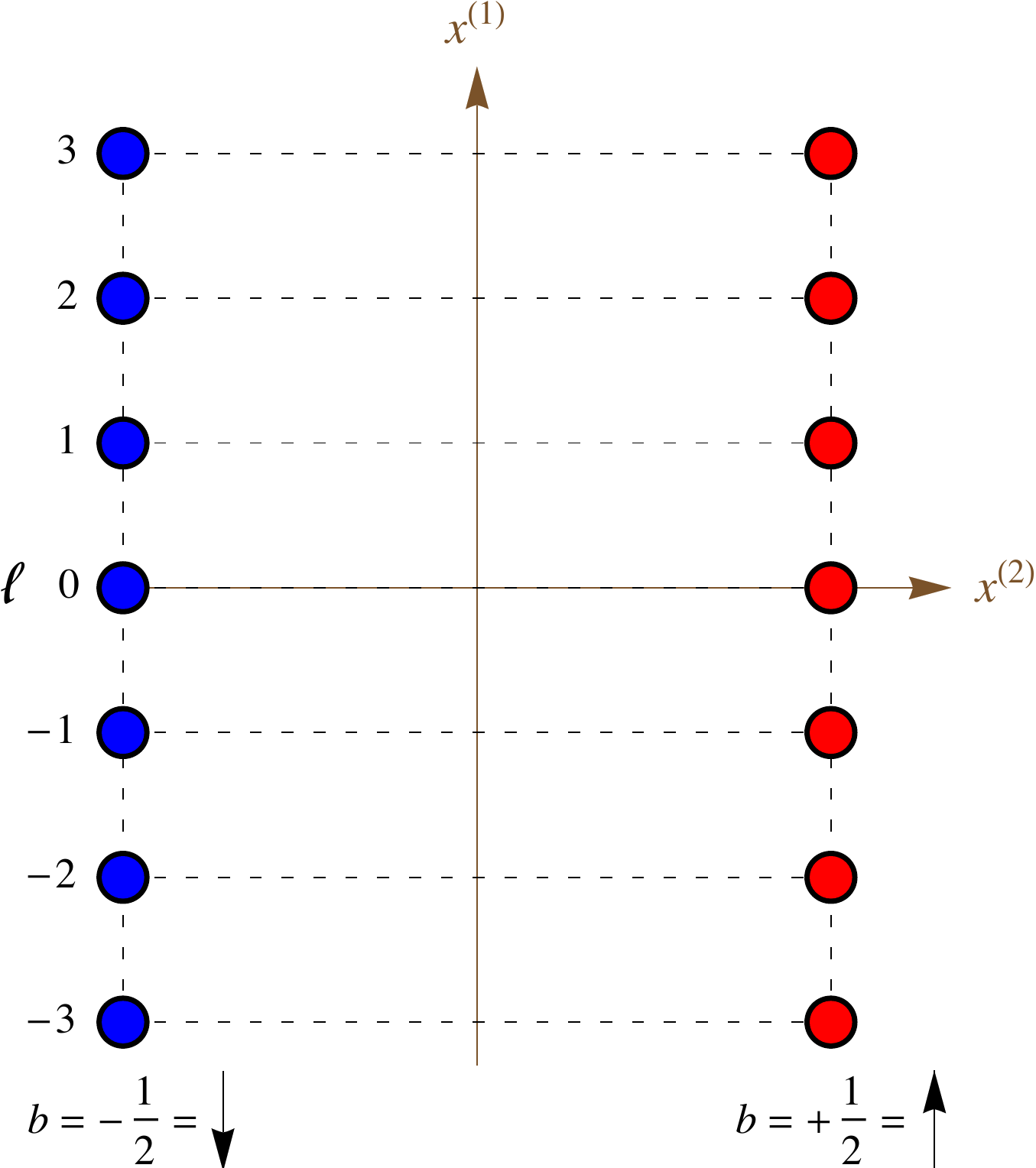}
\caption{In the new coordinates $x^{(1)}$ and $x^{(2)}$, the positions of the
sites constitute a one-dimensional lattice with two sites per unit cell.  Here $b=-\frac{1}{2}$ for the ``down"($\downarrow$) chain and $b=\frac{1}{2}$ for the ``up"($\uparrow$) chain.}
\label{fig-Unwound}
\end{figure}

Next we define a lattice index $\ell$, which specifies the cell (altitude on the double-helix) and chain index $b$, which specifies the basis site, where $b=-\frac{1}{2}$ for the ``down"($\downarrow$) chain and $b=\frac{1}{2}$ for the ``up"($\uparrow$) chain, as shown in Fig.~\ref{fig-Unwound}.   Using these variables, the coordinates can be written as
\begin{equation}
(x^{(1)},x^{(2)})=(\ell a^{(1)}, b a^{(2)})   \;
\end{equation}
where
\begin{equation}
\ell = 0,\pm 1, \pm2, \dots ; \enskip b = \pm\frac{1}{2}   \; .
\end{equation}
Thus, although the sites on the DNA molecule do not constitute a periodic lattice in
real space, they do constitute a lattice in an appropriately-defined
coordinate space (see Fig.~\ref{fig-Unwound}).  As we will see, however, this
choice of coordinates will make the form of the interaction potential
more complicated as a result.

The use of $(x^{(1)},x^{(2)})$ instead of the cylindrical coordinates $(\phi,z)$
indicates a more fundamental shift in our description of the DNA
double-helix.  The two-dimensional surface on which the helices lie is a
cylinder of radius $R_{DNA}$, and the helices inherit the cylinder's geometric properties.  The geometry of the cylinder, however, is locally indistinguishable from the geometry of the flat plane.  
One common consequence of this is that it
is possible to smoothly wrap a flat sheet of paper around a cylinder. 
In contrast, it is not possible to smoothly wrap a sheet of paper around
a sphere; this problem is well-known because of the geometrical
distortions that occur in flat maps of a spherical Earth.  Maps of a
cylindrical surface, however, have no such distortions.  This
geometrical difference is quantified by the Riemannian curvature tensor,
which vanishes for both the cylinder and the plane, but not for the
sphere\cite{mtw73}.  This means that the local geometry of the
cylinder behaves in exactly the same way as the local geometry of the
plane, so that a helix on a cylinder is geometrically equivalent to a
line on a plane.  

Our choice of coordinates is simply a map of the
cylindrical surface that reduces the double-helix to two parallel lines, as shown in Fig.~\ref{fig-Unwound}.  The result of this map is that we have a linear lattice with two phosphate sites per cell,  with the
 cells  labeled from $\ell=-{\mathcal N}$ to ${\mathcal N}$.
  All $2{\mathcal N}+1$ cells are identical, and the lattice can be imagined to satisfy periodic boundary conditions.

The analogy describing DNA as a ladder wrapped around a cylinder of
radius $R_{\text{DNA}}$ then has a true mathematical basis, because the
local structure of the double-helix is equivalent to the structure of
a ``ladder" --- a one-dimensional chain with a unit cell containing one
site from each strand.  If interactions are ignored, then the problem is described simply by this one-dimensional lattice.

\section{DNA Interaction Potential}
\label{sec-DNApotential}

The geometrical structure of the double helix is important in determining the electrostatic energy $U_{\text{int}}$ of
the sites (with and without dimers) interacting with one another.  The full Hamiltonian must contain these contributions and can be written as
\begin{equation}
{\mathcal
H}=\sum_{\ell,b}\sum_{\gamma}\varepsilon_\gamma  n_{\ell,b}^{(\gamma)}
+U_{\text{int}}   \;   ,
\label{eq-HwithU}
\end{equation}
where the interaction energy $U_{\text{int}}$ is the total interaction energy between all the charges on all the sites.  If 
${\mathcal V}_{(\ell_1,b_1),(\ell_2,b_2)}$  is the screened
Coulomb energy between a charge $q_{\gamma_1}$ at site $\ell_1$ on chain $b_1$ with another charge $q_{\gamma_2}$ at site $\ell_2$ on chain $b_2$, then $U_{\text{int}}$ can be written as
\begin{equation}
 U_{\text{int}}=\frac{1}{2} \sum_{\gamma_1,\gamma_2} \sum_{(\ell_1,b_1),(\ell_2,b_2)}
q_{\gamma_1} n_{\ell_1,b_1}^{(\gamma_1)} 
{\mathcal V}_{(\ell_1,b_1),(\ell_2,b_2)}
n_{\ell_2,b_2}^{(\gamma_2)} q_{\gamma_2}    \;  ,
\label{eq-UintwithVell1b1ell2b2}
\end{equation}
where $n_{\ell,b}^{(\gamma)}$ is the number of particles of species $\gamma$ on the lattice site at $(\ell,b)$ and $q_\gamma$ is the charge of that species, in units of the magnitude $e$ of the charge of an electron.  The factor of $1/2$ is to ensure that we are not double counting when we sum over all lattice sites, and we implicitly exclude same-site interactions $(\ell_1 , b_1) \neq (\ell_2 , b_2)$.  The total charge on a given site is given by summing all the charges on that site, so that the total charge on the site $(\ell,b)$ is
\begin{equation}
 \rho_{\ell,b} =\sum_{\gamma}  q_\gamma n_{\ell,b}^{(\gamma)}    \;  .
\end{equation}
The interaction potential can then be written in terms the the total charge on each site as
\begin{equation}
U_{\text{int}} = \frac{1}{2} \sum_{(\ell_1,b_1),\\(\ell_2,b_2)}
 \rho_{(\ell_1,b_1)}
 {\mathcal V}_{(\ell_1,b_1),(\ell_2,b_2)}
\rho_{(\ell_2,b_2)}   \; ,
\label{eq-UintUsingrho}
\end{equation}
where ${\mathcal V}$ is the electrostatic interaction energy between the sites.  This screened electrostatic energy between charges $q_{\gamma_1}$ and  $q_{\gamma_2}$ at positions $(\ell_1,b_1)$ and $(\ell_2,b_2)$, (in SI units) is
\begin{equation}
\mathcal{V}_{(\ell_1,b_1),(\ell_2,b_2)} 
=\frac{e^2}{4 \pi \epsilon }  \frac{e^{-q_s d_{b_1 b_2}(\ell_1-\ell_2)}}
{d_{b_1 b_2}(\ell_1-\ell_2)} 
 \;   ,
\label{eq-screenedCoulombOfD}
\end{equation}
where $\epsilon$ is the electric permittivity of the medium between the two charges, and $d_{b_1 b_2}(\ell_1-\ell_2)$ is the straight-line distance between the two charges.  Distances along the chain are invariant under translations by a lattice spacing, and so  $d_{b_1 b_2}(\ell_1-\ell_2)$ depends only on the difference between the lattice site indices $\ell_1-\ell_2$.  As in the noninteracting lattice-gas model, we have three species of ``particles" on the lattice sites, vacancies of charge $q_v=-1$, parallel ($\parallel$) dimers of charge $q_\parallel=0$, and perpendicular ($\perp$) dimers of charge $q_\perp=+1$.

Under physiological conditions, the presence of monovalent salt ions such as Na$^+$
leads to screening of the bare charges.
Traditionally, screening is treated by modeling the ions as a
continuous density, resulting in the nonlinear Poisson-Boltzmann
equation \cite{ck96, gns02}. The Poisson-Boltzmann equation is not
analytically solvable, so it is often further approximated by linearization.  The
resulting Thomas-Fermi model is analytically solvable and gives the
screened Coulomb (or Yukawa) potential between two charges given in Eq.~(\ref{eq-screenedCoulombOfD}) where $\epsilon=78.5\epsilon_0$ is the permittivity of water (with $\epsilon_0$ the permittivity of free space) and $q_s$ is the magnitude of the screening wave vector \cite{am76}.  Various names are ascribed to both the nonlinear and linearized equations, including Debye-H{\"u}ckel, Thomas-Fermi, and Poisson-Boltzmann, but we will always refer to the
Poisson-Boltzmann equation when we mean the nonlinear form and the (linearized)
Thomas-Fermi equation when we mean the linearized form. 

One must be cautious in using the screened Coulomb potential for screened
interactions.  The continuous density approximation from which the nonlinear
Poisson-Boltzmann equation was derived constitutes a form of mean-field
theory\cite{gns02}, which fails to describe any of the effects due to
correlations between the ions such as charge inversion and condensation.  Thus
we cannot model screening of the DNA molecule by the dimers using the
Thomas-Fermi model, or even the Poisson-Boltzmann equation.  Instead, we must
treat the dimers as individual particles and compute their interactions so that
we do not ignore correlation effects.  

For the screening due to the monovalent salt, however, the valence
involved is small enough that a mean-field treatment may still be
accurate \cite{gns02}.  Thus, we will treat the interaction energy $U_{\text{int}}$ in
Eq.~(\ref{eq-UintUsingrho}) using the screened Coulomb potential for the dimer-dimer
interactions, with the screening vector $q_s$ given in the Thomas-Fermi model as
\cite{am76}
\begin{equation}
q_s = \sqrt{2 n   \frac{\beta e^2}{\epsilon}}
=\sqrt{8\pi n \ell_B},
\label{eq-ScreeningVector}
\end{equation}
where $\beta = 1/{k_B}T$ is the inverse temperature, $n$ is the
concentration of ions, and $\ell_B=\beta e^2/(4\pi \epsilon)$ is the Bjerrum length
\cite{kn02}.  The Bjerrum length is a characteristic length scale
equal to the distance at which two proton charges interact with
energy $k_B T$.  It should be noted that these expressions differ
slightly in the literature because of various authors' choices of units
for the electrostatics.  Here, we use SI units.

By choosing a coordinate system based on the arc lengths around the
surface of the cylinder, we have managed to align the sites into a
regular lattice, but the potential $\mathcal{V}_{(\ell_1,b_1),(\ell_2,b_2)}$
depends on the {\it straight-line} distance between the sites rather
than the surface arc-length connecting them.  We must then express
Eq.~(\ref{eq-screenedCoulombOfD}) in terms of the straight-line distance
function $d_{b_1 b_2}(\ell_1-\ell_2)$ between the sites at 
$(\ell_1 a^{(1)},b_1 a^{(2)})$ and $(\ell_2 a^{(1)},b_2 a^{(2)})$.  The straight-line distance 
in cylindrical coordinates between two points on the surface of a cylinder of radius
$R_{\text{DNA}}$ is
\begin{eqnarray}
&&{\mbox{distance}}(\phi_1,z_1,\phi_2,z_2)=
 \nonumber   \\
&&\left\{ 2 R_{\text{DNA}}^2 \left[1-\cos\left(\phi_1-\phi_2\right)\right]
 +\left(z_1-z_2\right)^2 \right \} ^{\frac{1}{2}},
 \label{eq-distance}
\end{eqnarray}
and applying our change of variables (\ref{eq-coords1},\ref{eq-coords2})
gives the straight-line distance between sites as
\begin{eqnarray}
d_{b_1 b_2}(\ell) &=& R_{\text{DNA}} 
\left\{ 2 \left[1 -
\right. \right. \nonumber \\
  & & \left. \left. - \cos\left(\frac{a^{(1)}}{R_{\text{DNA}}}\ell \cos  \psi
 +\frac{a^{(2)}}{R_{\text{DNA}}} b \cos  \alpha \right)\right]
\right. \nonumber \\
&&+\left.\left[\frac{a^{(1)}}{R_{\text{DNA}}} \ell \sin  \psi+\frac{a^{(2)}}{R_{\text{DNA}}} b \sin 
\alpha\right]^2
\right\}^{\frac{1}{2}}  \; , \nonumber   \\
\label{eq-gendist}
\end{eqnarray}
where $\ell=\ell_1 - \ell_2$ and $b=b_1-b_2$.  Because of the translational invariance of the lattice, the distance in Eq.~(\ref{eq-gendist}) depends 
only on the differences between the lattice and basis indices, and not their values individually.  
From this relation, we see the symmetry property of the distance,
\begin{subequations}
	\label{eq-dSymmetry}
	\begin{align}
	d_{b_1 b_2}(\ell)  &= d_{b_2 b_1}(-\ell)   
	\\
	d_{\uparrow \uparrow}(\ell)  &= d_{\downarrow \downarrow}(-\ell)   \; .
	\end{align}
\end{subequations}

\begin{figure}
\includegraphics[width=3in]{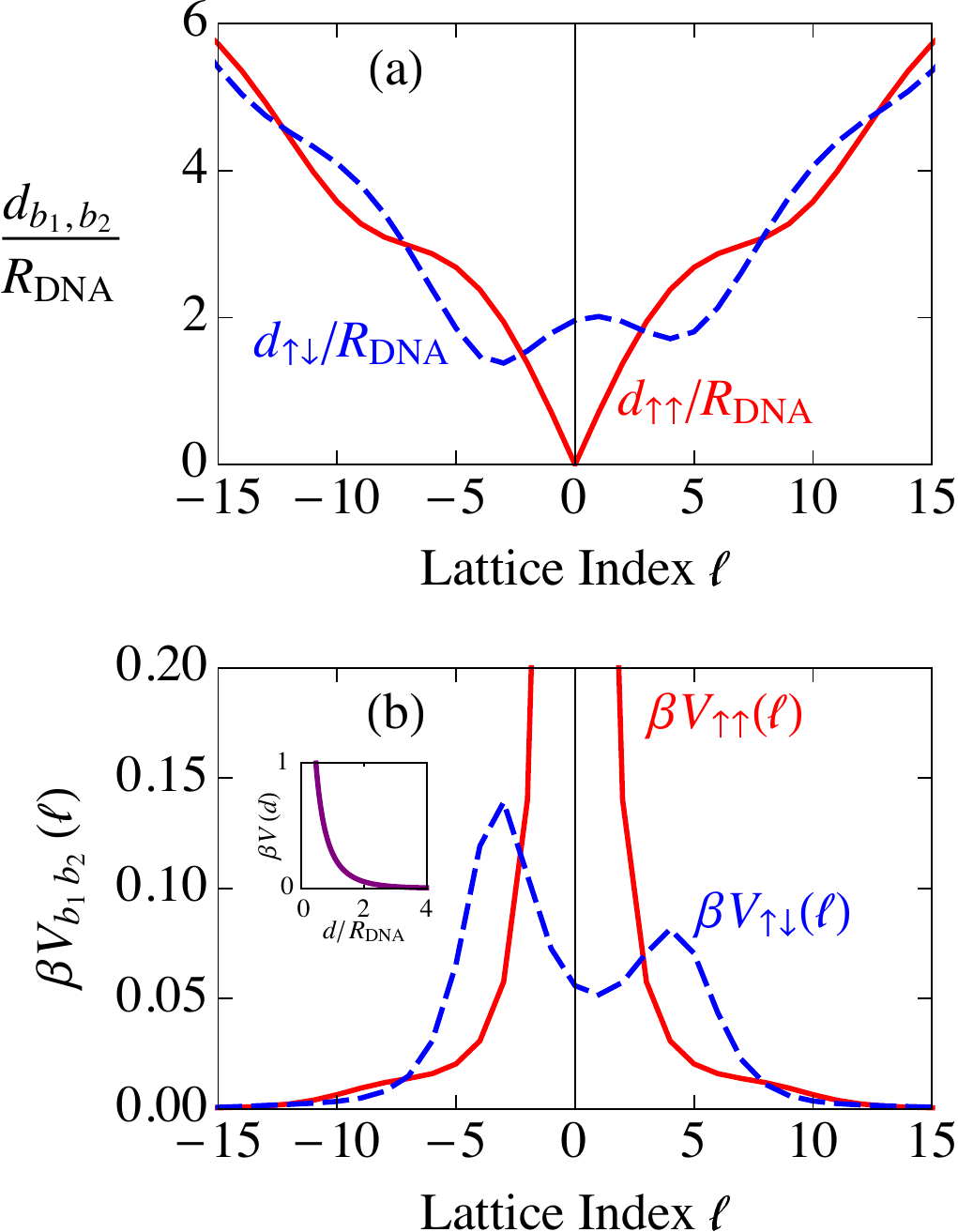}
\caption{(a) Distance $d_{b_1 b_2}(\ell)/R_{\text{DNA}}$ between sites and
(b) screened Coulomb potential $V_{b_1b_2}(\ell_1-\ell_2)$ as a function of lattice index $\ell$, within the same chain ($\uparrow \uparrow$) and between the two chains ($\uparrow \downarrow$).
The red curve ($\uparrow\uparrow$),
representing same-chain repulsion, is an even function, and the blue curve
($\uparrow\downarrow$), representing opposite-chain repulsion, is neither even
nor odd.   The point $\ell=0$ for the same chain repulsion is omitted because it would correspond to a site interacting with itself.  The inset in (b) is the screened Coulomb interaction as a function of the straight-line distance $d$ between charges.
At the physiological temperature of $T=37^\circ$~C~(300.15 K), $\beta \equiv 1/k_\text{B} T \approx$~1/(27 meV). }
\label{fig-CombodVofL}
\end{figure}

Since $b_1$ and $b_2$ are either $+\frac{1}{2}$ or $-\frac{1}{2}$, $b=b_1-b_2$ takes the three values $b=0$ for interactions of one site on a single chain with all other sites on the same chain, $b=1$ for interactions between a site on the ``up" chain with with all the sites on the ``down" chain, and $b=-1$ for interactions of one site on the ``down" chain with all the sites on the ``up" chain.  We will write an up arrow $\uparrow$ will indicate $b_{1,2}=\frac{1}{2}$, and a down arrow $\downarrow$ will indicate $b_{1,2}=-\frac{1}{2}$.
Plots of $d_{\uparrow \uparrow}(\ell)/R_{\text{DNA}}$ and 
$d_{\uparrow \downarrow}(\ell)/R_{\text{DNA}}$ are shown in Fig.~\ref{fig-CombodVofL}a.  There are wiggles in the curve because, as the chain wraps around the cylinder (see Fig.~\ref{fig-DimerAdsorption}), the distances between lattice sites can become larger or smaller than they would be on a linear chain.

In order to understand the consequences of the screened Coulomb potential (\ref{eq-screenedCoulombOfD}) depending on the straight-line distances $d$ between sites, we define a new notation
for the potential, which is expressed in terms of the difference in lattice sites indices $\ell_1-\ell_2$,
\begin{equation}
V_{b_1b_2}(\ell_1-\ell_2)\equiv \mathcal{V}_{(\ell_1,b_1),(\ell_2,b_2)} 
\label{eq-VmatrixInTermsofFunction}
\end{equation}
This potential decays exponentially as a function of the straight-line distance $d$ (inset of Fig.~\ref{fig-CombodVofL}(b)). However, as a function of the lattice index $\ell$ as in
Fig.~\ref{fig-CombodVofL}(b), there are corresponding ``wiggles" in the decay.  This occurs
because the distance between the sites decreases slightly as the helix completes
a turn.  These wiggles are present for every turn the helix makes, but the effect on the potential becomes negligibly small after about the first turn.  Thus, by using the indices $\ell$ and $b$ to describe the relationships between sites, we have reduced the DNA double-helix structure to a regular one-dimensional lattice with two sites per cell at the cost of an irregular potential due to the geometrical structure of the DNA molecule.  The potentials for other combinations of $b_1$ and $b_2$ can be related to those in Fig.~\ref{fig-CombodVofL}(b) through the symmetry relations
\begin{subequations}
	\label{eq-VellSymmetry}
	\begin{align}
	V_{b_1b_2}(\ell) &= V_{b_2 b_1}(-\ell)   
	\\
	V_{\uparrow \uparrow}(\ell) &= V_{\downarrow \downarrow}(-\ell)   \; .
	\end{align}
\end{subequations}

Fourier transforming  in the site index $\ell$ block diagonalizes this matrix into  $2 \times 2$ blocks, corresponding to the chain indices $b_1$ and $b_2$. 
Explicitly, the Fourier transform is
\begin{equation}
{\tilde {\mathcal V}}  = 
{\mathcal F}^\dagger {\mathcal V} {\mathcal F}  \; ,
\end{equation}
where  the elements of ${\mathcal F}$  are independent of $b_1$ and $b_2$ and are given by
\begin{equation}
{\mathcal F}_{(\ell,b_1),(k,b_2)}=
\frac{1}{\sqrt{2{\mathcal N}+1}}
e^{i k \ell}               \; ,
\label{eq-FourierTransformMatrix}
\end{equation}
and the dagger denotes the adjoint (complex conjugate transpose).
We use periodic boundary conditions, so that the wave vector $k$  appearing here, which is dimensionless, takes the $2 {\mathcal N} +1$ values 
\begin{equation}
   k=-{\mathcal N}\Delta k \cdots {\mathcal N}\Delta k \; ,
\end{equation}
where
\begin{equation}
   \Delta k=\left(\frac{2 \pi}{2 {\mathcal N} +1}\right)\;  .
\end{equation}
If ${\mathcal N}$ is large, the range of $k$ is essentially continuous from $-\pi$ to $\pi$.

The matrix elements of this Fourier transform are
\begin{eqnarray}
{\tilde {\mathcal V}}_{(k_1,b_1),(k_2,b_2)}&=& 
\frac{1}{2{\mathcal N}+1}
\sum_{\ell_1, \ell_2 = -{\mathcal N}}^{\mathcal N}
 e^{-i k_1 \ell_1}
\times \nonumber \\   &&\times
{\mathcal V}_{(\ell_1b_1),(\ell_2,b_2)}e^{i k_2 \ell_2}    \; .    
\end{eqnarray}
Substituting for ${\mathcal V}$ from
Eq.~(\ref{eq-VmatrixInTermsofFunction}), we have
\begin{eqnarray}
{\tilde {\mathcal V}}_{(k_1,b_1),(k_2,b_2)}&=&
\frac{1}{2{\mathcal N}+1}
\sum_{\ell_1, \ell_2 = -{\mathcal N}}^{\mathcal N}
e^{-i (k_1  \ell_1 - k_2 \ell_2)}
\times \nonumber \\ &&\times
V_{b_1 b_2}(\ell_1-\ell_2)
   \; .   
\end{eqnarray}

Since we regard the chain as having periodic boundary conditions, changing summation indices to $\ell=\ell_1 - \ell_2$ and $\ell_2$ gives
\begin{eqnarray}
{\tilde {\mathcal V}}_{(k_1,b_1),(k_2,b_2)}&=&
\frac{1}{2{\mathcal N}+1}
\sum_{\ell= -{\mathcal N}}^{\mathcal N}
e^{-i k_1 \ell } V_{b_1 b_2}(\ell)
\times \nonumber \\ &&\times
 \sum_{\ell_2= -{\mathcal N}}^{\mathcal N}
e^{-i(k_1- k_2 )\ell_2}
   \; ,   
\end{eqnarray}
where the sum over $\ell_2$ becomes
\begin{equation}
\sum_{\ell_2= -{\mathcal N}}^{\mathcal N}
e^{-i(k_1- k_2 ) \ell_2}=(2{\mathcal N}+1) \delta_{k_1,k_2}  \; .
\label{eq-sumexpikell}
\end{equation}

\begin{figure}
\includegraphics[width=3in]{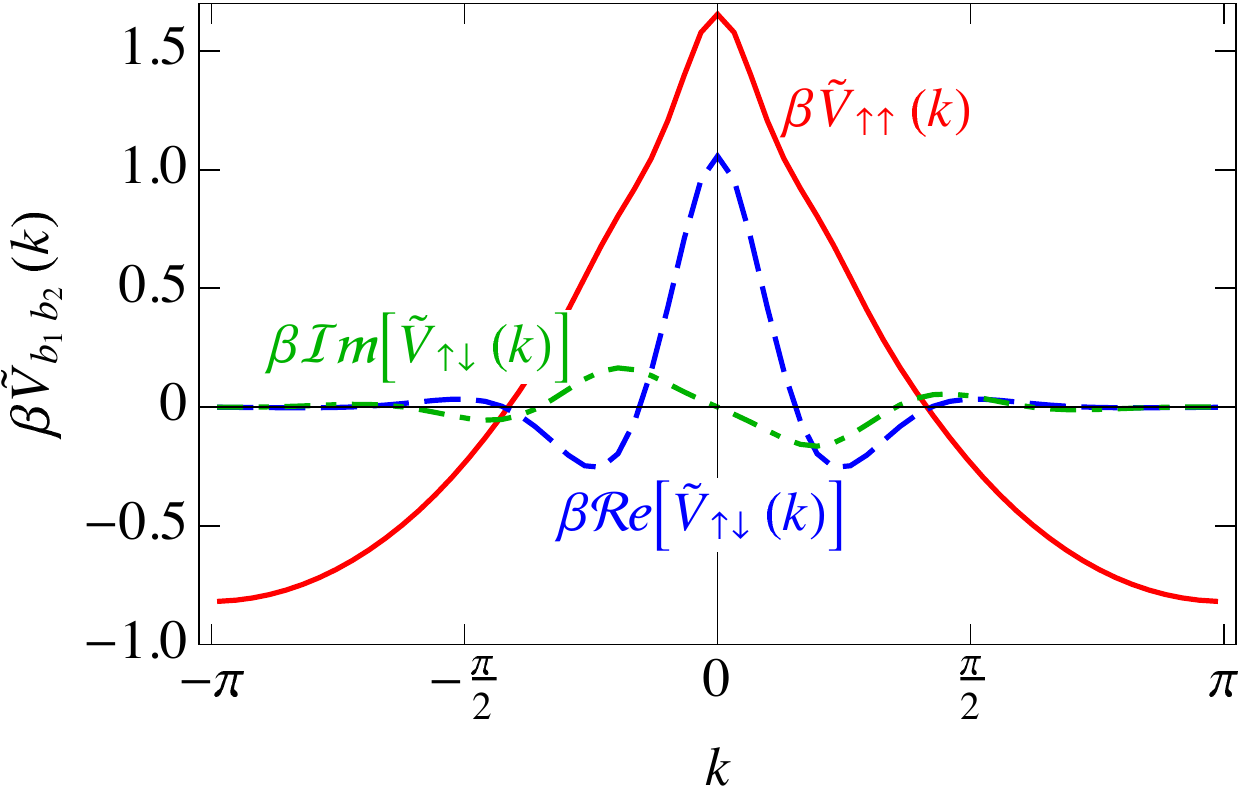}
\caption{Plots of the Fourier transforms $\tilde V_{\uparrow\uparrow}(k)$ (solid red curve) and
$\tilde V_{\uparrow\downarrow}(k)$ against the dimensionless wavevector $k$.  The symbols ${\mathscr Re}$ and ${\mathscr Im}$ correspond to the real (blue dashed curve) and imaginary parts (green dot-dashed curve) of the complex Fourier transform, respectively.}
\label{fig-VofK10}
\end{figure}

We can now write the Fourier transform of $V_{b_1 b_2}(\ell)$ as
\begin{equation}
  {\tilde V_{b_1 b_2}(k)}=
\sum_{\ell= -{\mathcal N}}^{\mathcal N}
e^{- i k \ell }V_{b_1 b_2}(\ell)  \;.
\label{eq-FourierTransform}
\end{equation}
Substituting this back into the matrix, we have
\begin{equation}
{\tilde {\mathcal V}}_{(k_1,b_1),(k_2,b_2)}=
\delta_{k_1,k_2}{\tilde V_{b_1 b_2}(k_1)}
   \; .  
\end{equation}
Thus, the Fourier transform ${\tilde {\mathcal V}}$ of ${\mathcal V}$ is diagonal in the $k$ indices 
and contains $2\times 2$ blocks
\begin{equation}
\tilde V(k) = \left[ \begin{array}{cc} 
\tilde V_{\downarrow\downarrow}(k)    &  \tilde V_{\downarrow\uparrow}(k)   \\
\tilde V_{\uparrow\downarrow}(k)  &  \tilde V_{\uparrow\uparrow}(k) \\
\end{array} \right]  
\label{eq-2by2Blocks1}  \;.
\end{equation}
The symmetries (\ref{eq-VellSymmetry}) of the screened Coulomb potential on the DNA lattice are reflected in the Fourier transforms (\ref{eq-FourierTransform}) as well:
\begin{subequations}
	\begin{align}
\tilde V_{b_1 b_2}(k) &= \tilde V_{b_2 b_1}(k)^* =\tilde V_{b_2 b_1}(-k)  
\label{eq-FourierSymmetry}
	\\
	\tilde V_{\uparrow\uparrow}(k) &= \tilde V_{\downarrow\downarrow}(k) \;  .
	\label{eq-SameChainFourierSymmetry}
	\end{align}
\end{subequations}
Also, because $V_{\uparrow\uparrow}(\ell)$ is an even function of $\ell$,  $\tilde V_{\uparrow\uparrow}(k)$ is real.
The Fourier transforms (\ref{eq-FourierTransform}) are plotted in
Fig.~\ref{fig-VofK10}.

As can be seen in Fig.~\ref{fig-VofK10}, the Fourier transforms $\tilde V(k)$ have regions of $k$ that are negative,
which is a result of transforming with respect to our helical coordinate system. 
When the screened Coulomb potential (\ref{eq-screenedCoulombOfD}) is Fourier transformed with
respect to the straight-line distance $d$, the function is positive definite. 
We have instead Fourier transformed the potential a function of the lattice index $\ell$,
resulting in the ``wiggles" in Fig.~\ref{fig-CombodVofL}(b) caused by the turns of the
helix.  These deviations from the exponential decay of $V(d)$ result in the
deviations here from the positive-definite Fourier transform.

\begin{figure}
\includegraphics[width=3in]{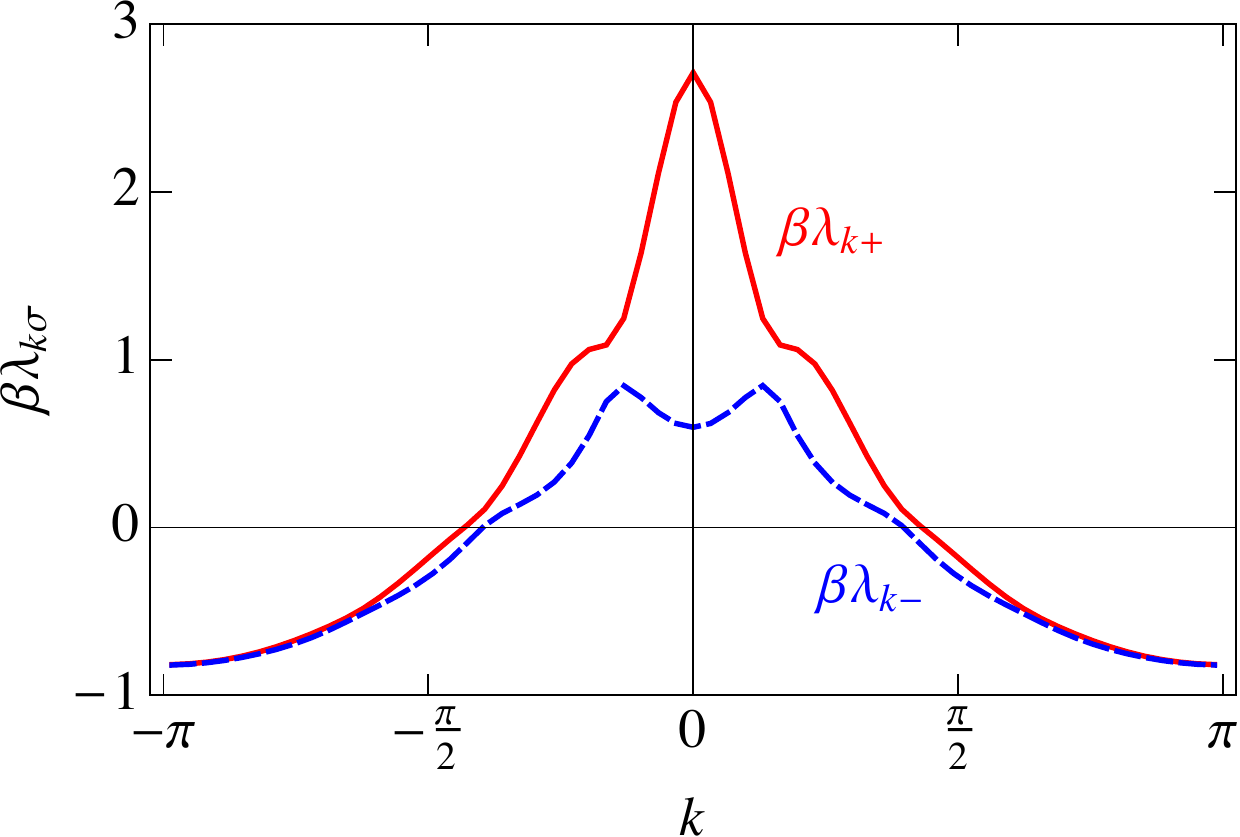}
\caption{Plots of the eigenvalues $\lambda_{k\pm}$ of the $2\times 2$ blocks of $\tilde
V(k)$.}
\label{fig-Eigenvalues}
\end{figure}

Using the symmetry operations in Eqs. (\ref{eq-FourierSymmetry}) and (\ref{eq-SameChainFourierSymmetry})
obeyed by the Fourier transforms allows us to simplify the $2\times 2$ blocks $\tilde V(k)$
in Eq.~(\ref{eq-2by2Blocks1}) to
\begin{equation}
\tilde V(k) = \left[ \begin{array}{cc} 
\tilde V_{\uparrow\uparrow}(k)    &  \tilde V_{\uparrow\downarrow}^*(k)   \\
\tilde V_{\uparrow\downarrow}(k)  &  \tilde V_{\uparrow\uparrow}(k) \\
\end{array} \right]  
\label{eq-2by2Blocks2}.
\end{equation}
This matrix is easily diagonalized, yielding the real eigenvalues
\begin{equation}
\lambda_{k,\pm}=\tilde V_{\uparrow\uparrow}(k) \pm 
\vert\tilde V_{\uparrow\downarrow}(k)\vert,
\label{eq-Eigenvalues}
\end{equation}
which are plotted in Fig.~\ref{fig-Eigenvalues}.
The transformation that diagonalizes $\tilde V(k)$ is a unitary matrix
$\xi(k)$ whose columns are the eigenvectors of $\tilde V(k)$.   Choosing the $\lambda_{k+}$ eigenvector to be first, this transformation matrix is given by
\begin{equation}
\xi(k)=\frac{1}{\sqrt{2}}
\left[ \begin{array}{cc} 
1    &  1   \\
\frac{\tilde V_{\uparrow\downarrow}(k)}{\vert \tilde
V_{\uparrow\downarrow}(k)\vert} & 
-\frac{\tilde V_{\uparrow\downarrow}(k)}{\vert \tilde
V_{\uparrow\downarrow}(k)\vert} \\
\end{array} \right]  
\label{eq-Eigenvectors}.
\end{equation}
This diagonalizes the $2\times 2$ matrix $\tilde V(k)$ according to
\begin{equation}
\left[\begin{array}{cc}
\lambda_+(k) &   0         \\
      0      & \lambda_-(k) \\
\end{array}\right]
= \xi^\dagger (k) \tilde V (k) \xi (k).
\end{equation}

We can write the combined process of Fourier transforming ${\mathcal V}$ into
block-diagonal form and then diagonalizing the $2\times 2$ blocks $\tilde V(k)$
as a single unitary transformation ${\mathcal M}$ given by
\begin{equation}
{\mathcal M}_{(\ell, b),(k, \sigma)}=\frac{\xi_{b,\sigma}(k)}{\sqrt{2{\mathcal N}+1}}
e^{i k \ell},
\end{equation}
which completely diagonalizes ${\mathcal V}$ such that 
\begin{equation}
 \Lambda={\mathcal M}^\dagger {\mathcal V} {\mathcal M}   .
\end{equation}
Here $\Lambda$ is a diagonal matrix, with each element along the diagonal given by:
\begin{equation}
\Lambda_{(k_1,\sigma_1), (k_2,\sigma_2)}=\delta_{k_1,k_2}\delta_{\sigma_1,\sigma_2}\lambda_{k_1 \sigma_1}  \;  ,
\end{equation}
where $\sigma$ is either $+$ or $-$.

Although we have diagonalized the potential with a unitary transformation, the fact that the transformation matrix is complex means that the charge density in the diagonal basis could also be complex, which is physically undesirable.  However, the potential matrix in the position basis is real and symmetric, which means that it should be diagonalizable by a real orthogonal transformation matrix ${\mathcal W}$.  In fact, the transformation matrix is not unique, since there is a degeneracy in the eigenvalues, since $\lambda_{k,\sigma}=\lambda_{-k,\sigma}$.  This means that we can choose eigenvectors that are real by taking two orthogonal linear combinations of the eigenvectors of the unitary transformation $\cal{M}$ for $k$ and $-k$.  Those two new eigenvectors, which represent the columns $(k,\sigma)$ for $k>0$ and $k<0$ of the new transformation matrix ${\mathcal W}$, are given by
\begin{eqnarray}
{\mathcal W}_{(\ell,b),(k,\sigma)} &=&
 \frac{1}{\sqrt {2}}
\left\{ {\mathcal M}_{(\ell, b),(k, \sigma)}+
{\mathcal M}_{(\ell, b),(-k, \sigma)} \right\}  \; ,
\nonumber  \\ &&k > 0    \; ,
\end{eqnarray}
and
\begin{eqnarray}
{\mathcal W}_{(\ell,b),(k,\sigma)}
&=& i  \frac{1}{\sqrt {2}}
\left\{ {\mathcal M}_{(\ell, b),(k, \sigma)} -
{\mathcal M}_{(\ell, b),(-k, \sigma)} \right\} \; ,  
\nonumber \\ &&k < 0    \; .
\end{eqnarray}
Since $\xi_{b \sigma}(-k)=\xi_{b \sigma}^*(k)$, these can be written in terms of real and imaginary parts as
\begin{eqnarray}
{\mathcal W}_{(\ell,b),(k,\sigma)}&=&
\frac{\sqrt 2}{\sqrt {(2{\mathcal N}+1)}}
\text{Re}\left\{ e^{ ik \ell} \xi_{b \sigma}(k) \right\}\; ,  
\nonumber \\ &&k > 0    \; .
\label{eq-Wtranskgt0}
\end{eqnarray}
and
\begin{eqnarray}
{\mathcal W}_{(\ell,b),(k,\sigma)}
&=&  -\frac{\sqrt 2}{\sqrt {(2{\mathcal N}+1)}}
\text{Im}\left\{ e^{ ik \ell} \xi_{b \sigma}(k) \right\}\; ,  
\nonumber \\ &&k < 0    \; .
\label{eq-Wtransklt0}
\end{eqnarray}
When $k=0$, the orthogonal transformation is the same as the unitary transformation, that is
\begin{equation}
{\mathcal W}_{(\ell, b),(0, \sigma)}={\mathcal M}_{(\ell, b),(0, \sigma)}=\frac{\xi_{b,\sigma}(0)}{\sqrt{2{\mathcal N}+1}}  \; ,
\label{eq-Wtranskeq0}
\end{equation}
where
\begin{equation}
\xi(0)=\frac{1}{\sqrt{2}}
\left[ \begin{array}{cc} 
1    &  1   \\
1    & -1   \\   
\end{array} \right]   \; .
\label{eq-Eigenvectorskeq0}
\end{equation}
Here the columns are labeled by the eigenvalue $\sigma=\pm$, and the rows are labeled by the chain $b=-1/2$ for the first row and $b=1/2$ for the second row.

This transformation can be written in a more compact form by introducing the unitary transformation matrix $\mathcal{Z}$, whose elements are defined by
\begin{equation}
{\mathcal{Z}}_{(k \sigma),(k'  \sigma')}=\delta_{\sigma,\sigma'}
\left\{
 \begin{array}{cc} 
\frac{i}{\sqrt{2}}(\delta_{k,k'} - \delta_{k,-k'})    &  k<0   \\
\delta_{k,k'}    & k=0   \\
\frac{1}{\sqrt{2}}(\delta_{k,k'} + \delta_{k,-k'})       & k>0   
\end{array} \right.       \; \;\;   .
\label{eq-ZTransformDef}
\end{equation}
Then the orthogonal transformation is
\begin{equation}
\mathcal{W} = \mathcal{M} \mathcal{Z}    \; .
\label{eq-WdefMZ}
\end{equation}

In deriving the thermodynamic quantities for the interacting lattice gas model, it will be convenient to be able to use this orthogonal transformation to diagonalize the potential.

\section{Partition Function for Interacting DNA  Chains}
\label{sec-PartitionFunctionInteracting}

For the interacting lattice gas, the partition function is similar to the noninteracting one in Eqs.~(\ref{eq-ZG1}) and (\ref{eq-ZNI1}).  However, here the full Hamiltonian ${\mathcal H}$ from Eq.~(\ref{eq-HwithU}) is used, and then the partition function is given by
\begin{equation}
Z_G = \sum_{\text{configurations}} e^{-\beta\left({\mathcal
H}-\sum_{\ell,b}\sum_{\gamma}\mu_{\gamma}n_{\ell,b}
^{(\gamma)}\right)}      \; .
\label{eq-ZGen2}
\end{equation}
The full Hamiltonian, including the interaction term from Eq.~(\ref{eq-UintUsingrho}) written in matrix form, is
\begin{equation}
{\mathcal H}=\sum_{\ell,b}\sum_{\gamma}\varepsilon_\gamma n_{\ell,b}^{(\gamma)}
+\frac{1}{2} \rho^T {\mathcal V} \rho
\label{eq-HwithUMatrixForm}
\end{equation}
In Sec.~\ref{sec-DNApotential}, we showed that the orthogonal transformation matrix ${\mathcal W}$ diagonalizes ${\mathcal V}$.  In order to streamline the calculation, it will be useful to insert the identity matrix in the form ${\mathcal W}{\mathcal W}^T$, to the left and right of of ${\mathcal V}$ in the interaction term of the Hamiltonian, which gives
\begin{equation}
{\mathcal H}=\sum_{\ell,b}\sum_{\gamma}\varepsilon_\gamma n_{\ell,b}^{(\gamma)}
+\frac{1}{2} (\rho^T {\mathcal W})({\mathcal W}^T{\mathcal V}{\mathcal W})({\mathcal W}^T \rho)   \;,
\label{eq-HwithUMatrixFormtransform}
\end{equation}
where we have grouped factors to show the transformed quantities.  

The Hamiltonian, with the interaction term written in the diagonal basis, can then be written as
\begin{eqnarray}
{\mathcal H}&=&\sum_{\ell,b}\sum_{\gamma}\varepsilon_\gamma n_{\ell,b}^{(\gamma)}
+\frac{1}{2} {\tilde \rho}^2 \Lambda 
\nonumber \\
&=&  \sum_{\ell,b}\sum_{\gamma}\varepsilon_\gamma n_{\ell,b}^{(\gamma)}
+\frac{1}{2} \sum_{k,\sigma} (\tilde \rho_{k,\sigma})^2 \lambda_{k,\sigma}  \;  ,
\label{eq-HwithUMatrixFormDiag}
\end{eqnarray}
where $\tilde \rho={\mathcal W}^T \rho$ is the transformed charge density.  We have not transformed the first term in the Hamiltonian to the diagonal basis because it is more convenient later in the calculation to have it in the position basis.  Substituting this form into the partition function and writing the resulting expression as two separate exponentials, we have
\begin{eqnarray}
Z_G &=& \sum_{\text{configurations}} 
e^{-\beta\
\sum_{\ell,b}\sum_{\gamma}(\varepsilon_\gamma - \mu_{\gamma})n_{\ell,b}^{(\gamma)}}
\times \nonumber \\   &&   \; \; \; \;  \; \; \;  \; \; \; \;  \; \; \;  \times
 e^{-\frac{\beta}{2}\sum_{k,\sigma} \tilde \rho_{k,\sigma}^2 \lambda_{k,\sigma}}   \; .
\label{eq-ZdiagU}
\end{eqnarray}
We can write the exponential of the sum in the interaction term as a product of exponentials as
\begin{eqnarray}
Z_G &=& \sum_{\text{configurations}} 
e^{-\beta\
\sum_{\ell,b}\sum_{\gamma}(\varepsilon_\gamma - \mu_{\gamma})n_{\ell,b}^{(\gamma)}}
\times \nonumber \\ &&     \; \; \; \;  \; \; \;  \; \; \; \;  \; \; \;   \times 
 \prod_{k, \sigma} e^{-\frac{\beta}{2} \tilde \rho_{k,\sigma}^2 \lambda_{k,\sigma}} 
\;  ,
\label{eq-ZintExpProd}
\end{eqnarray}
From this expression, we can see that the mean site occupancy for the interacting lattice-gas model can be obtained in the same way as in the noninteracting case (\ref{eq-ngammaderivZ}), 
\begin{equation}
\langle n^{(\gamma)} \rangle  
=\frac{1}{{\mathcal{N}}_{\text{sites}} \beta Z_{\text{G}} }
\frac{\partial Z_{\text{G}} }{\partial \mu_\gamma}
   \;  ,
\label{eq-ngammaderivZinteracting}
\end{equation}
by differentiating the partition function with respect to the chemical potential.

In the noninteracting case, since all the information about the configuration was contained in $n_{\ell,b}^{(\gamma)}$, we were able to decouple the sum over configurations from the sum over lattice sites because the activity was independent of position, and $n_{\ell,b}^{(\gamma)}$ only appeared in its exponent.  This is because only linear factors of $n_{\ell,b}^{(\gamma)}$ were in the exponent in the partition function.  While it is still true that $n_{\ell,b}^{(\gamma)}$ contains the configuration dependence, it now appears quadratically in the exponent of the partition function, since $\tilde \rho_{k,\sigma}$ contains linear factors of the $n_{\ell,b}^{(\gamma)}$, and the exponent in the interaction term contains $\tilde \rho_{k,\sigma}^2$.  This makes it impossible to decouple the sum over configurations from the sum over lattice sites.  However, we can use an integral identity, known as the Hubbard-Stratonivich transformation, to replace the term quadratic in $\tilde \rho_{k,\sigma}$ in the exponential with a term linear in $\tilde \rho_{k,\sigma}$, meaning that there will only be linear terms in $n_{\ell,b}^{(\gamma)}$.  This requires introducing the integral over an auxiliary field $\tilde \Delta_{k,\sigma}$. Then the sum over configurations is possible to do exactly, although at the cost of having to integrate over the auxiliary fields.
  
The integral identity, which is a complicated way of writing unity, as is shown in Appendix \ref{sec-GaussianIntegralIdentity}, is given by
\begin{equation}
 \int \sqrt{\frac{\beta \lambda}{-2 \pi}}
d\tilde\Delta \ 
e^{\frac{\beta}{2}\lambda  \tilde\Delta^2  - \beta \lambda  \tilde\Delta \tilde\rho}
e^{\frac{\beta}{2}\lambda \tilde\rho^2}  =  1   \; ,
\label{eq-GaussianIntegralIdentitytext}
\end{equation}
where the integration path is over the real axis from $-\infty$ to $\infty$ when $\lambda<0$ and over the imaginary axis from $-i \infty$ to $i \infty$ when $\lambda>0$. In the previous work by Bishop and McMullen\cite{bm06}, the problem was done in the position basis, and this integral identity, known as the Hubbard-Stratonovich transformation, was written with the matrix version of the potential in the exponent, as given in Negele and Orland\cite{no88}.  This form creates a dilemma when there are negative and possibly zero eigenvalues of the potential, and it is difficult to determine the path of integration, since it changes depending on the sign of the eigenvalues.  Zero eigenvalues would make the determinant of the matrix in that formula zero, and one then finds that there is a division by zero in the formula.  By transforming to the diagonal basis, all these difficulties are avoided, since there is a separate integral for each eigenvalue, and if a particular eigenvalue is zero, the identity is not used at all.

To use the identity in Eq.~(\ref{eq-GaussianIntegralIdentitytext}), we make the identifications that 
$\lambda = \lambda_{k,\sigma}$,
$\tilde \rho= \tilde \rho_{k,\sigma}$, and 
$\tilde\Delta = \tilde \Delta_{k,\sigma}$.  Substituting this ``1" into the partition function, we have
\begin{eqnarray}
Z_G &=& \sum_{\text{configurations}} 
 \prod_{k, \sigma} 
\int \sqrt{\frac{\beta \lambda_{k,\sigma}}{-2 \pi}}
d\tilde\Delta_{k,\sigma}   \times  \nonumber \\   &&\times
e^{\frac{\beta}{2}\lambda_{k,\sigma}  \tilde\Delta_{k,\sigma}^2 
 - \beta \lambda_{k,\sigma}  \tilde\Delta_{k,\sigma} \tilde\rho_{k,\sigma}}
e^{\frac{\beta}{2}\lambda_{k,\sigma} \tilde\rho_{k,\sigma}^2} 
\times  \nonumber  \\   &&\times
e^{-\beta
\sum_{\ell,b}\sum_{\gamma}(\varepsilon_\gamma - \mu_{\gamma})n_{\ell,b}^{(\gamma)}}
\times \nonumber \\ &&\times 
e^{-\frac{\beta}{2}\tilde \rho_{k,\sigma}^2 \lambda_{k,\sigma}} 
\;  .
\label{eq-ZintExpHubbStrat}
\end{eqnarray}
Now we see that the two exponentials containing $\tilde \rho_{k,\sigma}^2$ cancel, as we planned.  Of course, we have gained this convenience by introducing the auxiliary field ${\tilde \Delta}_{k,\sigma}$, and we will have to do the integral over this variable at a later stage.  However, since ${\tilde \Delta_{k,\sigma}}$ does not depend on the configuration of the system, we can bring the sum over configurations and the leading exponential factors in the partition function inside the integral, and the expression for the partition function becomes
\begin{eqnarray}
Z_G &=& 
\int{ \left(\prod_{k, \sigma} \sqrt{\frac{\beta \lambda_{k,\sigma}}{-2 \pi}}
d\tilde\Delta_{k,\sigma}   \right)} \sum_{\text{configurations}}
\times \nonumber  \\   &&   \times
{\exp{\left\{-\beta             
\sum_{\ell,b}\sum_{\gamma}             
(\varepsilon_\gamma - \mu_{\gamma}) n_{\ell,b}^{(\gamma)}
\right.}} +
\nonumber \\ 
&& {{ \left.    
+\sum_{k, \sigma} \left[\frac{\beta}{2} \lambda_{k,\sigma} 
{\tilde \Delta}_{k,\sigma}^2 
- \beta  \lambda_{k,\sigma}  {\tilde \rho}_{k,\sigma} {\tilde \Delta}_{k,\sigma}
\right] \right\} }}
  \;  ,   
\label{eq-ZintconfigInside}
\end{eqnarray}
It will be useful to define an action ${\mathscr S}$ such that the partition function can be written in the form
\begin{equation}
Z_G = \int{ {\mathcal D}[{\tilde \Delta}]\sum_{\text{configurations}}
e^{-{\mathscr S}[{\tilde \Delta}]}}     ,
\end{equation}
where
\begin{eqnarray}
{\mathscr S}[{\tilde \Delta}] &=& \beta \left\{            
\sum_{\ell,b}\sum_{\gamma}             
(\varepsilon_\gamma - \mu_{\gamma}) n_{\ell,b}^{(\gamma)} 
\right.      \nonumber \\  &&\left.  
+ \sum_{k_,\sigma} \left[ -\frac{\beta}{2} \lambda_{k,\sigma} 
{\tilde \Delta}_{k,\sigma}^2        
+ \beta  \lambda_{k,\sigma}  {\tilde \rho}_{k,\sigma} 
{\tilde \Delta}_{k,\sigma}              
 \right] \right\}
 \nonumber \\
\label{eq-Action1}
\end{eqnarray}
and 
\begin{equation}
{\mathcal D}[{\tilde \Delta}]=
 \left(\prod_{k, \sigma} \sqrt{\frac{\beta \lambda_{k,\sigma}}{-2 \pi}}
d\tilde\Delta_{k,\sigma}   \right)    \; .
\label{DnormDef}
\end{equation}
Note that the factors of $\beta \lambda_{k,\sigma}$ are included here in the grand differential 
${\mathcal D}[{\tilde \Delta}]$, rather than incorporating them in the action ${\mathscr S}[{\tilde \Delta}]$, as was done in Bishop and McMullen\cite{bm06} and in Sievert\cite{mds07}.  When these factors appear in the action, they introduce a term of the form $\ln\sqrt{ \det(\beta v)}$.  This adds a large constant value to the mean field results, which is subtracted out when the fluctuation terms are included.  It actually has no real physical meaning and is part of the normalization of the integral\cite{as10}.

We see that by diagonalizing first, our auxiliary fields are in the diagonal basis of the potential.  If we transform the terms containing $\varepsilon_\gamma$ and $\mu_\gamma$ to this basis, they will no longer be diagonal.  Therefore, we will leave those terms in the position basis.  It will also be convenient to have the term containing 
$\sum_{k_,\sigma} \lambda_{k,\sigma}  {\tilde \rho}_{k,\sigma} 
{\tilde \Delta}_{k,\sigma}$ in the position basis also.  Therefore, the expression for the transformed charge density is
\begin{eqnarray}
{\tilde \rho}_{k,\sigma}
&=&\sum_{\ell,b} {\mathcal W^T_{(k,\sigma),(\ell,b)}}\sum_\gamma 
\rho_{\ell,b}^{(\gamma)}
\nonumber  \\
&=&\sum_{\ell,b} {\mathcal W^T_{(k,\sigma),(\ell,b)}}\sum_\gamma  q_\gamma 
n_{(\ell,b)}^{(\gamma)}   \;  ,
\label{rhoDiagonalBasis}
\end{eqnarray}
in terms of the quantities in real space.

Substituting this into the action, we have
\begin{eqnarray}
{\mathscr S}[{\tilde \Delta}] &=& \beta \left\{              
\sum_{\ell,b}\sum_{\gamma}             
(\varepsilon_\gamma - \mu_{\gamma}) n_{(\ell,b)}^{(\gamma)} 
\right.      \nonumber \\  && \left.      
-\sum_{k_,\sigma} \left[\frac{1}{2} \lambda_{k,\sigma} 
{\tilde \Delta}_{k,\sigma}^2 
\right. \right. \nonumber  \\  &&  \left. \left. 
+  {\tilde \Delta}_{k,\sigma}  \lambda_{k,\sigma}  
\sum_{\ell,b} {\mathcal W^T_{(k,\sigma),(\ell,b)}}\sum_\gamma  q_\gamma 
n_{(\ell,b)}^{(\gamma)}
 \right]  \right\}   \; .
 \nonumber \\
\label{eq-Action2}
\end{eqnarray}
It is convenient to identify the last term in this expression as a self energy of species $\gamma$,
\begin{equation}
\tilde \Sigma_{\ell,b}^{(\gamma)}[{\tilde \Delta}]=
\left(\sum_{k,\sigma}  \tilde \Delta_{k,\sigma} \lambda_{k,\sigma}
\mathcal W^T_{(k,\sigma),(\ell,b)} \right) q_\gamma    \; .
\label{eq-SelfEnergyDiagBasis}
\end{equation}
By transforming $\tilde \Delta_{k,\sigma}$ to the position basis as 
\begin{equation}
\tilde \Delta_{k,\sigma}=
\sum_{\ell',b'} \Delta_{\ell',b'} 
\mathcal W_{(\ell',b'),(k,\sigma)}  
    \; ,
\label{eq-DeltaInverseTransform}
\end{equation}
this self-energy can also be written in the position basis as
\begin{eqnarray}
\Sigma_{\ell,b}^{(\gamma)}[\Delta] &=&
\tilde\Sigma_{\ell,b}^{(\gamma)}[\tilde\Delta]
\nonumber \\
&=& q_{\gamma} \sum_{\ell',b'} \Delta_{\ell',b'} 
\left(  \sum_{k,\sigma} \mathcal W_{(\ell',b'),(k,\sigma)}   
\lambda_{k,\sigma}
\mathcal W^T_{(k,\sigma),(\ell,b)} \right) 
\nonumber \\
&=&  q_{\gamma} \sum_{\ell',b'} \Delta_{\ell',b'} 
\mathcal V_{(\ell',b'),(\ell,b)}
    \; .
\label{eq-SelfEnergyPositionBasis}
\end{eqnarray}
This will be a useful form to use when we discuss the saddle point, or mean-field, approximation.

Keeping the first term in the position basis and the second term in the diagonal basis, we now write the action in a compact form, combining the self energy term with the first term in the action to obtain
\begin{eqnarray}
{\mathscr S}[{\tilde \Delta}] &=& \beta \left\{              
\sum_{\ell,b}\sum_{\gamma}             
(\varepsilon_\gamma + \Sigma_{\ell,b}^{(\gamma)}[\Delta]-\mu_{\gamma}) n_{(\ell,b)}^{(\gamma)} 
\right.  \nonumber \\  && \left. 
-\sum_{k_,\sigma} \left[\frac{1}{2} \lambda_{k,\sigma} 
{\tilde \Delta}_{k,\sigma}^2 
\right]  \right\}
\label{eq-Action3}
\end{eqnarray}

Analogous to the noninteracting case, we define the activity as
\begin{equation}
\tilde a_{\ell,b}^{(\gamma)}[{\tilde \Delta}] =
a_{\ell,b}^{(\gamma)}[{\Delta}] =
e^{-\beta (\varepsilon_\gamma
+\Sigma_{\ell,b}^{(\gamma)}[ \Delta]-\mu_{\gamma})}
\label{eq-Activity2}
\end{equation} 
With this definition, the partition function becomes
\begin{eqnarray}
Z_G &=& \int{ {\mathcal D}[{\tilde \Delta}]\sum_{\text{configurations}}
\left( \prod_{\ell,b}\prod_{\gamma}  
\{ a_{\ell,b}^{(\gamma)}[ \Delta]\}^{n_{(\ell,b)}^{(\gamma)}}  \right)  }
 \times \nonumber \\  && \times
e^{\beta \sum_{k_,\sigma} \left[\frac{1}{2} \lambda_{k,\sigma}
{\tilde \Delta}_{k,\sigma}^2  \right] }    ,
\label{eq-ActionwithActivity}
\end{eqnarray}

The trace over configurations is now done over each site separately exactly as we did for the noninteracting case, where $n_{(\ell,b)}^{(\gamma)}=1$ for one and only one of $\gamma = \parallel, \perp$,  or $v$, and zero otherwise.  Thus, performing the trace gives
\begin{equation}
\sum_{\substack{{\text{single-site}} \\ {\text{configurations}}  }}
 \prod_\gamma
\{\tilde a_{\ell,b}^{(\gamma)}[{\tilde \Delta}]\}^{n_{(\ell,b)}^{(\gamma)}} 
= \sum_{\gamma} \tilde a_{\ell,b}^{(\gamma)}[{\tilde \Delta}]
\end{equation}
and the grand partition function becomes
\begin{equation}
Z_G = \int{ {\mathcal D}[{\tilde \Delta}]
e^{-{\mathscr S}_{\text{eff}}[{\tilde \Delta}]}}    ,
\label{eq-PartitionFunctionExact}
\end{equation}
where the effective action is
\begin{equation}
{\mathscr S}_{\text{eff}}[{\tilde \Delta}] 
= -\frac{\beta}{2} \sum_{k_,\sigma}  \lambda_{k,\sigma} 
\tilde \Delta_{k,\sigma}^2   
- \sum_{\ell,b} \ln{\left(\sum_\gamma 
\tilde a_{\ell,b}^{(\gamma)}[\tilde \Delta]\right)}  \; .
\label{eq-Seff}
\end{equation}

This is the general result, which is in principle exact.  To understand
what the Hubbard-Stratonovich transformation has accomplished for us,
recall that the interaction energy $U_{\text{int}}$ posed two
difficulties: the additional configuration dependence
and the coupling between the sites.  By introducing auxiliary fields
through the Hubbard-Stratonovich transformation
(\ref{eq-GaussianIntegralIdentitytext}), we managed to separate these two complications
so that one factor
$\exp\left[-\frac{\beta}{2} \sum_{k_,\sigma} \lambda_{k,\sigma} 
{\tilde \Delta}_{k,\sigma}^2 \right]$ 
contains interactions between field fluctuations
but no configuration dependence, and another factor
$\exp\left[\beta \sum_{k,\sigma}\lambda_{k,\sigma}
{\tilde \rho}_{k,\sigma} {\tilde \Delta}_{k,\sigma} \right]$ that is
configuration-dependent (through ${\tilde \rho}_{k,\sigma}$) but decoupled.  This allowed us to define a
modified activity (\ref{eq-Activity2}) that incorporated all the
configuration dependence, making it possible to evaluate the sum over
configurations directly, as in the noninteracting case.

What we have done in using the Hubbard-Stratonovich transformation is replace
the interaction part of the partition function with its functional integral
representation.  From the form of Eq.~(\ref{eq-ActionwithActivity}), we
see that the largest contribution to the partition function comes from the values of
${\tilde \Delta}_{k,\sigma}^2$ for which the effective action ${\mathscr S_\text{eff}}$ is
stationary, {\em i.e.}, at a saddle point.  Thus a Taylor series expansion of the effective action
(\ref{eq-Seff}) about the saddle point will yield an
order-by-order approximation to the exact partition function
(\ref{eq-PartitionFunctionExact}).

\section{Expansion in Powers of the Auxiliary Field}
\label{sec-expansion}

Since it is not possible to evaluate the integral in the partition function of Eq.~(\ref{eq-PartitionFunctionExact}) exactly, we will approximate it by expanding the action in Eq.~(\ref{eq-Seff}) about a mean field $\Delta_c$, and include fluctuations to lowest order about this mean field.  This mean field is determined by the saddle point of the integrand in Eq.~(\ref{eq-PartitionFunctionExact}), which is the largest contribution to the integral.

We can write the effective action then as an expansion about this uniform
saddle point. As an aid to keeping track of the order of the expansions, we add a multiplicative factor $m$ in front of the effective action. as
\begin{eqnarray} 
&&m{\mathscr S}_{\text{eff}}
\left({\tilde \Delta}={\tilde \Delta}_c + 
\delta {\tilde \Delta}/\sqrt m\right)  =
\nonumber \\
&=& m {\mathscr S}_{\text{eff}}[{\tilde \Delta}_c]
+\left. \sum_{k,\sigma} 
\frac{\delta{\tilde \Delta}_{k,\sigma}}{\sqrt m} m
\frac{\partial{\mathscr S}_{\text{eff}}[{\tilde \Delta}]}
{\partial {\tilde \Delta}_{k,\sigma}}
\right|_{{\tilde \Delta}_c}+
  \nonumber \\
&+& \left. \frac{1}{2}\sum_{k_1,\sigma_1,k_2,\sigma_2}
\frac{\delta{\tilde \Delta}_{k_1,\sigma_1}}{\sqrt m} m
\frac{\partial^2{\mathscr S}_{\text{eff}}[{\tilde \Delta}]}
{\partial {\tilde \Delta}_{k_1,\sigma_1}
\partial {\tilde \Delta}_{k_2,\sigma_2}}
\right|_{{\tilde \Delta}_c}
\frac{\delta{\tilde \Delta}_{k_2,\sigma_2}}{\sqrt m} 
\nonumber   \\
&&        +\dots   \;  .
\label{eq-SaddlePtExpansion}
\end{eqnarray} 
Here the expansion parameter 
$\delta {\tilde \Delta}$
is scaled by $m^{-1/2}$
by writing
$\delta {\tilde \Delta_{k,\sigma}}\equiv \sqrt m
[{\tilde \Delta}_{k,\sigma}-{\tilde \Delta}_{k,\sigma,c}]$.
The factors of $m$ cancel in the quadratic term, so that it is of order 
$m^0$.

In evaluating the integral in the partition function, the largest contribution comes from the region near a saddle point, and the terms beyond the saddle point give corrections to the integral.  Physically, the saddle point solution corresponds to a mean field approximation, and we begin with that.  The saddle point is found by setting the first derivative in the expansion to zero, that is
\begin{equation}
\left.\frac{\partial{\mathscr S}_{\text{eff}}[{\tilde \Delta}]}
{\partial {\tilde \Delta}_{k,\sigma}}
\right|_{{\tilde \Delta}_c}
=0   \;  .
\label{eq=saddleptfirstderivzero}
\end{equation}

The second derivatives evaluated at the saddle point are proportional to the components of a matrix $\tilde D^{-1}$, which is the inverse propagator for the fluctuations of the auxiliary Hartree field $\Delta$ that we introduced to decouple the interparticle interactions.
Because we want to absorb the normalization factor in the grand differential ${\mathcal D}[{\tilde \Delta}]$, we have defined each of the elements of  $\tilde{D}^{-1}$ by the second derivative of 
${\mathscr S}_{\text{eff}}[{\tilde \Delta}]$ 
divided by  $-\beta \lambda_{k,\sigma}$ as 
\begin{equation}
(\tilde D^{-1})_{(k_1,\sigma_1),(k_2,\sigma_2)} 
=
- \left( \frac{1}{\beta \lambda_{k,\sigma}} \right)
\left. \frac{\partial^2{\mathscr S}_{\text{eff}}[{\tilde \Delta}]}
{\partial {\tilde \Delta}_{k_1,\sigma_1}
\partial {\tilde \Delta}_{k_2,\sigma_2}}
\right|_{{\tilde \Delta}_c}   .
 \\
\label{eq-DtldeInvDef}
\end{equation}
We will show in Sec.~\ref{sec-HartreePropagator} that, for a spatially uniform saddle point, this matrix is diagonal, that is,
${\mathscr S}_{\text{eff}}[{\tilde \Delta}]$ 
is diagonal in the $\delta {\tilde \Delta}_{k_1,\sigma_1}$'s, that is,
\begin{equation}
(\tilde D^{-1})_{(k_1,\sigma_1),(k_2,\sigma_2)} 
= (\tilde D^{-1})_{(k_1,\sigma_1),(k_1,\sigma_1)}
\delta_{k_1,k_2} \delta_{\sigma_1,\sigma_2} \;  .
\end{equation}
Although the Gaussian integral can still be done if $\tilde D^{-1}$ is not diagonal, the argument becomes much easier to follow if we
assume at this point that $\tilde D^{-1}$ is diagonal because then the Gaussian integral can be done straightforwardly.

With this assumption that $\tilde D^{-1}$ is diagonal, the expansion of the effective action reduces to
\begin{eqnarray} 
&&m{\mathscr S}_{\text{eff}}
\left({\tilde \Delta}={\tilde \Delta}_c + 
\delta {\tilde \Delta}/\sqrt m\right) =
\nonumber \\
&\simeq& m {\mathscr S}_{\text{eff}}[{\tilde \Delta}_c] -
  \nonumber \\
&-& \frac{1}{2}\sum_{k,\sigma}   
\beta \lambda_{k,\sigma} 
(\tilde D^{-1})_{(k,\sigma),(k,\sigma)}     
(\delta{\tilde \Delta}_{k,\sigma})^2   \;  .
\label{eq-SaddlePtExpansionReduction}
\end{eqnarray} 
The grand canonical partition function can then be written as
\begin{eqnarray}
Z_G &\simeq&  e^{- m \mathscr S_{\text{eff}}[{\tilde \Delta}_c]}
\int{ \left(\prod_{k, \sigma} \sqrt{\frac{\beta \lambda_{k,\sigma}}{-2 \pi}}
d(\delta \tilde\Delta_{k,\sigma})   \right) }  
\times    \nonumber \\  &&  {  \times
e^{\frac{1}{2}   \beta  \sum_{k,\sigma}\lambda_{k,\sigma}             
(\tilde D^{-1})_{(k,\sigma),(k,\sigma)}     
(\delta{\tilde \Delta}_{k,\sigma})^2 }}  \; .
 \nonumber \\          
\end{eqnarray}
This can be written as a product of Gaussian integrals as
\begin{eqnarray}
Z_G &\simeq&  
e^{ - m \mathscr S_{\text{eff}}[{\tilde \Delta}_c]}
\prod_{k,\sigma}
\int{ \left( \sqrt{\frac{\beta \lambda_{k,\sigma}}{-2 \pi}}
d( \delta \tilde\Delta_{k,\sigma} )  \right)}
\times    \nonumber \\  &&  \times
e^{\frac{1}{2}  \beta \lambda_{k,\sigma}              
(\tilde D^{-1})_{(k,\sigma),(k,\sigma)}     
(\delta{\tilde \Delta}_{k,\sigma})^2 }  \; ,
 \nonumber \\ 
 \label{eq-partitonfctproduct}         
\end{eqnarray}
and each Gaussian integral can be evaluated using the same change of variable and path of integration as was used in the Hubbard-Stratonovich transformation in Eq.~(\ref{eq-GaussianIntegralIdentitytext}) that originally was used to define the auxiliary fields, and as was discussed in Appendix \ref{sec-GaussianIntegralIdentity}.  Each Gaussian integral produces a result of the same form, regardless of whether $(\tilde D^{-1})_{(k,\sigma),(k,\sigma)}$ is positive or negative.  This quantity can never be zero, since we would not have defined an auxiliary field for that case, which is the same as the case in which an eigenvalue of the potential is zero.  Therefore, the integral can always be evaluated, and the partition function can be written as
\begin{eqnarray}
Z_G &\simeq&  
e^{ - m \mathscr S_{\text{eff}}[{\tilde \Delta}_c]}
\prod_{k,\sigma}
\frac{1}
{\sqrt{(\tilde D^{-1})_{(k,\sigma),(k,\sigma)}}} \; , 
\label{ZGOneLoop}       
\end{eqnarray}
Note that all the factors of $\beta \lambda_{k,\sigma}$ have cancelled out due to our choices of the form of the effective action ${\mathscr S}_{\text{eff}}$ and the inverse  Hartree field fluctuation propagator $\tilde D^{-1}$.  Choosing these quantities carefully allows for the correct normalization of the integral\cite{as10}.
The product can now be brought inside the square root, and since $\tilde D^{-1}$ is diagonal, this gives the determinant of the matrix inside the square root, and so the partition function assumes the form
\begin{eqnarray}
Z_G &\simeq&  
e^{ - m \mathscr S_{\text{eff}}[{\tilde \Delta}_c]}
\frac{1}{\sqrt{\det(\tilde D^{-1})}} \; ,        
\end{eqnarray}
and the partition function can be written as a single exponential,
\begin{eqnarray}
Z_G &\simeq&  
e^{ - \{ m \mathscr S_{\text{eff}}[{\tilde \Delta}_c]
+\frac{1}{2} \ln \det(\tilde D^{-1}) \} }    \; .  
\label{eq-ZGSingleExp}  
\end{eqnarray}

The grand canonical potential is then
\begin{equation}
\Omega_G = - \frac{1}{\beta} \ln Z_G \simeq \frac{1}{\beta}
\{ m \mathscr S_{\text{eff}}[{\tilde \Delta}_c] 
+\frac{1}{2} \ln \det({\tilde D}^{-1})  \}
\end{equation}
If we formally write the expansion in terms of orders in $m$, we have
\begin{equation}
\Omega_G \simeq  m \Omega_{m=1} +  \Omega_{m=0} + 
\frac{1}{\sqrt{m}} \Omega_{m=-\frac{1}{2}} 
+ \frac{1}{m} \Omega_{m=-1} + \cdots        \; ,
\label{eq-formalexpansion}
\end{equation}
The leading term in the expansion is therefore given by the saddle point value
\begin{equation}
m \Omega_{m=1}  =  m \Omega_c  =
\frac{m}{\beta} \mathscr S_{\text{eff}}[{\tilde \Delta}_c] 
\; ,
\label{eq-leadingexpansionterm}
\end{equation}
and what is known as the ``one-loop" correction term is 
\begin{equation}
\Omega_{m=0} =\frac{1}{2} \ln \det({\tilde D}^{-1}) = \frac{1}{2} {\text{Tr}} \ln \tilde D^{-1}     .
\label{eq-oneloopcorrection}
\end{equation}

\section{The Saddle-Point or Mean Field Approximation}
\label{sec-saddlepoint}

The first task in this section is to find an equation that determines the value of the auxiliary field at the saddle-point or mean-field value of the the effective action.   
In order to find this saddle point, we must set the first derivative of the effective action in Eq.~(\ref{eq-Seff}) with respect to the auxiliary field ${\tilde \Delta}_{k,\sigma}$ to zero.   That derivative is given by
\begin{eqnarray}
\frac{\partial{\mathscr S}_{\text{eff}}[{\tilde \Delta}]}
{\partial {\tilde \Delta}_{k,\sigma}}
&=&-\beta \lambda_{k,\sigma}{\tilde \Delta}_{k,\sigma} -
\nonumber   \\
&-&\beta \sum_{\ell,b}   
\frac{1 }
 {\left( \sum_{\gamma '} \tilde a_{\ell,b}^{(\gamma ')}[{\tilde \Delta}] \right)}
\sum_{\gamma}  \frac{\partial\tilde a_{\ell,b}^{(\gamma)}[{\tilde \Delta}]}
{\partial {\tilde \Delta}_{k,\sigma}}    \; .
\nonumber  \\
\label{eq-firstDerivActionDelt}
\end{eqnarray}
The activity $a_{\ell,b}^{(\gamma)}[{\tilde \Delta}]$ for species $\gamma$ is given by Eq.~(\ref{eq-Activity2}), and its derivative is
\begin{equation}
\frac{\partial\tilde a_{\ell,b}^{(\gamma)}[{\tilde \Delta}]}
{\partial {\tilde \Delta}_{k,\sigma}} =
-\beta \tilde a_{\ell,b}^{(\gamma)}[{\tilde \Delta}] 
 \frac{\partial\tilde\Sigma_{\ell,b}^{(\gamma)}[{\tilde \Delta}]}
{\partial {\tilde \Delta}_{k,\sigma}}    \;,
\label{eq-firstDerivActivity}
\end{equation}
where from Eq.~(\ref{eq-SelfEnergyDiagBasis}), the derivative of the self-energy is
\begin{equation}
\frac{\partial\tilde\Sigma_{\ell,b}^{(\gamma)}[{\tilde \Delta}]}
{\partial {\tilde \Delta}_{k,\sigma}}
= q_\gamma  \lambda_{k,\sigma}
 {\mathcal W}^T_{(k,\sigma),(\ell,b)}   \; .
 \label{eq-firstDerivSigma}
\end{equation} 
Substituting this into the derivative of the action, we have
\begin{eqnarray}
\frac{\partial{\mathscr S}_{\text{eff}}[{\tilde \Delta}]}
{\partial {\tilde \Delta}_{k,\sigma}}
&=&-\beta \lambda_{k,\sigma}{\tilde \Delta}_{k,\sigma} +
\nonumber   \\
&+&\beta \lambda_{k,\sigma}  \sum_{\ell,b} 
{\mathcal W}^T_{(k,\sigma),(\ell,b)}   
\frac{\sum_{\gamma} q_\gamma \tilde a_{\ell,b}^{(\gamma)}[{\tilde \Delta}] }
 {\sum_{\gamma '}      
\tilde a_{\ell,b}^{(\gamma ')}[{\tilde \Delta}] }
  \;,
\nonumber  \\
\label{eq-FirstDerivSeff}
\end{eqnarray}

Setting this first derivative to zero, we obtain an equation whose solution gives the auxiliary field in the saddle point approximation,
\begin{equation}
{\tilde \Delta}_{(k,\sigma)} =
\sum_{\ell,b} 
{\mathcal W}^T_{(k,\sigma),(\ell,b)}   
\left(\frac{\sum_{\gamma}\ q_\gamma \tilde a_{\ell,b}^{(\gamma)}[{\tilde \Delta}] }
 {\sum_{\gamma '}      
\tilde a_{\ell,b}^{(\gamma ')}[{\tilde \Delta}] } \right)    \;  .
\label{eq-DeltaCDiagBasisFinal}
\end{equation} 
We see that on the right-hand side of the equation is the transformation to the diagonal basis of the potential, as given in Eq.~(\ref{eq-DeltaInverseTransform}).  We could therefore write the auxiliary field in the position basis as
\begin{equation}
\Delta_{(\ell,b)} =
\frac{\sum_{\gamma} q_\gamma  a_{\ell,b}^{(\gamma)}[\Delta]}
{\sum_{\gamma '}      
a_{\ell,b}^{(\gamma ')}[\Delta] }    \; , 
\label{eq-DeltaCPositionBasis}
\end{equation}
where $a_{\ell,b}^{(\gamma)}[\Delta]$ is the activity with the self energy in the diagonal basis
$\tilde \Sigma_{\ell,b}^{(\gamma)}[{\tilde \Delta}]$, which by Eq.~(\ref{eq-SelfEnergyPositionBasis}) has been replaced by the self energy in the position basis, $\Sigma_{\ell,b}^{(\gamma)}[\Delta]$.  Either of these two equations, Eq.~(\ref{eq-DeltaCDiagBasisFinal}) or (\ref{eq-DeltaCPositionBasis}), may be considered the equation giving the saddle-point, or mean-field, value of the auxiliary field, and we will refer to this as the mean field equation.  In this expression $a_{\ell,b}^{(\gamma)}[\Delta_c]$ is a function of all the 
$\Delta_{(\ell,b),c}$ elements, and so, for ${\cal{N}}=30$, which we have used in our calculations, there are ${\cal{N}}_{\text{sites}}=2(2{\cal{N}}+1)=122$ coupled nonlinear equations.  Therefore, for simplicity, we make the assumption that the mean field is spatially uniform, which means that
\begin{equation}
\Delta_{(\ell,b),c}=\Delta_c    \; ,
\label{eq-DeltaCeqUniformPositionBasis}
\end{equation}
where $\Delta_c$ is constant.  This is still a nonlinear equation in $\Delta_c$, and so this must be solved numerically with an iterative approach.

Before solving the mean field equation, it is useful to have a physical interpretation of the mean auxiliary field 
$\Delta_c$.  The first step in this direction is the second task of this section, which is to find the mean site occupancies of species $\gamma$ at the mean field level.  This is done as in the noninteracting case, by taking the derivative of the partition function with respect to $\mu_\gamma$.  At this mean field level, the $\mu_\gamma$ dependence is contained in ${\mathscr S}_{\text{eff}}[{\tilde \Delta}_c]$ in Eq.~(\ref{eq-ZGSingleExp}), neglecting the term containing the inverse propagator $D^{-1}$.  The expression
for the mean occupancy can be gotten using Eq.~(\ref{eq-ngammaderivZinteracting}) as
\begin{equation}
\langle n^{(\gamma)} \rangle_c  
=\frac{1}{{\mathcal{N}}_{\text{sites}} \beta}
\frac{\partial }{\partial \mu_\gamma}
\ln[e^{ - \mathscr S_{\text{eff}}[{\tilde \Delta}_c]}]
  \;  .
\label{eq-meanOccInt1}
\end{equation}
Taking the derivative of the exponential, we can cancel out the partition function in the denominator and we are left with the derivative of the effective action as
\begin{equation}
\langle n^{(\gamma)} \rangle_c  
= - \frac{1}{{\mathcal{N}}_{\text{sites}} \beta  }
\frac{\partial }{\partial \mu_\gamma}    \mathscr S_{\text{eff}}[{\tilde \Delta}_c]  \;  .
\label{eq-meanOccInt2}
\end{equation}
Substituting the effective action from Eq.~(\ref{eq-Seff}), we have
\begin{equation}
\langle n^{(\gamma)} \rangle_c  
=  \frac{1}{{\mathcal{N}}_{\text{sites}} \beta  }
\sum_{\ell,b} \frac{1}{\left(\sum_{\gamma'} 
 a_{\ell,b}^{(\gamma')}[\Delta_c]\right)}
\frac{\partial a_{\ell,b}^{(\gamma)}[ \Delta_c]}{\partial \mu_\gamma}
  \;  .
\label{eq-meanOccInt3}
\end{equation}
The derivative of the activity is
\begin{equation}
\frac{\partial a_{\ell,b}^{(\gamma)}[ \Delta_c]}{\partial \mu_\gamma}
=\beta a_{\ell,b}^{(\gamma)}[ \Delta_c]
  \;  ,
\label{eq-activitDerivmu}
\end{equation}
and so the mean occupancy becomes
\begin{equation}
\langle n^{(\gamma)} \rangle_c  
= \frac{1}{{\mathcal{N}}_{\text{sites}}  }
\sum_{\ell,b} \frac{ a_{\ell,b}^{(\gamma)}[ \Delta_c]}
{\left(\sum_{\gamma'}
 a_{\ell,b}^{(\gamma')}[ \Delta_c]\right)}
  \;  .
\label{eq-meanOccInt4a}
\end{equation}

The third task of this section is to relate the mean charge density per site to the mean occupancy.  The charge density has the same form as in Eq.~(\ref{eq-rhoNIsitedef}) for the noninteracting model, so that at the mean field level we obtain
\begin{equation}
\rho_c=\langle \rho\rangle_c = \sum_\gamma q_\gamma \langle n^{(\gamma)} \rangle_c 
 = \frac{ \sum_\gamma q_\gamma a_{\ell,b}^{(\gamma)}[ \Delta_c]}
{\sum_{\gamma'}
 a_{\ell,b}^{(\gamma')}[ \Delta_c]}
  \;  ,
\label{eq-MeanChargeDensityUniform}
\end{equation}
which, like the mean site occupancy, is independent of the site index because of our choice of spatially uniform mean field.
This is equivalent to the equation for the saddle-point value of the auxiliary field
because the right-hand side of the equation shows that
\begin{equation}
 \rho_c = \Delta_c   \; .
\label{eq-DeltaCeqrhoPositionBasis}
\end{equation}
on comparison with Eq.~(\ref{eq-DeltaCPositionBasis})

In the mean field or saddle point equation for the spatially uniform saddle point, the auxiliary field is the same for all lattice sites, 
and the self energy becomes spatially uniform also and can be written as
\begin{equation}
\Sigma_{\ell,b}^{(\gamma)}[\Delta_c]   \equiv  \Sigma_{c}^{(\gamma)} = \rho_c q_{\gamma}  
\mathbb S_{\text{lattice}} \; .
\end{equation}
where $\mathbb S_{\text{lattice}}$ is a lattice sum that is independent of the choice of lattice site $(\ell,b)$ for a long DNA double helix, and is given by
\begin{equation}
\mathbb S_{\text{lattice}} =  \sum_{\ell',b'} {\mathcal V}_{(\ell',b'),(\ell,b)}   \; .
\label{eq-SLattice}
\end{equation}
For the parameters used in the figures,  
$\beta \mathbb S_{\text{lattice}} \approx 2.71$.

Consequently, at the mean field level, the self-energy simply represents the electrostatic energy of charge $q_\gamma$ interacting with the rest of the
DNA lattice.  This term plays the role of the
self-energy $\Sigma$ in many-body quantum mechanics \cite{bm06,no88,as10}.  Since the self-energy is independent of lattice site, so is the activity, which can be written as
\begin{equation}
a_c^{(\gamma)}=a_{\ell,b}^{(\gamma)}[ \Delta_c]= 
e^{-\beta (\varepsilon_\gamma - \mu_{\gamma}
+\Sigma_c^{(\gamma)})} \; .
\label{eq-activityMF}
\end{equation}

The simplifications made possible by the assumption of a spatially uniform saddle point allow us to rewrite the mean field equation given by Eq.~(\ref{eq-DeltaCPositionBasis}) in an explicit form as
\begin{equation}
\rho_c
=  \frac{\sum_{\gamma}q_{\gamma}e^{-\beta\left(\varepsilon_{\gamma}
-\mu_{\gamma}
+q_{\gamma}\rho_c {\mathbb S}_{\text{lattice}}\right)}}
{\sum_{\gamma}  e^{-\beta\left(\varepsilon_{\gamma}
-\mu_{\gamma}
+q_{\gamma}\rho_c {\mathbb S}_{\text{lattice}}\right)}}  \;  .
\label{eq-RhoC}
\end{equation} 
Again, recall that $\varepsilon_v - \mu_v=0$, identically, $q_{\parallel}=0$, and  $\mu_{\parallel}=\mu_{\perp}=\mu_{\text{dimer}}$.  Therefore, when
$\rho_c =0$, we can see from Eq.~(\ref{eq-RhoC}) that
$q_{\perp} e^{-\beta (\varepsilon_{\perp}-\mu_{\text{dimer}})} +  q_v = 0$,
and since $q_{\perp}=+1$, and $q_v=-1$, it follows that   
$\varepsilon_{\perp}=\mu_{\text{dimer}}$.  Therefore, this is the point at which the charge density $\rho_c $ goes to zero.  When $\varepsilon_{\parallel}=2\varepsilon_{\perp}$, the charge density goes to zero at $\varepsilon_{\parallel}=2\mu_{\text{dimer}}$, which is the same place it went to zero for the noninteracting model, as shown in Fig.~\ref{fig-NIchargeDensities}.

\begin{figure}
\includegraphics[width=3in]{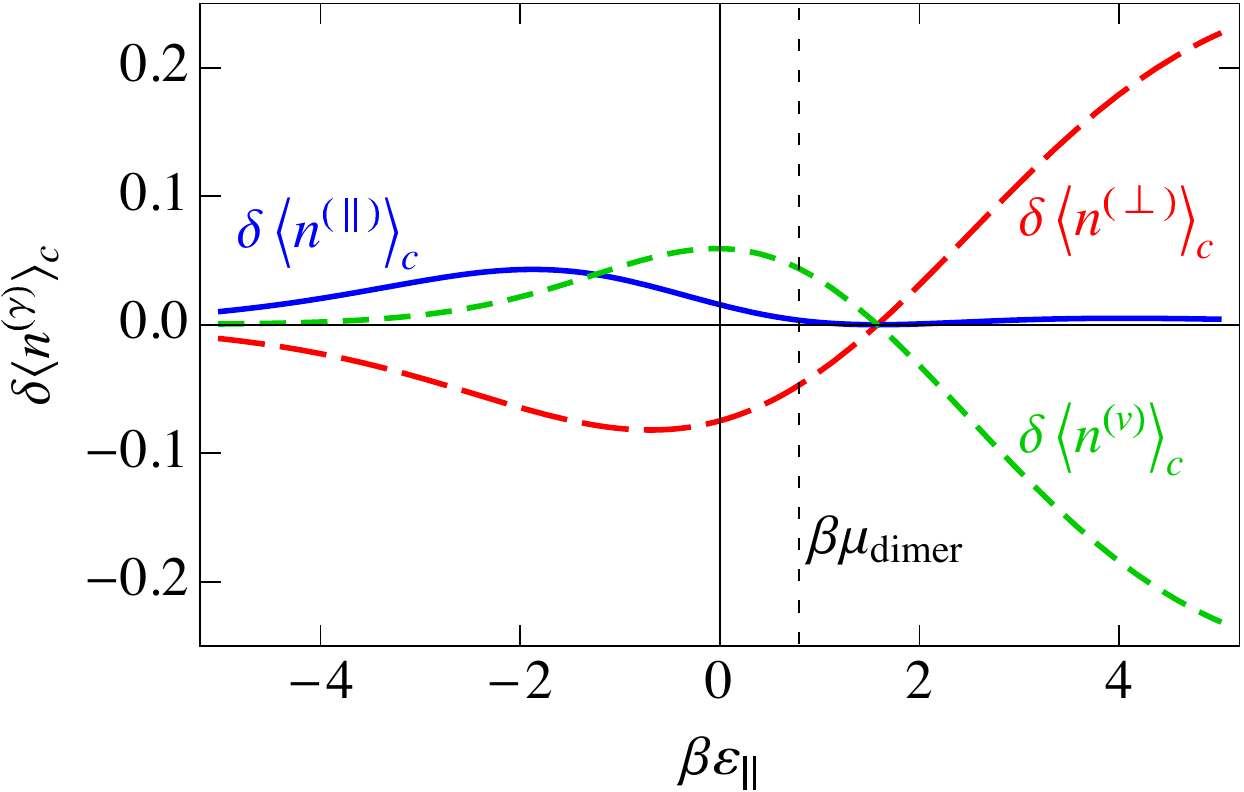}
\caption{A plot of the differences $\delta\langle n^{(\gamma)}\rangle_c$ in mean-field average occupation numbers from their noninteracting values  for $\parallel$ (blue, solid),$\gamma = \perp$ (red, long-dashed),  and $v$
(green, short-dashed). The parameters used in the calculation are
$\beta \mu_{\text{dimer}}=0.79$ and $\varepsilon_{\parallel}=2\varepsilon_{\perp}$. The noninteracting results were shown in Fig.~\ref{fig-NIOccupation}.}
\label{fig-OccupationsMF}
\end{figure}

\begin{figure}
\includegraphics[width=3in]{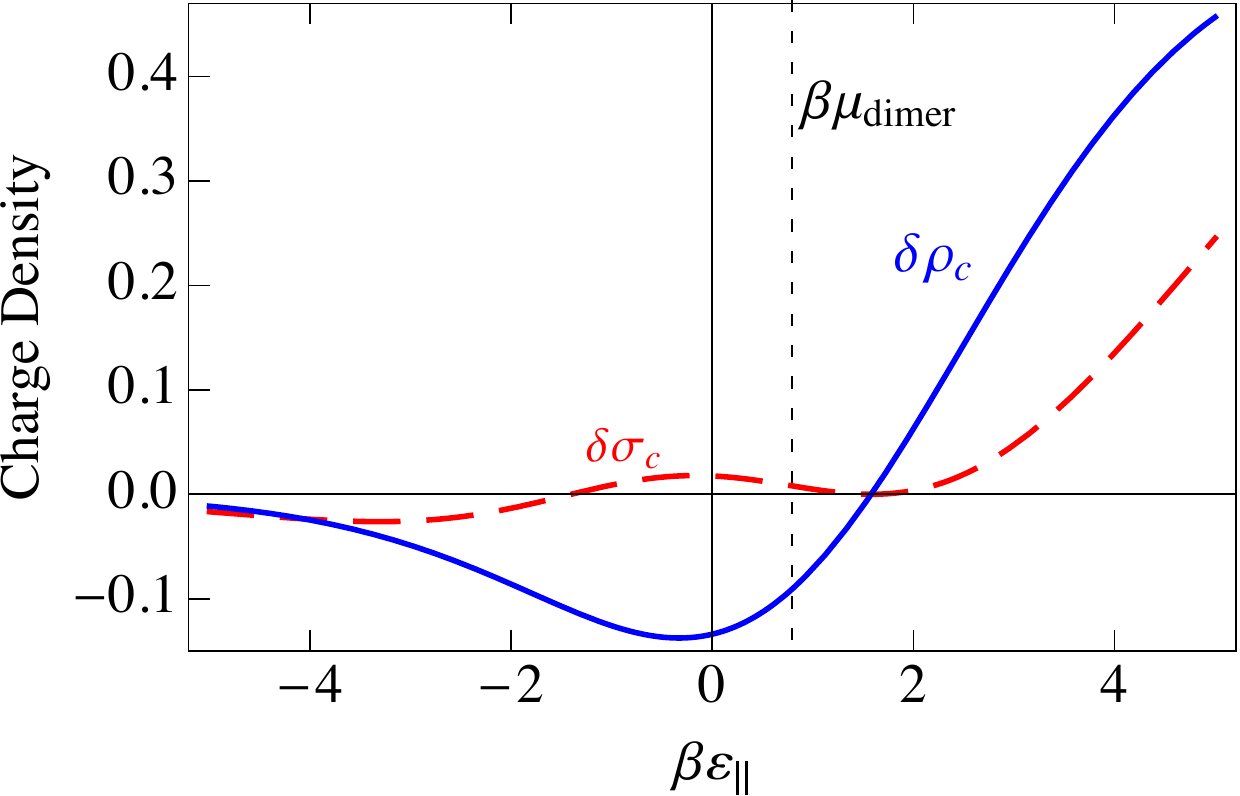}
\caption{Plots of the differences in the charge densities and standard deviations of the charge density for the mean field of the interacting lattice gas model and the noninteracting model. $\delta \rho_c=\rho_c-\rho_{\text{NI}}$  is shown as the solid blue curve, and $\delta \sigma_c=\sigma_c-\sigma_{\text{NI}}$ is shown as the dashed red curve. 
The parameters used in the calculation are
$\beta \mu_{\text{dimer}}=0.79$ and $\varepsilon_{\parallel}=2\varepsilon_{\perp}$.  The noninteracting results were shown in Fig.~\ref{fig-NIchargeDensities}.}
\label{fig-MFchargeDensities}
\end{figure}

Returning to the mean site occupancy in Eq.~(\ref{eq-meanOccInt4a}), the quantity inside the sum over $\ell$ and $b$ is constant and the sum yields 
${\mathcal{N}}_{\text{sites}}$, and so the average occupancy per species becomes
\begin{equation}
\langle n^{(\gamma)} \rangle_c  
=\frac{ a_c^{(\gamma)}}
{\left(\sum_{\gamma'}
 a_c^{(\gamma')}\right)}
  \;  .
\label{eq-meanOccInt4b}
\end{equation}
Using the mean-field value $\rho_c=\Delta_c$ of the charge density, calculated from Eq.~(\ref{eq-RhoC}), we have calculated the average occupation numbers $\langle n^{(\gamma)} \rangle$ via Eq.~(\ref{eq-meanOccInt4b}) using 
Eq.~(\ref{eq-activityMF}) for the activities.   In Fig.~\ref{fig-OccupationsMF}, we have plotted $\delta \langle n^{(\gamma)} \rangle_c =\langle n^{(\gamma)} \rangle_c -
\langle n^{(\gamma)} \rangle_{\text{NI}} $ , which are the differences in the mean-field occupancies from the occupancies in noninteracting case shown in Fig.~\ref{fig-NIOccupation}.   The corresponding change in the charge density 
$\delta \rho_c =\rho_c - \rho_{\text{NI}}$  is shown as the solid blue curve in Fig.~\ref{fig-MFchargeDensities}. 

The shifts in occupancy plotted in Fig.~\ref{fig-OccupationsMF} can be primarily understood as a consequence of the electromagnetic ``self energy'' $\Sigma_{c}^{(\gamma)}= \rho_c q_{\gamma}  \mathbb S_{\text{lattice}}$ of each species $\gamma$ interacting with the total charge $\rho_c$ of the lattice.  Immediately, this implies that the parallel dimers are hardly modified at all, since they are electrically neutral and not directly affected by the total charge $\rho_c$ of the lattice.  For the vacancies and perpendicular dimers, which are charged, the mean field corrections plotted in Figs.~\ref{fig-OccupationsMF} and \ref{fig-MFchargeDensities} act to reduce the overall charge of the lattice.  Below $\beta \varepsilon_\parallel \approx 2$, the total charge of the lattice is positive, so the mean field corrections penalize the positively-charged perpendicular dimers and enhance the negatively-charged vacancies.  Above $\beta \varepsilon_\parallel \approx 2$, the total charge of the lattice is negative, and this effect is reversed.  Thus, the mean-field corrections tend to reduce (but not eliminate) the charge inversion at weak binding seen in Fig.~\ref{fig-NIOccupation}.  Note that the biggest effect of this electrostatic self energy is to heavily penalize the ``naked lattice'' of negatively-charged vacancies which was the favored ground state at large positive $\varepsilon_\parallel$ for the non-interacting Hamiltonian.

The fourth task of this section is to determine variance of the charge density from the mean field, similar to Eq.~(\ref{eq-sigmaNIdef}) for the noninteracting model, which gives a measure of the importance of spatial fluctuations in the charge density, can be written in the form
\begin{equation}
\sigma_c^2 =  \langle \rho^2 \rangle_c - \rho_c^2\; ,
\label{eq-VariancerhoMF}
\end{equation}
where the average $ \langle \rho^2 \rangle_c$ of the square of the charge density, similar to Eq.~(\ref{eq-rhosqsiteNI}) for the noninteracting model, is
\begin{equation}
 \langle \rho^2 \rangle_c = \sum_\gamma q_\gamma^2 \langle n^{(\gamma)} \rangle_c 
 = \frac{ \sum_\gamma q_\gamma^2 a_c^{(\gamma)}}
{\sum_{\gamma'}    a_c^{(\gamma')}}     \;  .
 \label{eq-rhosqavg}
 \end{equation}
The variance $\sigma_c^2$ will be useful in knowing the size of the charge fluctuations in the mean field, and this quantity will appear in the next level of approximation.  The difference in the standard deviation 
$\delta \sigma_c=\sigma_c-\sigma_{\text{NI}}$ is plotted as the red dashed curve in Fig.~\ref{fig-MFchargeDensities}, where $\sigma_{\text{NI}}$ for the noninteracting model was plotted as the red dashed curve in Fig.~\ref{fig-NIchargeDensities}.

Using a spatially uniform saddle point, which yields a charge density $\rho_c$ that is equal at the mean field level to the auxiliary field $\Delta_c$ in real space, it is interesting to write down the auxiliary field in the diagonal basis, given by
\begin{equation}
{\tilde \Delta}_{(k,\sigma),c} =
\sum_{\ell,b} 
{\mathcal W}_{(k,\sigma),(\ell,b)}^T   
\left.  \langle \rho_{\ell,b} \rangle \right|_c
=  \rho_c  \sum_{\ell,b} {\mathcal W}_{(k,\sigma),(\ell,b)}^T  \; .
\label{eq-DeltaCeqrhoDiagBasis}
\end{equation} 
From Eqs.~(\ref{eq-Wtranskgt0}) and (\ref{eq-Wtransklt0}), we see that when $k \ne 0$, ${\mathcal W}_{(k,\sigma),(\ell,b)}^T$ has the $\ell$ dependence of the form $e^{\pm i k \ell}$, which by Eq.~(\ref{eq-sumexpikell}) sums to zero.  The only surviving term is for $k=0$ which when summed over $\ell$ gives $2 {\mathcal N} +1$, and we have from Eqs.~(\ref{eq-Wtranskeq0}) and (\ref{eq-DeltaCeqrhoDiagBasis})
\begin{equation}
{\tilde \Delta}_{(k,\sigma),c} = \tilde \rho_{(k,\sigma),c}=
\delta_{k,0}  \rho_c \sqrt{2 {\mathcal N} +1}  \sum_b \xi_{b,\sigma}(0)  \; .
\label{eq-DeltaCDiagBasis}
\end{equation} 
There are two different results corresponding to the two different eigenvalue labels $\sigma=+$ and $\sigma=-$ in Eq.~(\ref{eq-Eigenvalues}).  Summing over $b$, these are
\begin{eqnarray}
{\tilde \Delta}_{(0,+),c} &=&\tilde \rho_{(0,+),c}= \rho_c \sqrt{ {\mathcal N}_{\text{sites}}}   
\nonumber \\
{\tilde \Delta}_{(0,-),c} &=& \tilde \rho_{(0,-),c}=0
  \; ,
\label{eq-DeltaCDiagBasispm}
\end{eqnarray} 
where ${\mathcal N}_{\text{sites}}=2(2{\mathcal N}+1)$.  The interpretation of Eq.~\eqref{eq-DeltaCDiagBasispm} can be understood from Eq.~(\ref{eq-Eigenvectorskeq0}).  The sum over $b$ in Eq.~(\ref{eq-DeltaCDiagBasis}) is a sum over the elements rows in Eq.~(\ref{eq-Eigenvectorskeq0}), which for $\sigma=+$ sums the mean charge densities on the two chains and for $\sigma=-$ subtracts the charge densities on the two chains.

\begin{figure}
\includegraphics[width=3in]{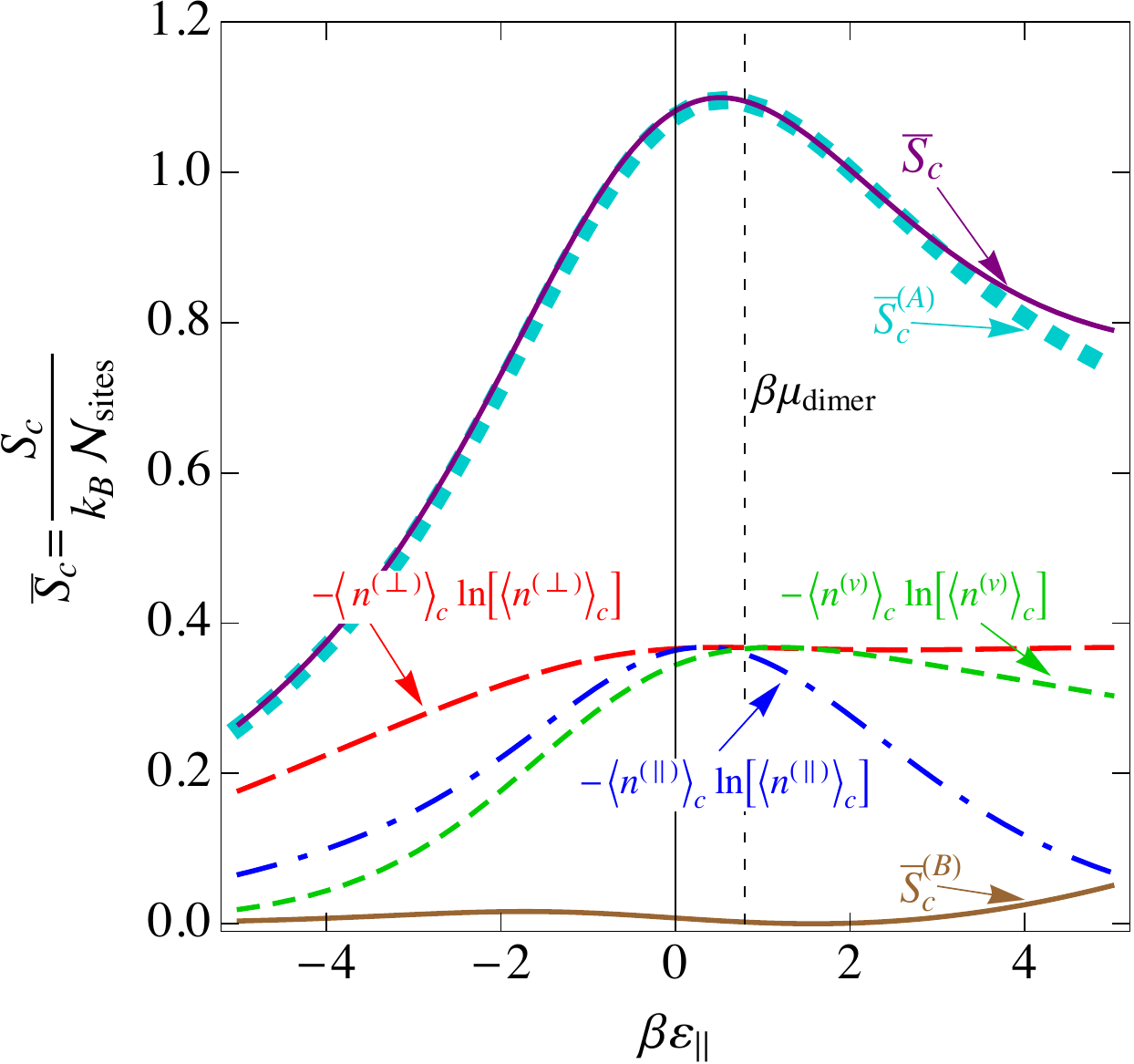}
\caption{The functional form of mean-field entropy 
$\frac{S^{{0}}}{k_B  {\mathcal{N}}_{\text{sites}}}$ for the interacting lattice gas model at the mean field level.  The constituent parts of the entropy are shown as well.}
\label{fig-EntropyMFconstituentPlot}
\end{figure}

\begin{figure}
\includegraphics[width=3in]{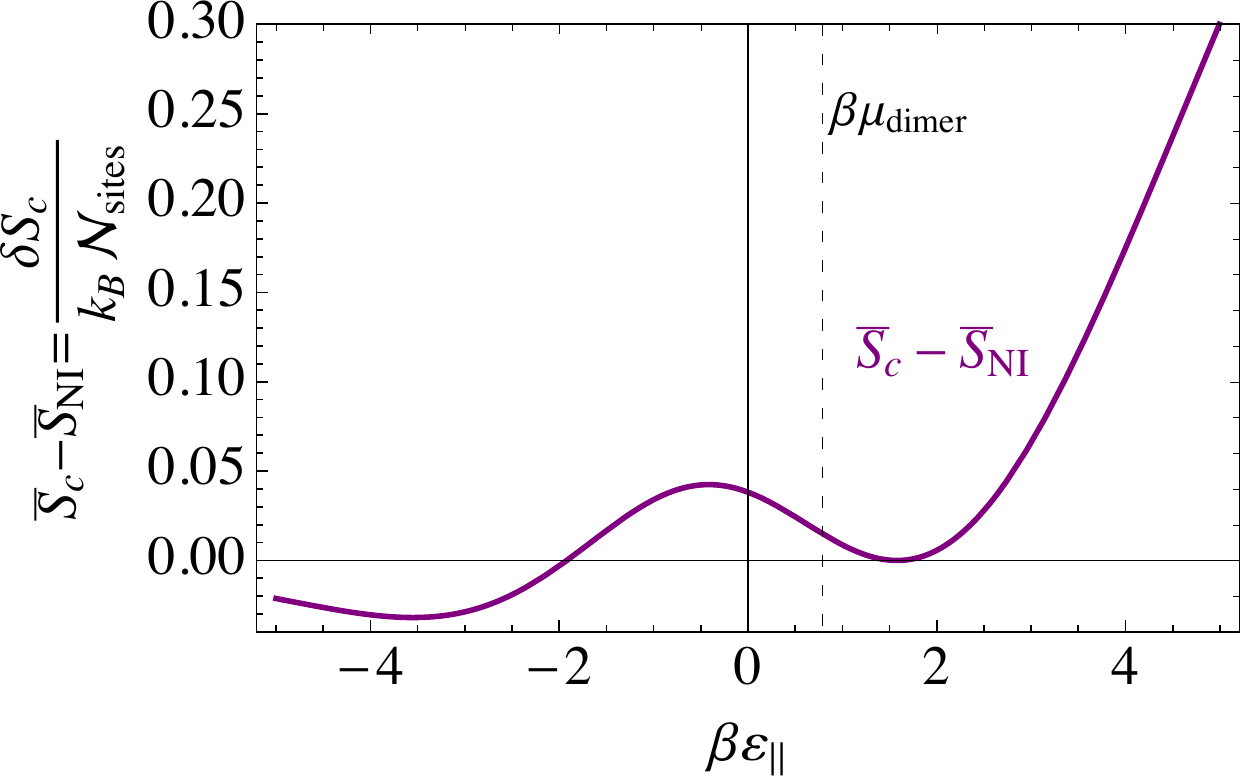}
\caption{The difference between the mean-field entropy for the interacting lattice gas model and the entropy for the noninteracting model.}
\label{fig-diffScSNIPlot}
\end{figure}

The fifth task of this section is to calculate the entropy for the spatially uniform saddle point, which is our mean field value of the entropy.  This requires the effective action at this saddle point, which becomes
\begin{eqnarray} 
{\mathscr S}_{\text{eff},c}
&=&
- {\mathcal{N}}_{\text{sites}}\ln\left(\sum_{\gamma}a_{c}^{(\gamma)}\right)
 \nonumber \\
&&
-\frac{\rho_{c}^2}{2} {\mathcal{N}}_{\text{sites}}\beta \,
\mathbb S_{\text{lattice}}
\; .
\label{eq-EffectiveActionMF}
\end{eqnarray} 
The grand canonical potential is then
\begin{equation}
\Omega_{m=1}  =  \Omega_c  =
\frac{1}{\beta} \mathscr S_{\text{eff,c}} 
\; .
\end{equation}
We can now compute the entropy from the derivative of the grand
thermodynamic potential with respect to the inverse temperature $\beta$, as in
Sec.~\ref{sec-NILatticeGas}.  This derivative becomes much more complicated with the
introduction of
$U_{\text{int}}$ because the screened Coulomb potential (\ref{eq-screenedCoulombOfD}) depends on $\beta$ through the
screening vector $q_s$ (\ref{eq-ScreeningVector}), where
\begin{equation}
\frac{\partial q_s}{\partial \beta} =\frac{q_s}{2\beta}   \; .
\label{eq-qsbetaderiv}
\end{equation} 
This introduces a $\beta$-dependence into
both terms of Eq.~(\ref{eq-EffectiveActionMF}), so that the mean field
entropy is given by the more complex expression
\begin{equation}
\bar{S}_c = \frac{S_c}{k_B  {\mathcal{N}}_{\text{sites}}} 
= \frac{S_c^{(A)}+S_c^{(B)}}{k_B  {\mathcal{N}}_{\text{sites}}}  \; ,
\label{eq-EntropyMF}
\end{equation}
where
\begin{equation}
\bar{S}_c^{(A)}=\frac{S_c^{(A)}}{k_B  {\mathcal{N}}_{\text{sites}}} 
= -\sum_{\gamma}\langle n^{(\gamma)}
\rangle_c \ln \langle n^{(\gamma)} \rangle_c 
 \; ,
\label{eq-EntropyMF00}
\end{equation}
is the mixing entropy of the three species using the mean field occupation numbers and
\begin{equation}
\bar{S}_c^{(B)}=\frac{S_c^{(B)}}{k_B  {\mathcal{N}}_{\text{sites}}} 
= \frac{\rho_c^2}{2}\beta^2\mathbb \, 
\frac{\partial\mathbb S_{\text{lattice}}}{\partial \beta}  
\label{eq-EntropyMF01}
\end{equation}
arises directly from the self energy $\Sigma \propto \mathbb S_{\text{lattice}}$.  The occupation numbers $\langle n^{(\gamma)} \rangle_c$ used in this expression are those of the mean-field level, given by Eq.~(\ref{eq-meanOccInt4b}) and shown in Fig.~\ref{fig-OccupationsMF}.  The mean-field entropy $\bar{S}^{(0)}$, given in Eq.~(\ref{eq-EntropyMF}), is plotted in Fig.~\ref{fig-EntropyMFconstituentPlot}, together with $-\langle n^{(\gamma)}
\rangle_c \ln \langle n^{(\gamma)} \rangle_c$ for each species.   Also shown are the two constituents 
$\bar{S}_c^{(A)}$ and $\bar{S}_c^{(B)}$ of the total entropy. This plot shows that, as with $S_{NI}$ in Fig.~\ref{fig-NIEntropy}, the mean-field entropy is
greatest for $\varepsilon_{\parallel}$ near $\mu_{\text{dimer}}$ and decreases outside this region.  The difference 
$\delta \bar{S}_c=\frac{S_c-S_{\text{NI}}}{k_B T}$ between this mean-field entropy for the interacting model and the entropy for the noninteracting model is plotted in Fig.~\ref{fig-diffScSNIPlot}.  Note that this curve has roughly the same shape as that of $\delta \sigma_c$ in Fig.~\ref{fig-MFchargeDensities}, with a large increase in disorder at large positive $\varepsilon_\parallel$ and small changes elsewhere.

By employing field-theoretic techniques borrowed from quantum mechanics, we have
included the interaction energy $U_{\text{int}}$ in the partition function and found a
mean-field approximation to the exact partition function.  However, these mean-field
level calculations assume the same average charge $\rho_c$ for all
DNA molecules in the same solution.  Thus, mean-field calculations would predict a
repulsive interaction between two DNA molecules, which would not lead to DNA
condensation.  Indeed, this repulsive interaction is all that this zeroth-order
calculation can predict.  It is the fluctuations of the charge from its mean-field
value that enable the net attraction observed experimentally; to understand the
attractive forces between DNA molecules in solution, it is necessary to extend
this treatment by at least one additional order.  The next order in the
expansion (\ref{eq-SaddlePtExpansion})  will yield Gaussian-type integrals, which are
analytically solvable for the lowest-order fluctuations in the average charge and other
parameters.  Furthermore, once lowest-order fluctuations have been included, it will be
possible to calculate thermodynamic properties of the system from the partition
function to meaningful levels, such as the entropy changes due to small fluctuations
in the local charge.

\section{Inverse-Hartree-field-fluctuation propagator}
\label{sec-HartreePropagator}

In order to calculate the inverse-Hartree-field-fluctuation propagator  $\tilde{D}^{-1}$,
we need to find the second derivatives of the effective action, evaluated at the saddle point.  Since  $\tilde{D}^{-1}$ is diagonal, from Eq.~(\ref{eq-DtldeInvDef}), we can write
\begin{eqnarray}
(\tilde D^{-1})_{(k_1,\sigma_1),(k_2,\sigma_2)} 
&=&
\left(- \frac{1}{\beta \lambda_{k_1,\sigma_1}} \right) \times
\nonumber \\
&& \times \left. { \frac{\partial^2{\mathscr S}_{\text{eff}}[{\tilde \Delta}]}
 {\partial {\tilde \Delta}_{k_1,\sigma_1} \partial {\tilde \Delta}_{k_2,\sigma_2}}}
\right|_{{\tilde \Delta}_c}  \; .
\label{eq-DtldeInvDiagForm}
\end{eqnarray}
Since we already calculated the first derivative in Sec.~\ref{sec-saddlepoint}, we need one more derivative, now with respect to 
${\tilde \Delta}_{k_2,\sigma_2}$. Substituting from Eq.~(\ref{eq-FirstDerivSeff})
\begin{eqnarray}
&&(\tilde D^{-1})_{(k_1,\sigma_1),(k_2,\sigma_2)} =
\nonumber  \\
&&=
\left(- \frac{1}{\beta \lambda_{k_1,\sigma_1}} \right) \times
\nonumber \\
&& \times  \left. \frac{\partial}{\partial {\tilde \Delta}_{k_2,\sigma_2}}
\left\{  -\beta \lambda_{k_1,\sigma_1}{\tilde \Delta}_{k_1,\sigma_1} +  \right. \right.
\nonumber   \\ 
&&\left.\left.+\beta \lambda_{k_1,\sigma_1}  \sum_{\ell,b}   
{\mathcal W}_{(k_1,\sigma_1),(\ell,b)}^T   
\frac{\sum_\gamma q_\gamma \tilde a_{\ell,b}^{(\gamma)}[{\tilde \Delta}] }    
{\sum_{\gamma '}      
\tilde a_{\ell,b}^{(\gamma ')}[{\tilde \Delta}] } \right\} 
\right|_{{\tilde \Delta}_c} \; .
\label{eq-DtldeInvCalc}
\end{eqnarray}
Performing the derivative, we have
\begin{eqnarray}
&&(\tilde D^{-1})_{(k_1,\sigma_1),(k_2,\sigma_2)} =
\nonumber  \\
&&=
\left. \left\{ 1   - \sum_{\ell,b}   
{\mathcal W}_{(k_1,\sigma_1),(\ell,b)}^T    
\frac{\sum_\gamma q_\gamma  \frac{\partial {\tilde a}_{\ell,b}^{(\gamma)}[{\tilde \Delta}] }
{\partial {\tilde \Delta}_{k_2,\sigma_2}} }   
{\sum_{\gamma '}      
{\tilde a}_{\ell,b}^{(\gamma ')}[{\tilde \Delta}] } + \right.\right.
\nonumber   \\ 
&&\left. \left.+ \sum_{\ell,b}   
{\mathcal W}_{(k_1,\sigma_1),(\ell,b)}^T   
\frac{\sum_\gamma q_\gamma  {\tilde a}_{\ell,b}^{(\gamma)}[{\tilde \Delta}] }
 { \left( \sum_{\gamma '}      
{\tilde a}_{\ell,b}^{(\gamma ')}[{\tilde \Delta}]  \right)^2}
\sum_{\gamma''} \frac{\partial {\tilde a}_{\ell,b}^{(\gamma'')}[{\tilde \Delta}] }
{\partial {\tilde \Delta}_{k_2,\sigma_2} }    \right\}
\right|_{{\tilde \Delta}_c}  \; .
\nonumber\\
\label{eq-DtldeInvCalc1}
\end{eqnarray}
Substituting the derivative of the activity from Eqs. (\ref{eq-firstDerivActivity}) and (\ref{eq-firstDerivSigma}), we have
\begin{eqnarray}
&&(\tilde D^{-1})_{(k_1,\sigma_1),(k_2,\sigma_2)} =
\nonumber  \\
&&=
\left. \left\{ 1 + \beta \lambda_{k_1,\sigma_1}
 \times \right. \right.
\nonumber   \\ 
&&\left. \left. \times 
 \sum_{\ell,b}   
 {\mathcal W}_{(k_1,\sigma_1),(\ell,b)}^T  
 {\mathcal W}_{(k_2,\sigma_2),(\ell,b)}^T
\times \right. \right.
\nonumber   \\ 
&&\left. \left. \times  \left[
\frac{\sum_\gamma q_\gamma^2   {\tilde a}_{\ell,b}^{(\gamma)}} 
{\sum_{\gamma '}      
{\tilde a}_{\ell,b}^{(\gamma ')}[{\tilde \Delta}] } -   
\left(  \frac{\sum_\gamma q_\gamma  {\tilde a}_{\ell,b}^{(\gamma)}[{\tilde \Delta}] }
 { \sum_{\gamma '}      
{\tilde a}_{\ell,b}^{(\gamma ')}[{\tilde \Delta}] }  \right)^2 \right] \right\}
\right|_{{\tilde \Delta}_c}
 \; .
\nonumber\\
\label{eq-DtldeInvCalc2}
\end{eqnarray}
The quantity in square brackets is just the constant $\sigma_c^2$, the variance of the charge density, given in Eq.~(\ref{eq-VariancerhoMF}). Since ${\mathcal W}$ from Eqs. (\ref{eq-Wtranskgt0}), (\ref{eq-Wtransklt0}), and (\ref{eq-Wtranskeq0}), is an orthogonal matrix,
\begin{equation}
 \sum_{\ell,b}   
 {\mathcal W}_{(k_1,\sigma_1),(\ell,b)}^T 
 {\mathcal W}_{(k_2,\sigma_2),(\ell,b)}^T
 =\delta_{k_1 k_2} \delta_{\sigma_1 \sigma_2}   \;,
\label{eq-Worthog}
\end{equation}
and this means that the matrix is diagonal.
With these substitutions, the inverse-Hartree-field-fluctuation propagator becomes 
\begin{equation}
(\tilde D^{-1})_{(k_1,\sigma_1),(k_2,\sigma_2)} 
=\delta_{k_1 k_2}\delta_{\sigma_1 \sigma_2}
\left( 1 + \sigma_c^2 \beta \lambda_{k_1,\sigma_1} \right) \; .
\label{eq-DtldeInvCalcFinal}
\end{equation}
This is the inverse propagator for the fluctuations of the auxiliary Hartree field that we introduced to decouple the interparticle interactions, expressed in the diagonal basis.

\begin{figure}
\includegraphics[width=3in]{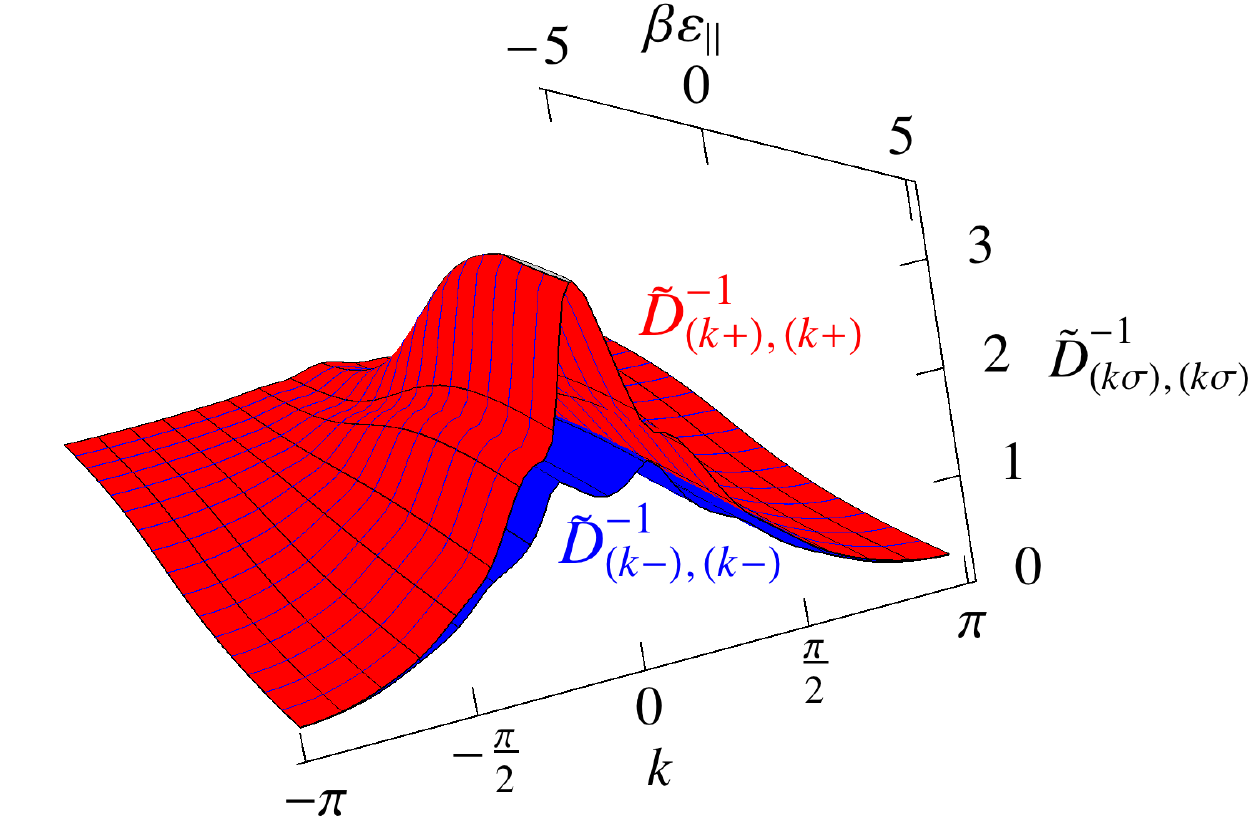}
\caption{The inverse propagator for the fluctuations of the auxiliary Hartree field
$(\tilde D^{-1})_{(k_1,\sigma_1),(k_2,\sigma_2)} $.}
\label{fig-Dtildeinverse3DPlot}
\end{figure}

\begin{figure}
\includegraphics[width=3in]{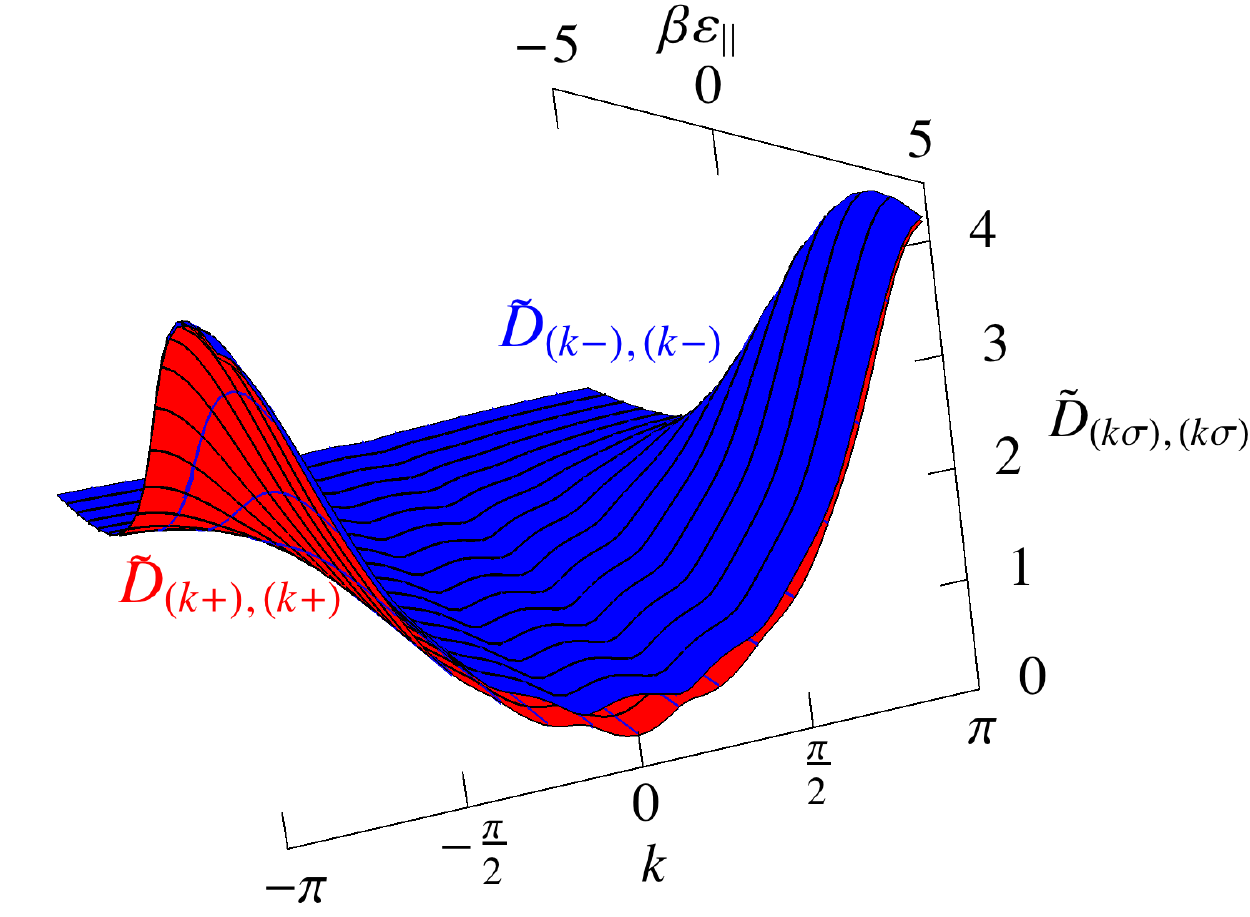}
\caption{The propagator for the fluctuations of the auxiliary Hartree field
$\tilde D_{(k_1,\sigma_1),(k_2,\sigma_2)} $.}
\label{fig-Dtilde3DPlot}
\end{figure}

The inverse propagator $(\tilde D^{-1})_{(k_1,\sigma_1),(k_2,\sigma_2)} $ in the diagonal basis is plotted in Fig.~\ref{fig-Dtildeinverse3DPlot} for the two eigenvalues $\lambda_{k\pm}$, shown in Fig.~\ref{fig-Eigenvalues}.  Notice that the shapes are those of the eigenvalues, but shifted upward.  They both have minima at $k=\pi$, with the one corresponding to $\lambda_{k+}$ having its maximum at $k=0$ and the one corresponding to $\lambda_{k-}$ having maxima near $k=0$.  
The corresponding plots of the propagator $\tilde D_{(k_1,\sigma_1),(k_2,\sigma_2)} $ are shown in  Fig.~\ref{fig-Dtilde3DPlot}, also in the diagonal basis.  
The propagator increases with $\beta \varepsilon_\parallel$ near $k=\pm \pi$, and this can be attributed, at least in part, to the increase in $\sigma_c$, shown in Fig.~\ref{fig-MFchargeDensities}, because of the dependence shown in Eq.~(\ref{eq-DtldeInvCalc2}).  Although the auxiliary field propagator normally describes the behavior of fluctuations of a physical field like the charge, its definition given by Eq.~(\ref{eq-partitonfctproduct}) shows that the field here have been rescaled by the square root of the eigenvalue.  This was discussed earlier in the context of entropy in the paragraph after Eq.~(\ref{DnormDef}), and so this form is convenient for the entropy calculation, our focus here, but its interpretation is not quite so direct.

\section{Inclusion of Lowest-Order Fluctuations}
\label{sec-Fluctuations}

\begin{figure}
\includegraphics[width=3in]{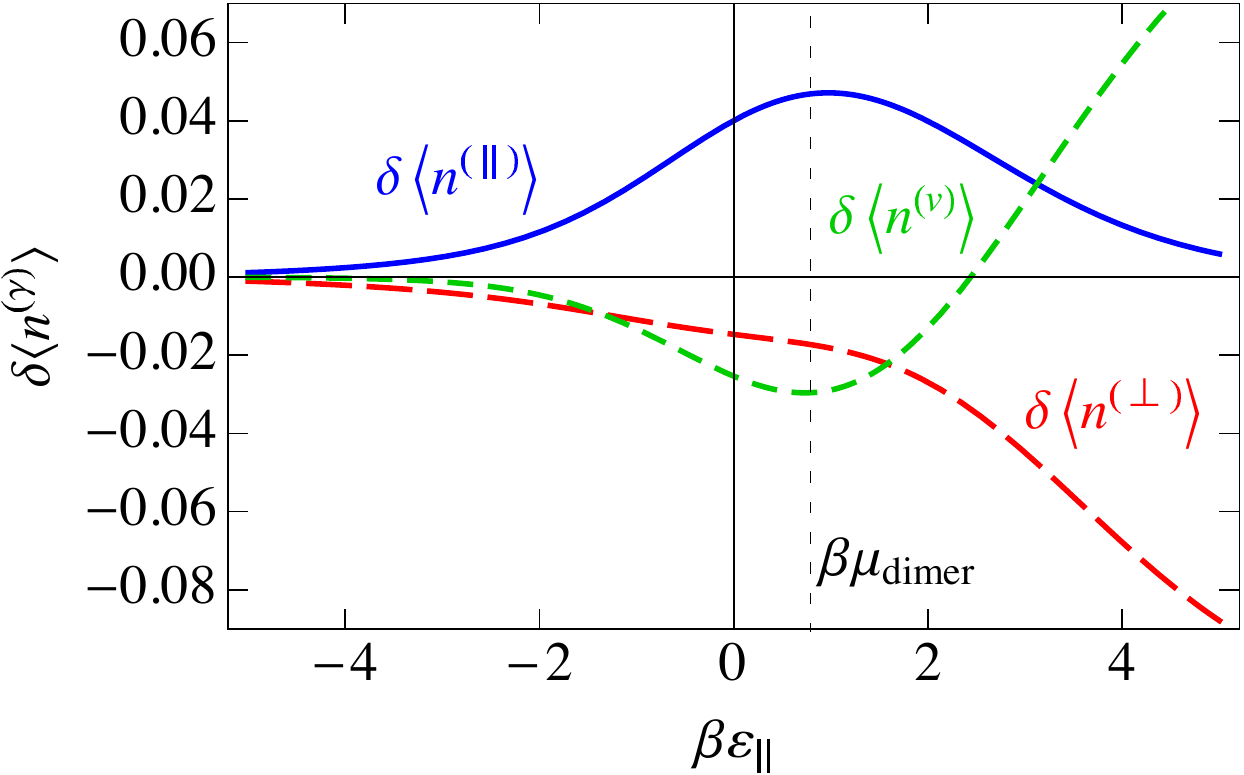}
\caption{The changes in the occupancies $ \delta\langle n^{(\gamma)}\rangle$ due to fluctuations.  These should be compared with the changes between occupancies of the mean-field and noninteracting systems shown in 
Fig.~ \ref{fig-OccupationsMF}.}
\label{fig-deltaOccupancieswithfluctuations}
\end{figure}

Going beyond mean field to one-loop order requires use of the full partition function given by 
Eqs.~(\ref{ZGOneLoop}) and (\ref{eq-ZGSingleExp}).  This has consequences for the site occupancies, the average charge per site, the charge variance, and the entropy.  The expression for site occupancy has the same form in terms of the partition function as for the noninteracting and mean field cases in Eqs.~(\ref{eq-ngammaderivZ}) and (\ref{eq-ngammaderivZinteracting}), and can be written explicitly for the one-loop order of approximation as
\begin{equation}
\langle n^{(\gamma)} \rangle_{\text{tot}}
=\frac{1}{{\mathcal{N}}_{\text{sites}} \beta Z_{\text{G}} }
\frac{\partial}{\partial \mu_\gamma}  
\left[\frac
{e^{ -  \mathscr S_{\text{eff}}[{\tilde \Delta}_c]}}
{\sqrt{\det(\tilde D^{-1})}} \right]      \; ,    
\label{eq-meanOccIntFlucdef}
\end{equation}
which is obtained from Eq.~(\ref{eq-leadingexpansionterm}) with $m$ set equal to one.  In that expression, $m$ was the parameter that was introduced in Sec.~\ref{sec-expansion} as an aid to identifying the various orders in the expansion.
This can be written as the sum of two terms,
\begin{equation}
\langle n^{(\gamma)} \rangle_{\text{tot}} 
=\langle n^{(\gamma)}\rangle_c +\delta\langle n^{(\gamma)}\rangle
  \; ,
\label{eq-meanOccIntFluctwoterms}
\end{equation}
where $\langle n^{(\gamma)}\rangle_c $ is the occupancy of species $\gamma$ at the mean field level, given by Eq.~(\ref{eq-meanOccInt4b}) ,
\begin{equation}
\langle n^{(\gamma)} \rangle_c  
=-\frac{1}{{\mathcal{N}}_{\text{sites}} \beta  }
\frac{\partial}{\partial \mu_\gamma}  
 \mathscr S_{\text{eff}}[{\tilde \Delta}_c]
  \; ,
\label{eq-meanOccIntFlucFirstTerm}
\end{equation}
and $\delta\langle n^{(\gamma)}\rangle$ is the correction to that mean site occupancy due to fluctuations, which is given by
\begin{equation}
\delta\langle n^{(\gamma)}\rangle
=\frac{1}{{\mathcal{N}}_{\text{sites}} \beta}
\frac{\partial}{\partial \mu_\gamma}\ln[ \det(\tilde{D}^{-1})]
  \; ,
\label{eq-deltameanOccFluc}
\end{equation}
with the inverse propagator $\tilde D_{(k_1,\sigma_1),(k_2,\sigma_2)}^{-1}$ from Eq.~(\ref{eq-DtldeInvCalcFinal}) written in matrix form as
\begin{equation}
\tilde D^{-1}=
I+ \sigma_c^2 \beta \Lambda \; ,
\label{eq-DiagonaltildeD}
\end{equation}
where $I$ is the identity matrix and $\Lambda$ is the (diagonal) matrix of eigenvalues of the potential.
Since $\tilde D^{-1}$  is diagonal, its determinant is given by the product of its diagonal elements as
\begin{equation}
\det \tilde D^{-1}=\prod_{k,\sigma} (1+\beta \lambda_{k,\sigma} \sigma_c^2)   \;  .
\label{eq-DettildeD-1}
\end{equation}
The logarithm of this determinant then gives the sum over logarithms of diagonal elements as
\begin{equation}
\ln(\det \tilde D^{-1})=\sum_{k,\sigma} \ln (1+\beta \lambda_{k,\sigma} \sigma_c^2)   \;  .
\label{eq-logDettildeD-2}
\end{equation}
The only $\mu_\gamma$ dependence is contained in $\sigma_c^2$, and its derivative is given by
\begin{equation}
\frac{\partial \sigma_c^2}{\partial \mu_\gamma}  
= \beta \langle n^{(\gamma)} \rangle_c  [q_\gamma^2 - \langle \rho^2 \rangle_c-2\rho_c(q_\gamma-\rho_c)]
 \;  ,
\label{eq-DettildeD-3}
\end{equation}
so that the derivative of $\ln(\det D^{-1})$ is given by the coefficient
\begin{equation}
\mathbb{C} = \frac{1}{2{\mathcal{N}}_{\text{sites}} }
\sum_{k,\sigma} \frac{\beta \lambda_{k \sigma}}{[1+\beta \lambda_{k,\sigma} \sigma_c^2]}
  \; .
\label{eq-coefficient}
\end{equation}
Then the mean fluctuation in occupancy
$\delta\langle n^{(\gamma)}\rangle$ becomes
\begin{equation}
 \delta\langle n^{(\gamma)}\rangle
=\mathbb{C}  \langle n^{(\gamma)} \rangle_c  [q_\gamma^2 - \langle \rho^2 \rangle_c-2\rho_c(q_\gamma-\rho_c)]
  \;  . 
\label{eq-deltameanOccFluc1}
\end{equation}
These changes in the mean field occupancies at this one loop order are plotted in Fig.~\ref{fig-deltaOccupancieswithfluctuations}.  These changes are not large, and the fluctuation corrections amount to no more than about 10\% of the total.
The most dramatic effects are seen in the region of large positive $\varepsilon_\parallel$, with the corrections to the vacancies and perpendicular dimers going in the opposite direction to the corrections due to the mean field in Fig.~\ref{fig-OccupationsMF}.  This can be interpreted as a relaxation of the rigid penalties imposed by the electrostatic self energies at the mean field level.  An enhancement of parallel dimers is also seen across the whole range of $\varepsilon_\parallel$ that peaks around $\varepsilon_\parallel = \mu_{\text{dimer}}$.

\begin{figure}
\includegraphics[width=3in]{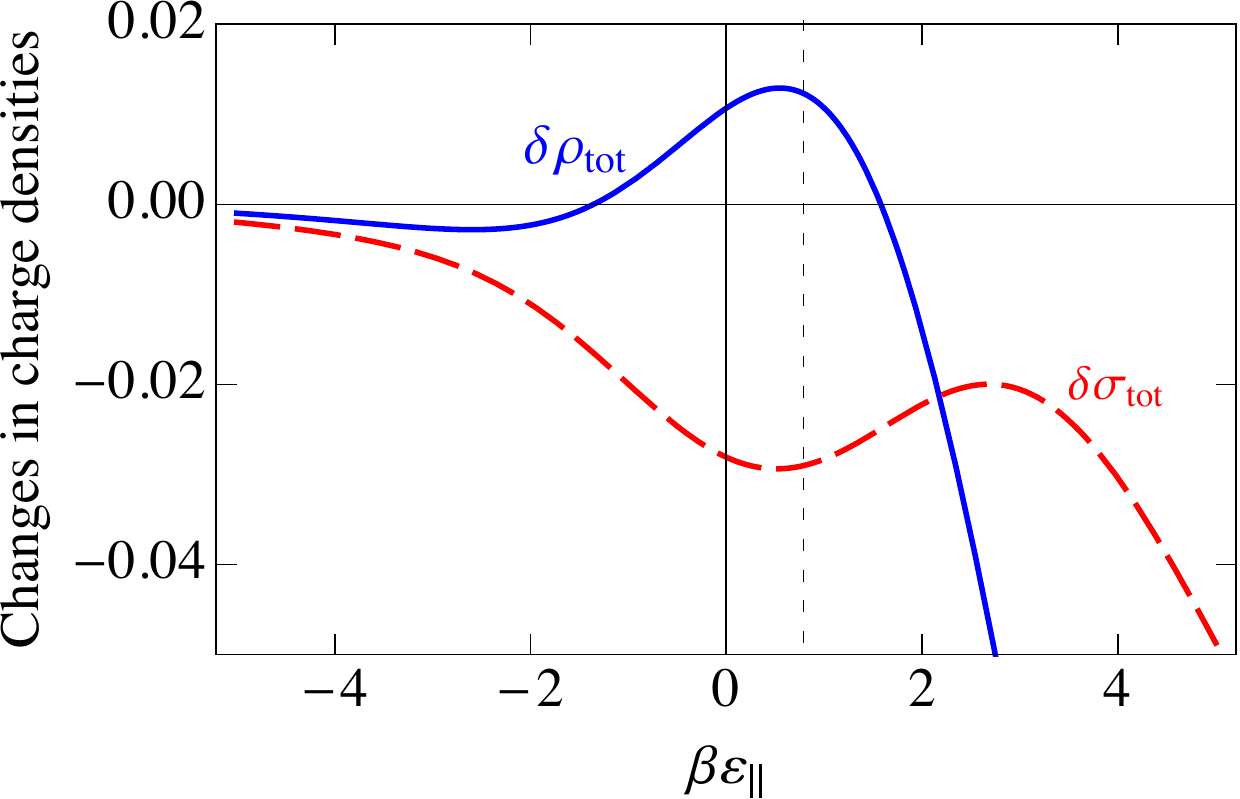}
\caption{The changes in the charge density $\delta\rho_{\text{tot}}$ and in the standard deviation $\delta\sigma_{\text{tot}}$ of the charge density due to fluctuations.  These are further corrections to the mean-field values of Fig.~\ref{fig-MFchargeDensities}.}
\label{fig-RhoFlucChanges}
\end{figure}

The mean charge density at this one-loop-correction level of approximation is given by the same form as Eqs.~(\ref{eq-rhoNIsitedef}) and (\ref{eq-MeanChargeDensityUniform}) for the noninteracting and mean field approximations and is given by
\begin{equation}
\rho_{\text{tot}} = \langle \rho \rangle_{\text{tot}}  =\sum_\gamma q_\gamma \langle n^{(\gamma)} \rangle_{\text{tot}}  \; ,
\label{eq-rhototdef}
\end{equation}
where now $\langle n^{(\gamma)} \rangle_{\text{tot}}$ is taken from Eqs.~(\ref{eq-meanOccIntFluctwoterms}) and (\ref{eq-deltameanOccFluc1}).
and the charge density becomes
\begin{equation} 
\rho_{\text{tot}} = \rho_c +  \delta\rho_{\text{tot}}
 \; ,
\label{eq-rhotot}
\end{equation}
where
\begin{equation} 
 \delta\rho_{\text{tot}}= \rho_c  \mathbb{C} (1- 3 \langle \rho^2 \rangle_c + 2\rho_c^2)
 \; ,
\label{eq-deltarhotot}
\end{equation}
The change in the charge density $\delta \rho_{\text{tot}} $ is shown as the solid blue curve in Fig.~\ref{fig-RhoFlucChanges}.
The most notable features are a negative contribution to the total charge above $\beta \varepsilon_\parallel \approx 2$, reflecting the addition of more negatively charged vacancies due to the fluctuations, and a positive contribution to the total charge around $\varepsilon_\parallel \approx \mu_{\text{dimer}}$.  Interestingly, this positive contribution is not associated with the increase of parallel dimer occupancy, since parallel dimers are electrically neutral.  Rather, it is due to the fact that, although both the negatively charged vacancies and positively charged perpendicular dimers are both reduced around $\varepsilon_\parallel \approx \mu_{\text{dimer}}$, the vacancies are decreased more.

The variance in the charge is given by
\begin{equation}
\sigma_{\text{tot}}^2  = \langle \rho^2 \rangle_{\text{tot}} - \rho_{\text{tot}}^2    \; .
\label{eq-sigmatotsqdef}
\end{equation}
The average of the square of the charge density is given by
\begin{equation}
\langle \rho^2 \rangle_{\text{tot}}   =\sum_\gamma q_\gamma^2 \langle n^{(\gamma)} \rangle_{\text{tot}}  \; ,
\label{eq-rhosqtotdef}
\end{equation}
and is written explicitly as
\begin{equation}
\langle \rho^2 \rangle_{\text{tot}}   = \langle \rho^2 \rangle_c +\delta\langle \rho^2 \rangle_{\text{tot}}  \; ,
\label{eq-rhosqtot}
\end{equation}
where $\langle \rho^2 \rangle_c$ is given by Eq.~(\ref{eq-rhosqavg}), and $\delta\langle \rho^2 \rangle_{\text{tot}}$ works out to be
\begin{equation}
\delta\langle \rho^2 \rangle_{\text{tot}}  =  \mathbb{C} [ \langle \rho^2 \rangle_c(1- \langle \rho^2 \rangle_c - 2\rho_c^2) - 2\rho_c^2]  \; .
\label{eq-rhosqf}
\end{equation}
The charge variance can then be written as
\begin{equation}
\sigma_{\text{tot}}^2  = \sigma_c^2  +  \sigma_f^2 \; ,
\label{eq-sigmatotsq}
\end{equation}
where the fluctuation correction is
\begin{eqnarray}
\sigma_f^2  &=&  - \langle \rho^2 \rangle_c^2 \mathbb{C} (1 + 9 \rho_c^2  \mathbb{C} )
\nonumber \\
&+&  \langle \rho^2 \rangle_c \mathbb{C} [(1+8 \rho_c^2)+6\rho_c^2\mathbb{C} (1+2\rho_c^2)]
\nonumber \\
&-&  \rho_c^2\mathbb{C}[4(1+\rho_c^2) - \mathbb{C} (1+4\rho_c^2+4\rho_c^4)]
  \; .
\label{eq-sigmafsq}
\end{eqnarray}
The standard deviation can then be written as
\begin{equation}
\sigma_{\text{tot}}  = \sigma_c  +  \delta\sigma_{\text{tot}}\; ,
\label{eq-sigmatotwithdeltasigma}
\end{equation}
The change $\delta\sigma_{\text{tot}}$ in the standard deviation of the charge densities due to fluctuations are given in Fig.~\ref{fig-RhoFlucChanges}, where the fluctuations account for changes typically of the order of 10\% of the mean field value.

Having obtained results for the effects of fluctuations on the mean occupancies and charge densities, we now consider the corresponding corrections to the entropy.  For that, we need the grand thermodynamic potential, which was expanded in Sec.~\ref{sec-expansion}, as Eqs.~(\ref{eq-formalexpansion})-(\ref{eq-oneloopcorrection}), with the mean field and lowest order fluctuation terms given by
\begin{equation}
\Omega_G  \simeq
\Omega_{c}+\frac{1}{2 \beta}\ln\det D^{-1}  \;  .
\label{eq-GrandF}
\end{equation} 
The corresponding expression for the interacting entropy is
\begin{equation}
S_{\text{tot}} = S_c + \delta S_{\text{tot}}   \;  ,
\label{eq-Entropyl}
\end{equation}
where the mean field entropy $S_c$ was derived and the results presented in Sec.~\ref{sec-saddlepoint} in Eqs.~(\ref{eq-EntropyMF})-(\ref{eq-EntropyMF01}).  The fluctuation contribution to the entropy $\delta S_{\text{tot}}$ is given by
\begin{equation}
 \delta S_{\text{tot}}= k_B \beta^2  \frac{\partial}{\partial \beta}\left\{\frac{1}{2\beta} \ln{[\det{(D^{-1})} ]}\right\}  ,
\label{eq-EntropyI2begin}
\end{equation}
which requires performing another derivative.

To evaluate this contribution to the entropy $ \delta S_{\text{tot}}$ in Eq.~(\ref{eq-EntropyI2begin}), we return to
Eq.~(\ref{eq-logDettildeD-2}) and substitute into the
fluctuation contribution to the entropy from Eq.~(\ref{eq-EntropyI2begin}), which after performing derivatives gives
\begin{equation}
 \delta S_{\text{tot}}= S_f^{(A)} +S_f^{(B)}   \;  ,
\label{eq-EntropyI21}
\end{equation} 
where
\begin{equation}
S_f^{(A)}=
 - \frac{k_B}{2} \sum_{k,\sigma} \ln{ (1 + \beta \lambda_{k,\sigma} \sigma_c^2)}\label{eq-EntropyI21A}
\end{equation}
and
 \begin{eqnarray}
S_f^{(B)}&=&
 \frac{k_B}{2} \sum_{k,\sigma} \frac{1}{(1 + \beta \lambda_{k,\sigma} \sigma_c^2)}
\times   \nonumber \\    &&\times
(\beta \lambda_{k,\sigma} \sigma_c^2  
+\sigma_c^2 \beta^2 \frac{\partial \lambda_{k,\sigma}}{\partial \beta}
+\beta \lambda_{k,\sigma} \beta \frac{\partial \sigma_c^2}{\partial \beta}) \; .
\nonumber \\
\label{eq-EntropyI21B}
\end{eqnarray}

The derivative of the eigenvalue $\lambda_{k,\sigma}$ of the potential is given by
\begin{equation}
\frac{\partial \lambda_{k,\sigma}}{\partial \beta} 
= \frac{\partial }{\partial \beta} \tilde{V}_{\uparrow\uparrow}(k)
+\sigma \frac{\partial }{\partial \beta} |\tilde{V}_{\uparrow\downarrow}(k)|
\; ,
\label{eq-lambdaDeriv}
\end{equation}
where $ \tilde{V}_{\uparrow\uparrow}(k)$ is the Fourier transform of the interaction potential of a chain with itself and $\tilde{V}_{\uparrow\downarrow}(k)$ is the Fourier transform of the interaction potential between the two chains.  Both of these potentials depend on the screening vector $q_s$, and the only $\beta$ dependence of the potential is contained in $q_s$, which is proportional to 
$\sqrt{\beta}$.  The derivative we need is then given by Eq.~\eqref{eq-qsbetaderiv}.  As a consequence, the derivative of the Fourier transform of the potential is
\begin{eqnarray}
\frac{\partial \tilde V_{b_1,b_2}(k)}{\partial \beta}
&=&  \frac{1}{2} \sum_{\ell=-\mathcal N}^{\mathcal N}
\left[ e^{-i k \ell} \frac {\partial V_{b_1,b_2}(\ell)}{\partial \beta}
\right.   \nonumber \\  
&&  \left. +e^{i k \ell} \frac{\partial{V_{b_2,b_1}(\ell)}}{\partial \beta} \right]
\;  ,
\label{eq-Vtilde}
\end{eqnarray}
where the derivative of the potential in the position basis is given by
\begin{equation}
\frac{\partial{V_{b_1,b_2}(\ell)}}{\partial \beta} 
=V_{b_1,b_2}(\ell)  \left(- \frac{q_s}{2 \beta} d_{b_1,b_2}(\ell)\right)
\;  .
\label{eq-Vrealspace}
\end{equation}
Substituting and using the symmetry relations in Eqs.~(\ref{eq-dSymmetry}) and (\ref{eq-VellSymmetry}), the derivative of the Fourier transform of the potential is
\begin{equation}
 \frac{\partial \tilde V_{b_1,b_2}(k)}{\partial \beta} 
= - \frac{q_s}{2\beta}  \sum_{\ell=-\mathcal N}^{\mathcal N}
e^{-i k \ell} V_{b_1,b_2}(\ell) d_{b_1,b_2}(\ell)
\;  .
\label{eq-VtildeDeriv}
\end{equation}
Using these relations, the derivative of the eigenvalues becomes
\begin{eqnarray}
\beta^2 \frac{\partial \lambda_{k,\sigma}}{\partial \beta} 
&=&  - \frac{q_s}{2} \sum_{\ell=-\mathcal N}^{\mathcal N} \left\{   
\beta V_{\uparrow \uparrow}(\ell) d_{\uparrow \uparrow}(\ell) \cos(k \ell)
 \right.   \nonumber \\   
&&  \left. + \sigma V_{\uparrow \downarrow}(\ell) d_{\uparrow \downarrow}(\ell)
\left[
\frac{{\text{Re}}[\tilde V_{\uparrow \downarrow}(k)] }{|\tilde V_{\uparrow \downarrow}(k)|}   
\cos(k \ell) 
\right.  \right.   \nonumber \\   
&&  \left.  \left.  
-\frac{{\text{Im}}[\tilde V_{\uparrow \downarrow}(k)] }{|\tilde V_{\uparrow \downarrow}(k)|}   
\sin(k \ell) 
\right] \right\}   \;  ,
\label{eq-lambdaDerivFinal}
\end{eqnarray}
where ${\text{Re}}[\tilde V_{\uparrow \downarrow}(k)]$ and ${\text{Im}}[\tilde V_{\uparrow \downarrow}(k)]$ are the real and imaginary parts of $\tilde V_{\uparrow \downarrow}(k)]$.

Eq.~(\ref{eq-EntropyI21}) also requires the derivative of the variance $\sigma_c^2$, where $\sigma_c^2$ is given by Eq.~(\ref{eq-VariancerhoMF}).  This contains the average charge density from Eq.~(\ref{eq-MeanChargeDensityUniform}) and the average of the square of the charge density from Eq.~(\ref{eq-rhosqavg}).  This derivative is
\begin{eqnarray}
\beta \frac{\partial\sigma_c^2}{\partial\beta} & = &
\frac{\rho_c}{1+ \sigma_c^2 \beta \mathbb S_{\text{lattice}}}
 \times
 \nonumber \\
&&\times \left\{ 
\left( [\rho {\cal{E}}]_\text{av} -
\rho_c {\cal{E}}_c \right) 
\left[ 2+(1-\langle\rho^2\rangle_c)
\beta \mathbb S_{\text{lattice}} \right] + \right.
\nonumber \\
&  &   + 
\left. 
\rho_c
\left(3 \sigma_c^2 + \rho_c^2 - 1\right) \beta^2  \frac{\partial \mathbb 
S_{\text{lattice}}}{\partial \beta}
\right\}
\nonumber \\
& &
-\left( [\rho^2{\cal{E}} ]_\text{av} -
\langle\rho^2 \rangle_c {\cal{E}}_c \right)
\label{eq-DerSigsq}
\end{eqnarray}
where
\begin{equation}
{\cal{E}} \equiv \beta(\varepsilon_\gamma - \mu_\gamma+ \rho_c
q_\gamma \mathbb S_{\text{lattice}} )
\; ,
\label{eq-ecaldef}
\end{equation}
\begin{equation}
{\cal{E}}_c=
\sum_{\gamma} 
{\cal{E}}\langle n^{(\gamma)}\rangle_c
\; ,
\label{eq-ecalavDef}
\end{equation}
\begin{equation}
[\rho {\cal{E}}]_\text{av} =
\sum_{\gamma} 
q_\gamma {\cal{E}}  
\langle n^{(\gamma)}\rangle_c  \; , 
\label{eq-rhoecalav}
\end{equation}
and
\begin{equation}
[\rho^2{\cal{E}} ]_\text{av}=
\sum_{\gamma} 
q_\gamma^2  {\cal{E}} \langle n^{(\gamma)}\rangle_c   \;  .
\label{eq-rhosqecalav}
\end{equation}
Here $[\cdots]_\text{av}$ is the weighted average of the specified quantity over the distribution given by the mean-field occupancies $\langle n^{(\gamma)}\rangle_c$ of the three species.

It is interesting to note that the combinations 
$[\rho^2{\cal{E}} ]_\text{av} -\langle\rho^2 \rangle_c {\cal{E}} _c $ and
$[\rho {\cal{E}}]_\text{av}  - \rho_c {\cal{E}}_c $  can be written in form similar to that of the zeroth-order entropy.  These forms are
\begin{eqnarray}
[\rho^2{\cal{E}} ]_\text{av} -\langle\rho^2 \rangle_c {\cal{E}} _c  & = &
- \sum_{\gamma} \left[  \langle n^{(\gamma)}\rangle_c
\ln{ \langle n^{(\gamma)}\rangle_c}    \times  \right.
\nonumber \\  
&& \left.  \times
\left(1 - \sum_{\gamma'} \langle n^{(\gamma')}\rangle_c
\ln{ \langle n^{(\gamma')}\rangle_c}  \right)  \right] 
 \nonumber \\
\label{eq-rhosqepsavg}
\end{eqnarray}
and
\begin{eqnarray}
[\rho {\cal{E}}]_\text{av}  - \rho_c {\cal{E}}_c & = &
 -\sum_{\gamma}  q_\gamma  \langle n^{(\gamma)}\rangle_c
\ln{ \langle n^{(\gamma)}\rangle_c}   + \nonumber \\  
&& +\rho_c  \sum_{\gamma} \langle n^{(\gamma)}\rangle_c
\ln{ \langle n^{(\gamma)}\rangle_c}    \;.
 \nonumber \\
\label{eq-rhoepsavg}
\end{eqnarray}

\begin{figure}
\includegraphics[width=3in]{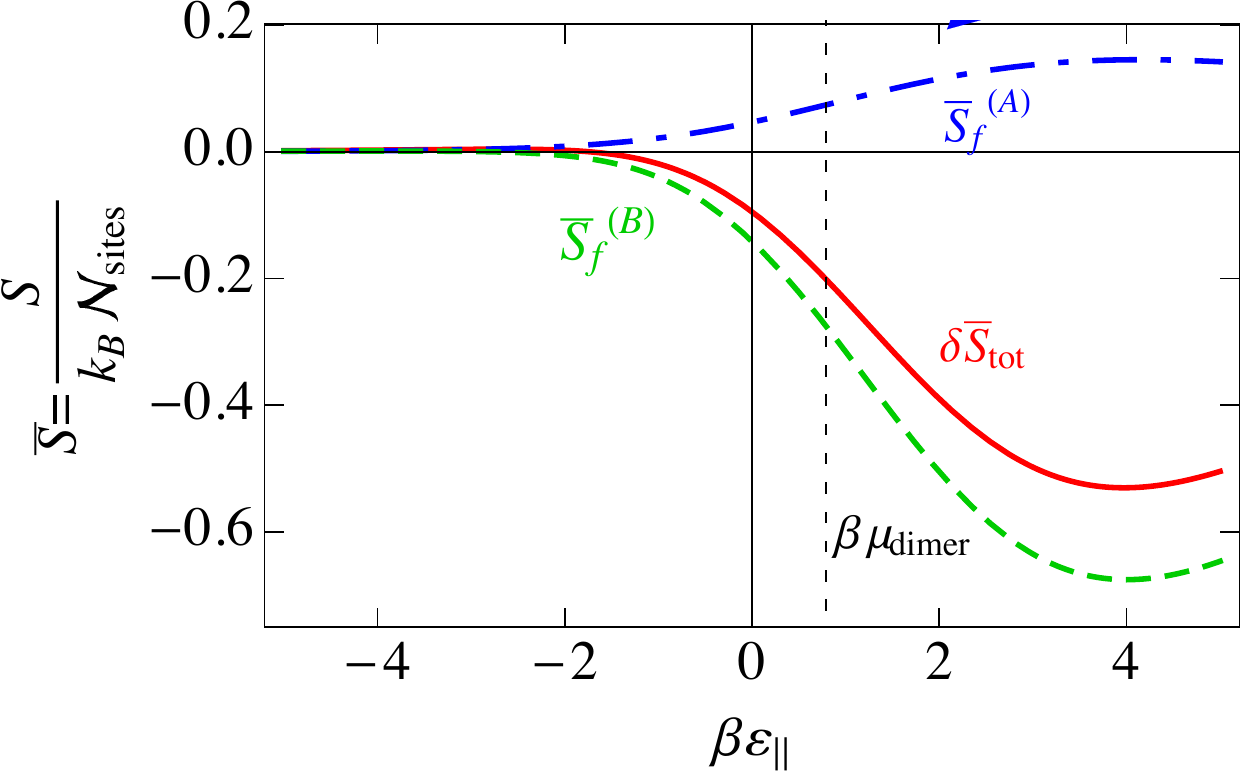}
\caption{The fluctuation contribution to the entropy $\bar{S}_f=\frac{S_f}{k_B N_\text{sites}}$ per site.   The two contributions, $\bar{S}_f^{(A)}$ and $\bar{S}_f^{(B)}$, to the total entropy are also shown, and they are defined similarly.}
\label{fig-EntropyS2Constituent}
\end{figure}

The fluctuation entropy contribution $S_f^{(A)}$ is easy to calculate, given the eigenvalues $ \lambda_{k,\sigma}$.  
Eqs.~(\ref{eq-lambdaDeriv})-(\ref{eq-rhoepsavg}) simply collect the results that allow us to calculate $S_f^{(B)}$.  
The results for the total fluctuation entropy $\delta S_{\text{tot}}$ and the two contributions to it are shown in Fig.~\ref{fig-EntropyS2Constituent} for $\varepsilon_\perp=\frac{1}{2}\varepsilon_\parallel$.
Like the mean field entropy, the fluctuation entropy becomes larger in magnitude as the binding becomes weak.  
However, the fluctuation contribution is actually negative in this region, suggesting that the fluctuations are leading to increasing order.  These corrections to the entropy are typically an order of magnitude larger than the difference between the mean field and noninteracting values of the entropy.

\section{Discussion}
\label{sec-discussion}

Entropy forms an important contribution to the free energies of many biological systems.  
Here we summarize the results for the entropies and particle densities for a lattice gas model representing the adsorption of molecules 
on double-stranded DNA.  
The entropies and particle densities are presented at three successive levels of approximation, first, for the noninteracting system
then at the mean-field level, and finally with the inclusion of fluctuations to one-loop order.
Those fluctuation corrections result from correlations induced by the electrostatic forces among the dimer molecules themselves and with the electrically
charged DNA substrate.

\begin{figure}
\includegraphics[width=3in]{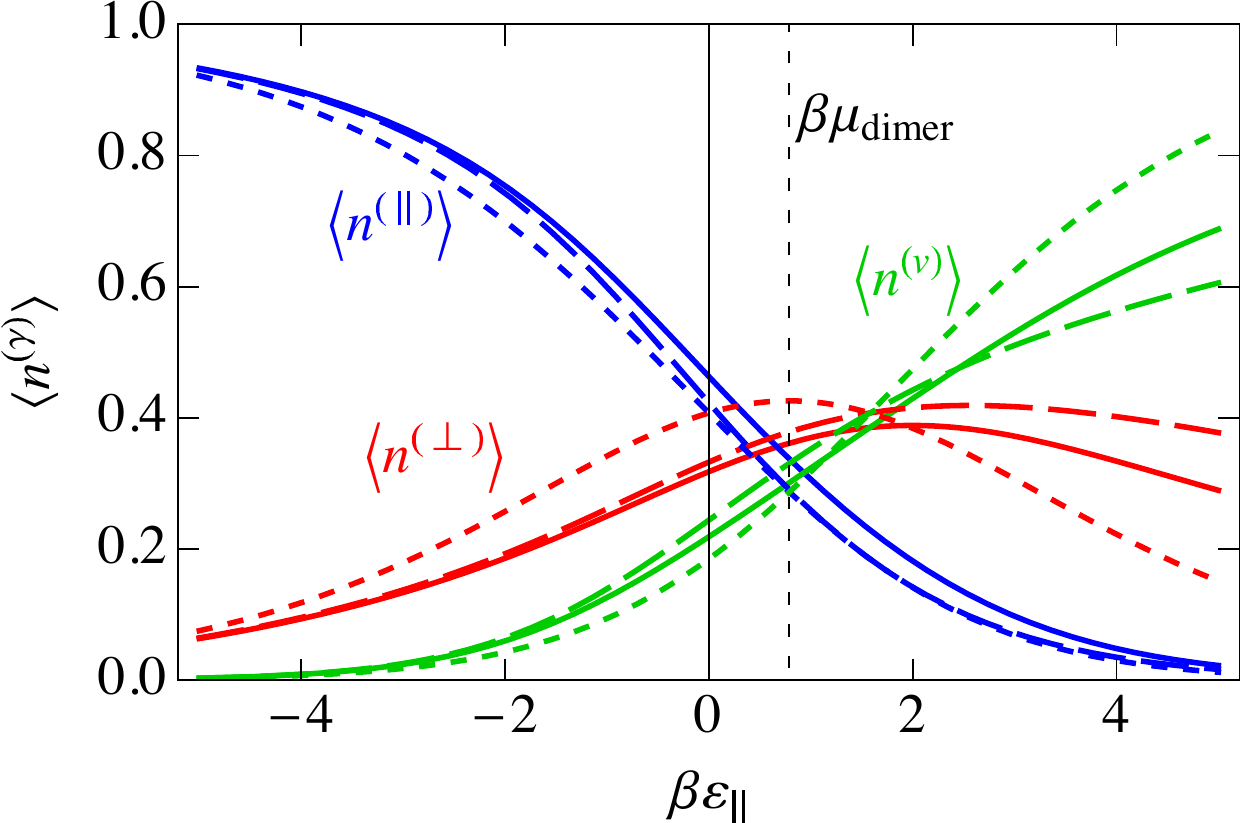}
\caption{The average occupation numbers in the noninteracting approximation ($\langle
n^{(\gamma)}\rangle_{NI}$), shown as dotted curves, the mean-field approximation ($\langle
n^{(\gamma)}\rangle_c$), shown as dashed curves, and the model including the lowest order 
corrections to mean field, shown as solid curves, plotted for $\gamma=\parallel$ (blue), $\perp$ (red), and $v$(green).  
Parameters used in the plot are $\beta \mu_\text{dimer}=0.79$ and
$\varepsilon_\parallel=2\varepsilon_\perp$.}
\label{fig-Compare3ModelOccupations}
\end{figure}

\begin{figure}
\includegraphics[width=3in]{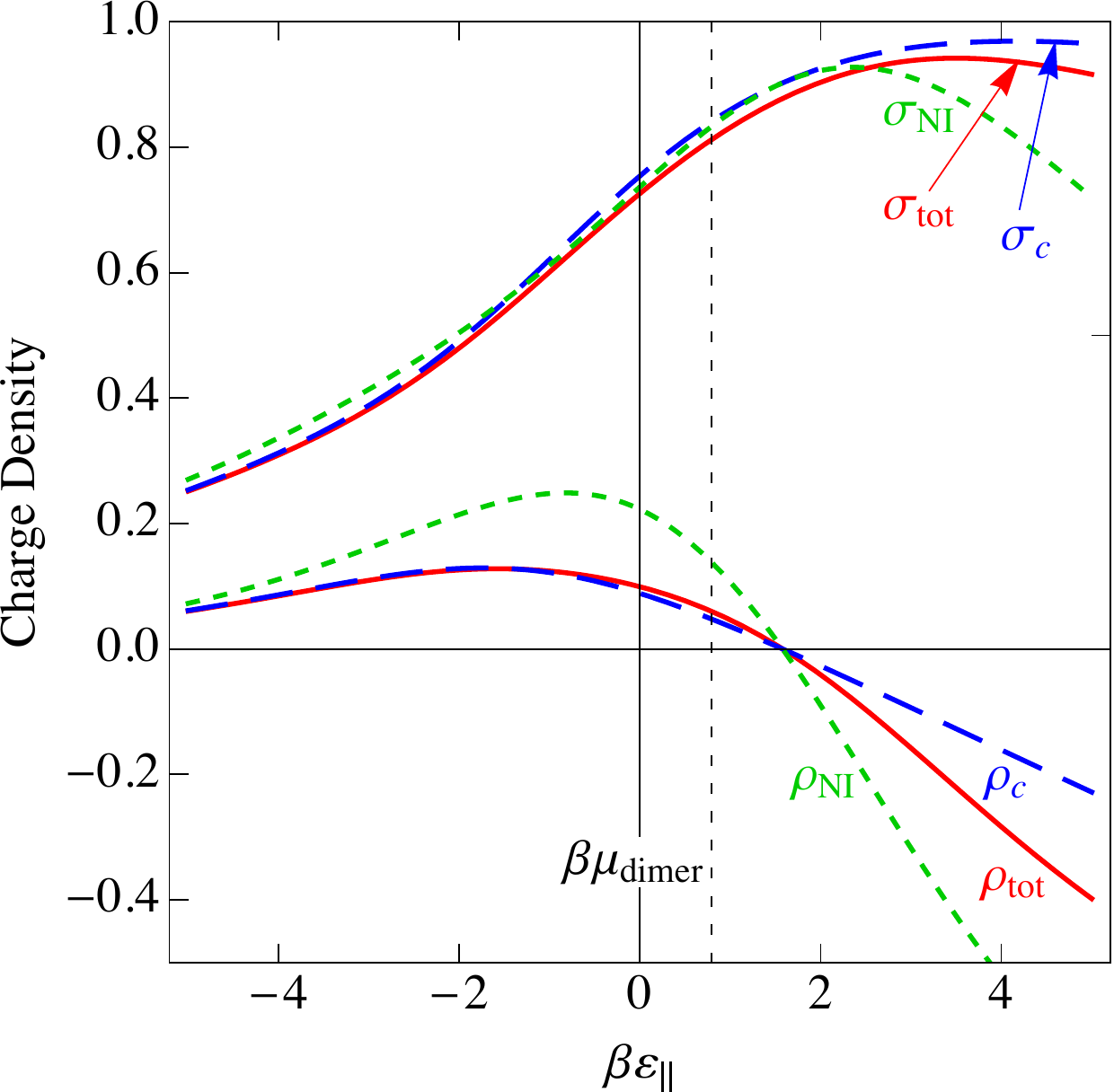}
\caption{The average charge per site in the noninteracting $\langle\rho\rangle_{NI}$ (green, short dashes), mean-field (blue, long dashes)  and inclusion of fluctuations (red, solid).  This plot assumes $\beta \mu_\text{dimer}=0.79$ and
$\varepsilon^{(\parallel)}=2\varepsilon^{(\perp)}$.}
\label{fig-RhoSigmaCompare}
\end{figure}

Perhaps the most transparent descriptors of the DNA system's behavior are the average
occupation numbers $\langle n^{(\parallel)} \rangle$, $\langle n^{(\perp)} \rangle$, and
$\langle n^{(v)} \rangle$ representing the thermal average of the numbers of parallel-adsorbed
dimers, perpendicular-adsorbed dimers, and vacancies, respectively, on each site.  These particle
densities are proportional to the probability that a given site would be found to be occupied by a particular
species.  These site-occupation numbers are shown in 
Fig.~\ref{fig-Compare3ModelOccupations} for the three levels of approximation.  The short-dashed curves are for the noninteracting level of approximation $ \langle n^{(\gamma)}\rangle_\text{NI}$,  
the long-dashed curves for the mean field level $\langle n^{(\gamma)}\rangle_c$, and the solid curves for the inclusion of fluctuations 
$\langle n^{(\gamma)}\rangle_{\text{tot}}$.  The corresponding charge densities and standard deviations of the charge density are shown in 
Fig.~\ref{fig-RhoSigmaCompare}.

The site occupancies exhibit similar qualitative behavior in all three levels of approximation.  Due to the energetics of binding $\varepsilon_\parallel = 2 \varepsilon_\perp$, parallel dimers occupy the greatest fraction of sites for all negative binding energies $(\varepsilon_\parallel,\varepsilon_\perp < 0)$,
followed the perpendicular dimers and then the vacancies, at least for the parameters used in these calculations.  
At weak binding ($\varepsilon_\parallel,\varepsilon_\perp\rightarrow 0$) in the noninteracting system, the difference between these energies becomes negligible, and both $n^{(\parallel)}$ and $n^{(\perp)}$ approach the
same value of $40\%$.  When the binding energy $\varepsilon_\parallel$ or $\varepsilon_\perp$ increases and reaches the chemical potential 
$\mu_\text{dimer}$ of the dimers in solution, they are repelled from the surface of the DNA and go into solution, leaving a ``naked'' lattice of negatively-charged vacancies, with the parallel dimers being more repelled than perpendicular ones.

In contrast to the case of noninteracting sites, the inclusion of interactions between sites gives a 
pronounced gap of about $25\%$ by which parallel adsorption exceeds perpendicular
adsorption in the weak binding limit.  This enhancement of parallel adsorption in the
mean-field approximation results from the inclusion of the charge on the perpendicular
dimers in an average or mean field sense, which penalizes the addition of positively-charged perpendicular dimers when the total charge on the DNA lattice is positive.  Fluctuations reduce this effect slightly.  Recall that the perpendicular dimers are positively charged because one end is not attached to a negative binding site, and this means that it is energetically unfavorable
to have a second one nearby.  Parallel binding, on the other hand, neutralizes the negative charge on the
sites, so that a DNA double strand completely covered in parallel-adsorbed dimers would be electrically neutral.  The decrease in perpendicular-dimer occupancy caused by this
electrostatic repulsion of like charges leads to an increase in sites occupied by vacancies. 
Vacancies are negatively charged and are energetically favored when the perpendicular dimers 
provide a positively-charged region. 
There may, in fact, be correlations resulting from the lower electrostatic energy of
a configuration in which sites alternate between perpendicular dimers and vacancies, i.e.,
between positively and negatively charged sites, in a Wigner-lattice-like state similar to those described in the
literature
\cite{ns02a,gns02,wmg00}.

The average charge $\rho_c$ per site depends only on the perpendicular dimers and vacancies since the parallel dimers neutralize sites, so that the charge density is written simply as
\begin{equation}
\rho = \sum_{\gamma}q_\gamma \langle n^{(\gamma)}\rangle = 
\langle n^{(\perp)}\rangle - \langle n^{(v)}\rangle      \;  ,
\end{equation}
and this connection can be seen in the plots of charge density $\rho$ in
Fig.~\ref{fig-RhoSigmaCompare}, when compared with 
Fig.~\ref{fig-Compare3ModelOccupations}.  For all approximations, the average charge is positive at negative 
and slightly positive binding energies, indicating charge
inversion.  This is a result of the
occupation numbers shown in Fig.~\ref{fig-Compare3ModelOccupations}, because
parallel binding, which neutralizes the charge, increasingly dominates in the
strong binding limit ($\varepsilon_\parallel=2\varepsilon_\perp \ll 0$).  Similarly, the decrease in the magnitude of the charge inversion from the
noninteracting to the mean-field approximations can be explained by the enhancement of parallel
adsorption and reduction of perpendicular adsorption due to like-charge repulsion.
The fluctuation corrections to the site occupancies are not large when compared with the mean-field values.
They only seem significant at positive binding energies where the site occupancies are shifted in the direction of the noninteracting system.  Of course, the site occupancies are single-particle properties and do not measure spatial correlations.

\begin{figure}
\includegraphics[width=3in]{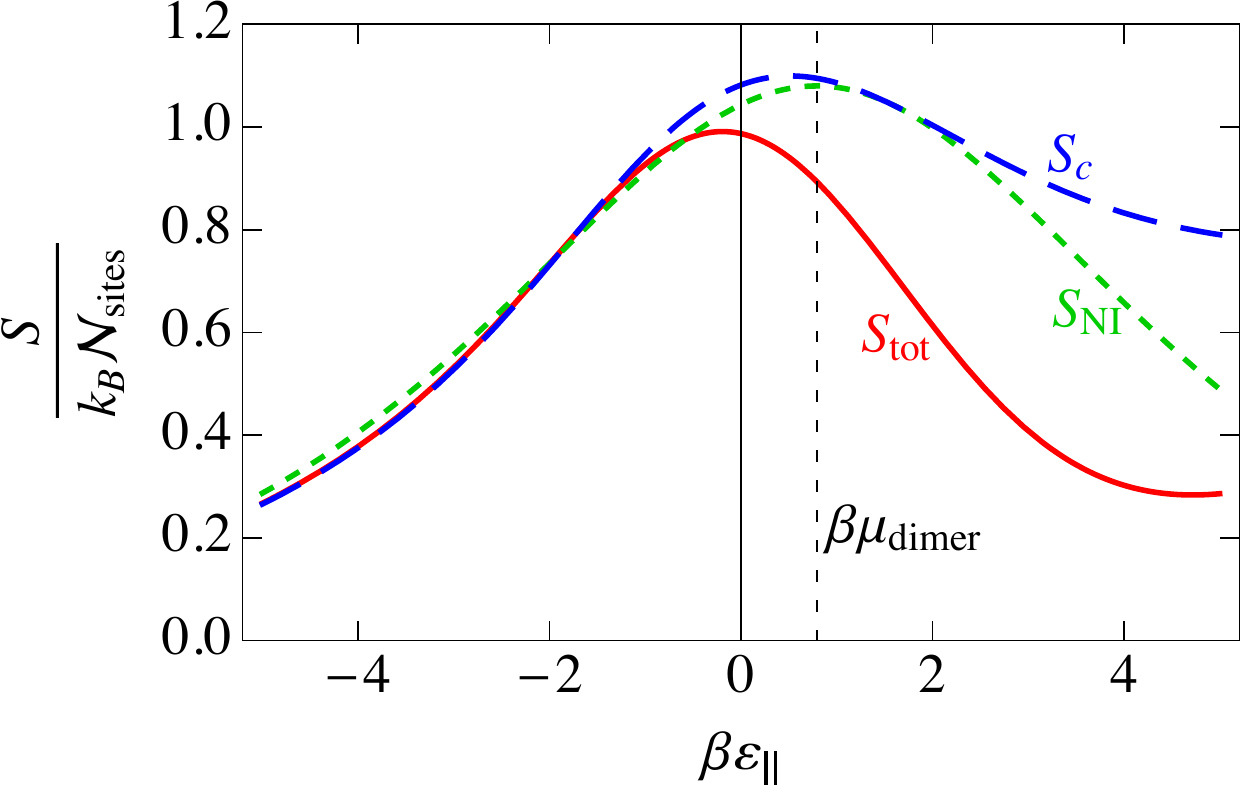}
\caption{The entropy $S$ per site, in units of the Boltzmann constant $k_B$, in the
noninteracting (green, short dashes), mean-field (blue, long dashes), and and inclusion of fluctuations (red, solid) approximations.
These curves assume $\beta \mu_\text{dimer}=0.79$ and
$\varepsilon_\parallel=2\varepsilon_\perp$.}
\label{fig-Entropies}
\end{figure}

The behavior of the entropy is shown in  Fig.~\ref{fig-Entropies},
in which  the entropy per site is plotted versus $\beta \varepsilon_\parallel$ for all three levels of approximation.  The  three curves are practically indistinguishable until the binding energy becomes quite weak, indicating little effect of
correlations induced by electrostatic repulsion on the entropy in that region. More pronounced 
differences are seen in the region of 
$\varepsilon_\parallel \approx  \mu_\text{dimer}$ and more positive binding energies ($\varepsilon_\parallel=2\varepsilon_\perp \gg 0$), where the 
system becomes more disordered, and where the dimers tend to leave the surface and enter the solution.
The fluctuation corrections to the entropy are pronounced in that region, and tend to reduce the entropy,
implying that the electrostatic correlations tend to increase the order in the lattice gas model.  Surprisingly, this decrease in entropy at large positive $\varepsilon_\parallel$ due to fluctuations is so significant that it {\textit{overcompensates}} for the entropy generated at the mean field level, leading to a total entropy $S_{\text{tot}}$ that is even lower than in the noninteracting case $S_{\text{NI}}$.

The form of the noninteracting entropy comes from the lattice gas model, in which parallel
adsorption is treated as ``local," only occupying one site.  When the geometrical blocking
effect is included, in which parallel-adsorbed dimers occupy two sites, we would expect the
entropy to differ significantly from the $n \ln n$ result even at the noninteracting level.  It
is unclear how large a contribution to the full entropy the blocking effect provides, but it is
certainly lower for dimers than for longer polymers, for which the nonlocality is greater. 
Because of this, it will be important in future work to develop a way of incorporating the
blocking effect into these types of field-theoretic models.  

Dimers are certainly the simplest example of polymers, although biological polymers are
typically considerably longer.   There is extensive work done on the
dimer model, beginning with the work of Fisher
\cite{mef61,fs63}.  Some of this work has been motivated by the analogy between
dimer models and quantum spin systems  \cite{mef66}. In this context,
Fisher has derived expressions for the contributions of hard-core crowding effects to the
partition function, including the geometrical blocking effect.  Fisher's results include an
expression of the noninteracting partition function for dimers on a 1D linear lattice.
In a future publication, we plan to adapt Fisher's approach to study the statistical
consequences these blocking effects\cite{jcb17}.

In addition, it has been suggested that methods can be adapted from analogous problems in particle physics\cite{ac03}
and condensed matter physics\cite{ss02}.
This involves using numerical codes like those of Adams
and Chandrasekharan\cite{ac03} that are used for simulations in lattice 
quantum chromodynamics.

Modeling the behavior of DNA and other biologically active polymeric molecules 
under physiological conditions is an area in which quantitative calculation is particularly
difficult.  This is because it is necessary to accommodate simultaneously the influences of
geometry, electrostatic forces, and charge correlations.  All of these
play roles of varying significance in determining the properties of these molecules.  
Despite these
challenges, considerable progress has been made in extending our physical understanding of these
systems, and some surprising new physics, such as charge inversion and condensation, have 
been discovered in the process.  

The work presented here focuses on electrostatic effects and especially on the role they play in the entropy, with the overall goal of determining the entropy as a contribution to the free energy and as a driving force in biochemical reactions.  For the DNA-based model that we consider, we find that, particularly in the low-coverage regime, a uniform mean field theory gives changes $\sim 20\%$ from the usual noninteracting entropy term.  We also find, though, that the fluctuation contributions are both considerably larger and in the opposite direction.  This establishes that the fluctuations are significant, and the opposite sign of their contribution suggests that they lead to additional order in the system.

\appendix

\section{A Gaussian Integral Identity}
\label{sec-GaussianIntegralIdentity}

In this appendix, we prove the integral identity that we use in the derivation of the partition function $Z_G$ for the interacting model.  The identity is
\begin{equation}
I = \int \sqrt{\frac{\beta \lambda}{-2 \pi}}
d\tilde\Delta \ 
e^{\frac{\beta}{2}\lambda  \tilde\Delta^2  - \beta \lambda  \tilde\Delta \tilde\rho}
e^{\frac{\beta}{2}\lambda \tilde\rho^2}  =  1
\label{eq-GaussianIntegralIdentity}
\end{equation}
The integration path is over the real axis from $-\infty$ to $\infty$ when $\lambda<0$ and over the imaginary axis from $-i \infty$ to $i \infty$ when $\lambda>0$. 
This integral can be evaluated by noticing that the exponent is a factor times a perfect square,
\begin{equation}
\frac{\beta}{2}\lambda  \tilde\Delta^2  - \beta \lambda  \tilde\Delta \tilde\rho
+\frac{\beta}{2}\lambda \tilde\rho^2  =
\frac{\beta}{2}\lambda(\tilde\Delta - \tilde\rho)^2
      \; .
\label{eq-GaussianCompleteSquare}
\end{equation}
Substituting, the integral becomes
\begin{equation}
I=\int \sqrt{\frac{\beta \lambda}{-2 \pi}}
 d\tilde\Delta e^{\frac{\beta}{2}\lambda(\tilde\Delta - \tilde\rho)^2}
    \; .
\label{eq-GaussianIntegralAfterCompleteSquare}
\end{equation}
Now it is clear why the path of integration must be chosen differently for positive versus negative $\lambda$, since it must be chosen such that the integral converges.  If $\lambda=0$, the identity fails, but in that case it is clearly unnecessary.  If $\lambda<0$, we can write $\lambda=-|\lambda|$, and the integral converges for 
$\tilde\Delta$ ranging over the real axis from $-\infty$ to $\infty$.  Substituting 
$y=\tilde\Delta - \tilde\rho$, with $y$ ranging from $-\infty$ to $\infty$ the integral can be evaluated as
\begin{eqnarray}
I &=&
 \int_{-\infty}^{\infty} \sqrt{\frac{\beta |\lambda|}{2 \pi}} dy \, e^{-\frac{\beta}{2}|\lambda| y^2 }
=\sqrt{\frac{\beta |\lambda|}{2 \pi}} \sqrt{\frac{2 \pi}{ \beta |\lambda|}} 
\nonumber \\ &=& 
1 
\; , \; \; (\lambda<0)   \; .
\label{eq-GaussianIntegralNegativelambda}
\end{eqnarray}

For $\lambda>0$, the situation is slightly more complicated.  In that case, in order to have a decaying Gaussian in the integrand, we must integrate $\tilde\Delta$ over the imaginary axis from $-i \infty$ to $i \infty$.  This is accomplished by substituting 
$y=-i\left(\tilde\Delta -\tilde\rho\right)$, for $y$ ranging from $-\infty$ to $\infty$.  In this case, we have  
\begin{eqnarray}
I &=&
i \int_{-\infty}^{\infty} \sqrt{\frac{\beta \lambda}{-2 \pi}} dy \, e^{-\frac{\beta}{2}\lambda y^2 }
= i \sqrt{\frac{\beta \lambda}{-2 \pi}} \sqrt{\frac{2 \pi}{ \beta \lambda}} 
\nonumber \\ &=& 
1 ,
\label{eq-GaussianIntegralPositivelambda}
\end{eqnarray}
where the $i$ the numerator cancels the $\sqrt{-1}$ in the denominator.  This proves the identity for positive $\lambda$.  Therefore, we have seen that this identity works for positive and negative values of $\lambda$, and, although it does not work $\lambda=0$, it is not needed in that case.

\section{Chemical Potential of Free Dimers}
\label{sec-ChemicalPotentialDimers}

\begin{figure}
\includegraphics[width=1.8in]{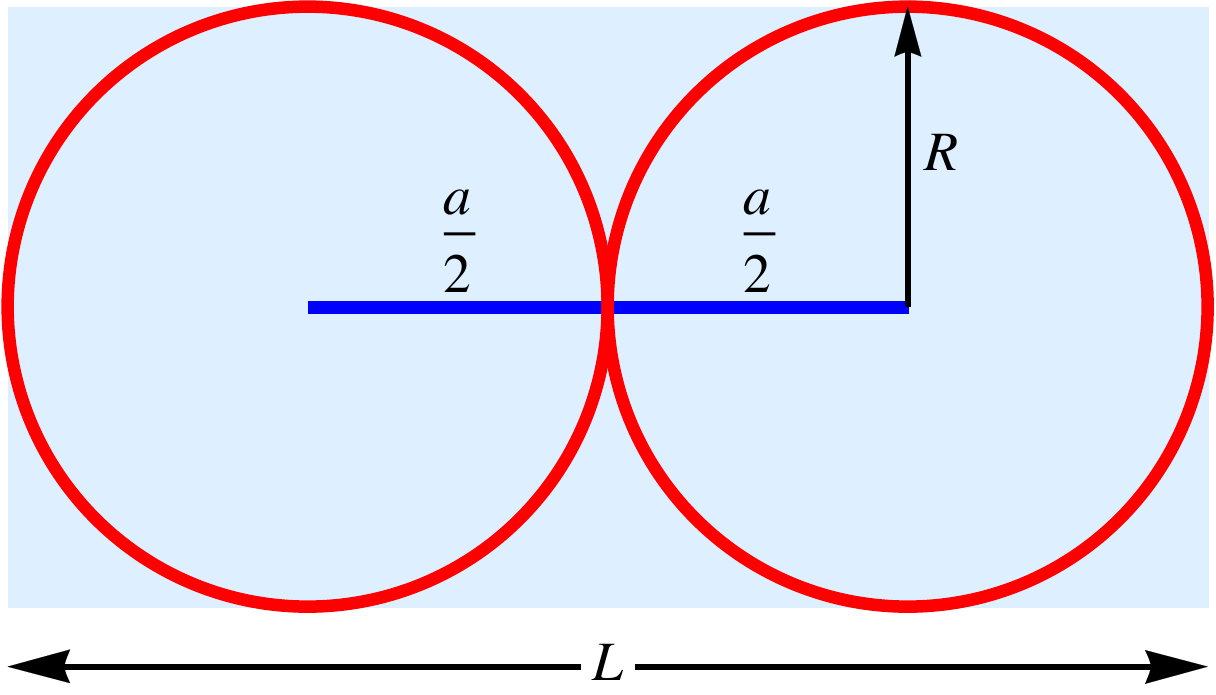}
\caption{A free dimer approximated by a cylinder, shown in cross section, where the dimer is designed to attach to two phosphates separated by lattice spacing $a$.  The length of the cylinder is $L=2a$ and the radius is $R=\frac{a}{2}$, making the ratio $\xi=\frac{L}{R}=4$.}
\label{fig-DimerCylinder}
\end{figure}

The chemical potential of a solution of free dimers is the derivative of the Helmholtz free energy of the dimers 
$F_{\text{dimer}}$ with respect to the number of dimers $N_{\text{dimer}}$, with temperature and volume held constant,
\begin{equation}
\mu_{\text{dimer}} = \left(\frac{\partial F_{\text{dimer}}}{\partial N_{\text{dimer}}}
 \right)_{T,V}
    \; .
\label{eq-ChemicalPotentialDef}
\end{equation}
Assuming that $F_{\text{dimer}}$ is due only to electrostatic interactions, in a dilute solution the free energy is just the number of dimers times the electrostatic self energy $E_{\text{dimer}}$ of a single dimer.  Then the chemical potential becomes  
\begin{equation}
\mu_{\text{dimer}} = \left(\frac{\partial (N_{\text{dimer}} E_{\text{dimer}})}
{\partial N_{\text{dimer}}} \right)_{T,V}  = E_{\text{dimer}} 
    \; ,
\label{eq-ChemicalPotentialElectrostatic}
\end{equation}
and so we need the electrostatic energy of a single dimer.  Suppose we model a dimer as a cylinder of cross-section radius $R$ and length $L$ of total charge $Q$, as shown in Fig.~\ref{fig-DimerCylinder}.  An exact solution to this problem has been provided by Cifta\cite{oc12}, and the derivation proceeds as follows. The volume charge density of the cylinder is given by
\begin{equation}
\rho_0  =  \frac{Q}{\pi R^2 L}  \; ,
\label{eq-DimerChargeDensity}
\end{equation}
and the Coulomb self-energy is given by
\begin{equation}
E_s  =  \frac{1}{4 \pi \epsilon} \left( \frac{\rho_0^2}{2}  \right)
\int_V d^3 r_1 \int_V d^3 r_2  \frac{1}{|\vec{r}_1 - \vec{r}_2|}
 \; ,
\label{eq-CoulombSelfEnergy}
\end{equation}
where $\epsilon$ is the dielectric permeability of the medium surrounding the cylinder.  In evaluating the integrals, the length to radius ratio is defined as
\begin{equation}
\xi  =  \frac{L}{R}  \; ,
\label{eq-LengthRadiusRatio}
\end{equation}
and the exact result, as calculated by Cifta\cite{oc12}, is given by
\begin{eqnarray}
E_s(\xi) &=&  \frac{1}{4 \pi \epsilon}  \left(  \frac{4Q^2}{R\xi^2} \right)
 \left\{ \frac{\xi}{16} [4 \ln(2\xi) - 3]  +\frac{32}{45 \pi}  \right.
\nonumber \\  &&
\left.  - \left(\frac{1}{16\xi}\right) {_4F_3}(\frac{1}{2},1,1,\frac{5}{2};2,3,4;-\frac{4}{\xi^2}) 
\right\} \; ,  \nonumber \\
\label{eq-ExactSelfEnergyCylinder}
\end{eqnarray}
where the function ${_pF_q}(a_1,a_2,\ldots,a_p;b_1,b_2,\ldots,b_p;z)$ is the generalized hypergeometric function whose series expansion is
\begin{eqnarray}
&& {_pF_q}(a_1,a_2,\ldots,a_p;b_1,b_2,\ldots,b_p;z) =
\nonumber \\  && \; \; \; \; \; \;
=\sum_{n=0}^\infty \frac{(a_1)_n(a_2)_n\ldots(a_p)_n}
{(b_1)_n(b_2)_n\ldots(b_q)_n} \frac{z^n}{n!} \; ,
\label{eq-HypergeometricSeriesExpansion}
\end{eqnarray}
where the variable $z$ can be either real or complex and $(c)_n$ is the Pochhammer symbol defined as
\begin{equation}
(c)_n  =  \frac{\Gamma(c+n)}{\Gamma(c)}  \; , \; n=1,2,\ldots    \; \;    . 
\label{eq-Pochhammer }
\end{equation}
Here $\Gamma(z)$ is the gamma function and $(c)_0=1$.  For a long, thin rod, the logarithm term would dominate, but a dimer approximated by a cylinder, pictured in Fig.~\ref{fig-DimerCylinder}, does not fall into this regime.  

As shown in the picture, for the dimer, the radius of the dimer is $R=2a$, where $a$ is the lattice spacing along a DNA helical chain.  The length of the dimer is $L=2a$, and so the length ratio is $\xi=\frac{L}{R}=4$. The charge is $Q=2e$, where $e$ is the magnitude of the charge of an electron, and the chemical potential becomes
\begin{eqnarray}
\mu_{\text{dimer}} &=& E_{\text{dimer}}  = 
\nonumber \\  &=&
 \frac{1}{4 \pi \epsilon}  \left(  \frac{e^2}{8a} \right)
 \left\{ \frac{1}{4} [4 \ln(8) - 3]  +\frac{32}{45 \pi}  \right.
\nonumber \\  &&
\left.  - \left(\frac{1}{128}\right) {_4F_3}(\frac{1}{2},1,1,\frac{5}{2};2,3,4;-\frac{1}{4}) 
\right\}   \; .
\label{eq-SelfEnergyEqDimer}
\end{eqnarray}
For the parameters given in the text, the chemical potential for the dimer, in units of $\beta=\frac{1}{k_B T}$ at $T=300K$ is given by $\beta \mu_{\text{dimer}} =  0.79$.  It is interesting to note that approximately the same value was obtained by Nguyen and Shklovskii\cite{ns02d} and Bishop and McMullen\cite{bm06} assuming that the dimer is an infinitely thin rod of length $L$ but that there were upper and lower screening cutoffs.  This exact solution avoids the use of those cutoffs.


\section*{Acknowledgement}

We are grateful to Dr. Kevin R. Ward, Director of Research, Department of
Emergency Medicine, University of Michigan, for introducing us to
the role of charge in biological systems and to the practical consequences
that could be achieved by controlling the charge distribution on biological
molecules.


%

\end{document}